\documentclass[aps, prl, twocolumn, 10pt, superscriptaddress, longbibliography]{revtex4-1}    
\usepackage{amssymb, amsmath, amsthm}
\usepackage{xcolor}
\usepackage{graphicx}
\usepackage{hyperref}
\usepackage{comment}
\usepackage{physics}

\usepackage{footnote}
\newcommand{\blue}[1]{{\color{blue} #1}}
 
\graphicspath{{Figures/}}

\newcommand{\bk}{{\bf k}}
\newcommand{\br}{{\bf r}}

\usepackage{mathtools}

\let\norm\undefined
\DeclarePairedDelimiter\norm{\lVert}{\rVert}

\newcommand{\mytitle}{Asymptotic Quantum Many-Body Scars}
 
\begin{document}

\title{\mytitle}

\author{Lorenzo Gotta}
\email{lorenzo.gotta@unige.ch}
\affiliation{Department of Quantum Matter Physics, University of Geneva, 24 Quai Ernest-Ansermet, 1211 Geneva, Switzerland}
\affiliation{Universit\'e Paris-Saclay, CNRS, LPTMS, 91405, Orsay, France}

\author{Sanjay Moudgalya}
\email{sanjaym@caltech.edu}
\affiliation{Department of Physics and Institute for Quantum Information and Matter,
California Institute of Technology, Pasadena, California 91125, USA}
\affiliation{Walter Burke Institute for Theoretical Physics, California Institute of Technology, Pasadena, California 91125, USA}

\author{Leonardo Mazza}
\email{leonardo.mazza@universite-paris-saclay.fr}
\affiliation{Universit\'e Paris-Saclay, CNRS, LPTMS, 91405, Orsay, France}

\date{\today}

\begin{abstract} 
We consider a quantum lattice spin model featuring exact quasiparticle towers of eigenstates with low entanglement at finite size,  known as quantum many-body scars (QMBS).
We show that the states in the neighboring part of the energy spectrum can be superposed to construct entire families of low-entanglement states whose energy variance decreases asymptotically to zero as the lattice size is increased.
As a consequence, they have a relaxation time that diverges in the thermodynamic limit, and therefore exhibit the typical behavior of exact QMBS although they are not exact eigenstates of the Hamiltonian for any finite size.
We refer to such states as \textit{asymptotic} QMBS.
These states are orthogonal to any exact QMBS at any finite size, and their existence shows that the presence of an exact QMBS leaves important signatures of non-thermalness in the rest of the spectrum; therefore, QMBS-like phenomena can hide in what is typically considered the thermal part of the spectrum.
We support our study using numerical simulations in the spin-1 XY model, a paradigmatic model for QMBS, and we conclude by presenting a weak perturbation of the model that destroys the exact QMBS while keeping the asymptotic QMBS.
\end{abstract}

\maketitle

\paragraph{\textbf{Introduction} ---}
Quantum Many-Body Scars (QMBS)~\cite{serbyn2020review, papic2021review,  moudgalya2021review, chandran2022review} in non-integrable quantum lattice models of any dimension are one of the paradigms for the weak violation of the Eigenstate Thermalization Hypothesis (ETH)~\cite{deutsch1991quantum, srednicki1994chaos}, according to which all local properties of energy eigenstates in the middle of the spectra of non-integrable models coincide with those of a thermal Gibbs density matrix at a suitable temperature~\cite{rigol2008thermalization, polkovnikov2011colloquium, d2016quantum, mori2018thermalization}.
QMBS are isolated energy eigenstates that are outliers in many respects, e.g., in the expectation value of a local observable or in the entanglement entropy.
Numerous instances of lattice models featuring exact towers of QMBS at finite size have been discovered~\cite{moudgalya2018exact, moudgalya2018entanglement, mark2020unified, schecter2019weak, mark2020unified, wildeboer2022quantum, yang1989eta, moudgalya2020eta, mark2020eta, pakrouski2020many, pakrouski2021group, yoshida2022exact, Gotta_2022, nakagawa2022exact}.
%
Most of these results have also been understood via unified frameworks or systematic construction recipes~\cite{shiraishi2017systematic, mark2020unified, moudgalya2020large, moudgalya2020eta,  pakrouski2021group, ren2020quasisymmetry, odea2020from,rozon2022constructing, moudgalya2022exhaustive, rozon2023broken, moudgalya2021review}.
A question that has been less explored is whether the presence of a finite-size QMBS affects the properties of the rest of the spectrum.
Ref.~\cite{lin2020slow} pointed out the existence of low-entanglement states in the PXP model which exhibit slow relaxation even though they are orthogonal to the known exact QMBS: 
%
the energy variance of such states is independent of system size and thus 
their fidelity relaxation time does not decrease~\cite{lin2019exact}.
This is a remarkable phenomenology to be contrasted with that of short-range correlated states, whose energy variance grows with  system size, whereas the fidelity relaxation time decreases.
Are there even more drastic examples of slowly relaxing states~\cite{banuls2020entanglement}, 
for instance with an energy variance decreasing with system size, which would lead to a relaxation time that \textit{diverges polynomially} in the thermodynamic limit (TL)?  
%
%
Slow relaxation of hydrodynamic origin is ubiquitous in systems with continuous symmetries, where it occurs at a diverging timescale known as the \textit{Thouless time}~\cite{thouless1977maximum, chan2018spectral, schiulaz2019thouless, dymarsky2022bound}, and is related to  diffusion or subdiffusion~\cite{chaikin1995principles, mukerjee2006statistical, lux2014hydrodynamic, gromov2020fracton, feldmeier2020anomalous, moudgalya2020spectral}. 
The interpretation of QMBS as an unconventional non-local symmetry~\cite{moudgalya2022exhaustive, buca2023unified} motivates the search for such slow relaxation.
%
Long-lived quasiparticles, e.g.~the phonons of a superfluid with Beliaev decay~\cite{stringari}, also induce slow relaxation.
QMBS are associated to quasiparticles with specific momenta and infinite lifetime~\cite{chandran2022review}, hence it is natural to look for long-lived quasiparticles at neighboring momenta.
In this letter we address these questions by considering the spin-1 XY model featuring exact QMBS at any finite size~\cite{schecter2019weak} and show that it is possible to construct slowly-relaxing low-entanglement initial states that exhibit QMBS-like features, but nevertheless are orthogonal to the exact QMBS.
They have an energy variance that goes to zero in the TL and asymptotically display the typical dynamical phenomenology of a QMBS, i.e.~the lack of thermalization; hence we refer to such initial states as \textit{asymptotic} QMBS.
Our work widens the range of initial states that qualitatively exhibit a non-thermalizing phenomenology and motivates the search for non-thermal features in regions of the spectrum where entanglement signatures do not make them evident.
\paragraph{\textbf{The model and the exact QMBS} ---}
We consider a one-dimensional spin-1 chain of length $L$ even, 
and consider a spin-1 XY model with external magnetic field and axial anisotropy:
\begin{align}
H = &J \sum\limits_{j}\left( S^x_{j} S^x_{j+1} + S^y_{j} S^y_{j+1} \right) + h \sum_{j} S_{j}^{z} \nonumber \\
&+ D \sum_j \left( S_j^{z} \right)^2 +J_3 \sum_{j} \left(S_{ j}^x S^x_{j+3}+S_{ j}^y S^y_{j+3}\right),
\label{Eq:Ham:1D}
\end{align}
where $S_{j}^\alpha$, with $\alpha = x,y,z$, are the spin-1 operators on site $j$. 
We use open boundary conditions (OBC) for the numerical simulations and periodic boundary conditions (PBC) for some of the analytical results. 
This model with OBC has been numerically shown to be non-integrable; the last term 
breaks a hidden non-local symmetry~\cite{kitazawa2003su2, schecter2019weak, chandran2022review}.
The Hamiltonian in Eq.~\eqref{Eq:Ham:1D} exhibits QMBS for any finite value of $L$~\cite{schecter2019weak}.
In order to see that, we define the fully-polarised state $\ket{\Downarrow} = \ket{-- \cdots --}$ with all spins in the eigenstate of $S^z_{j}$ with eigenvalue $-1$, and the operator
\begin{equation}
 J^+_{k} = \frac{1}{2} \sum_{j = 1}^L e^{i kj} \left( S^+_{j}\right)^2.
 \label{eq:Jplusdefn}
\end{equation}
The scar states read:
\begin{equation}
 \ket{n,  \pi} = \frac{1}{\sqrt{N_{n,  \pi}}}\left( J^+_{\pi} \right)^n \ket{\Downarrow},
 \label{Eq:Scar:Old:1D}
\end{equation}
where $N_{n, \pi}$ is a normalisation constant.
The state satisfies the energy eigenvalue equation $H \ket{n,  \pi}=\left(-Lh + 2nh+LD  \right)\ket{n, \pi}$ and for generic values of $h$ and $D$ it lies in the middle of the Hamiltonian spectrum. 
Its existence is related to quantum interference effects, similar to those that are responsible for the existence of $\eta$-pairing states in the Hubbard model~\cite{yang1989eta}.
Moreover, it is possible to consider the reduced density matrix $\rho_{A, n, \pi}$ of $\ket{n,\pi}$ defined on half the system  (conventionally, the region $A$ is  $1 \leq j < L/2 $), and to compute its entanglement entropy, $S_{n,\pi} = - \text{tr}[\rho_{A,n,\pi} \log \rho_{A,n,\pi}]$. 
The explicit calculation has been done in Ref.~\cite{schecter2019weak}, and it shows that it scales as $ \log L$, displaying a mild logarithmic violation of an entanglement area law, see Supplementary Materials (SM)~\cite{SM} and Ref.~\cite{vafek2017entanglement} for details.  
QMBS are easily found numerically by plotting the entanglement entropy $S_{E_i}$ of $\rho_{A, E_i}$, the reduced density matrix of the eigenstate $\ket{E_i}$, as a function of energy.
Indeed, almost all the eigenstates appear to satisfy the ETH and are characterised by an $S_{E_i}$ that is only a function of the energy $E_i$; they have a higher amount of entanglement than the QMBS states, which indeed violate ETH. 

\paragraph{\textbf{A family of states obtained by deforming the exact QMBS} ---}
We now consider other initial states for the dynamics of the model in Eq.~\eqref{Eq:Ham:1D}; they read as follows:
\begin{equation}
 \ket{n,  k} = \frac{1}{\sqrt{N_{n, k}}} J^+_{k} \left( J^+_{ \pi}\right)^{n-1} \ket{\Downarrow}, 
 \label{Eq:Scar:New}
\end{equation}
where $N_{n,k}$ is a normalisation constant, which coincide with the exact QMBS in Eq.~\eqref{Eq:Scar:Old:1D}.
%
When $ k \neq  \pi$ and is an integer multiple of $\frac{2\pi}{L}$, they are \textit{orthogonal} to the exact QMBS: the relation
$\langle n,  k \ket{n', \pi} = \delta_{n,n'} \delta_{ k,  \pi}$ for any $1 \leq n, n' \leq L-1$ is proved in the SM~\cite{SM}.
Models where such classes of multimagnon states are exact eigenstates have been studied in \cite{tang2021multimagnon}, however for $k \neq \pi$ these are \textit{not} eigenstates of the spin-1 XY model.
It is easy to show that the average energy of these states does not depend on $k$ and reads $\bra{n,k} H \ket{n,  k}= -Lh + 2nh+LD$~\cite{SM}.
Furthermore, the entanglement of the states in Eq.~\eqref{Eq:Scar:New} scales with system size as a sub-volume law.
For a quick proof, since $\ket{n,k} \propto J^+_k \ket{n-1, \pi}$, we note that $J^+_k$ can be straightforwardly expressed as a Matrix Product Operator (MPO) of bond dimension $\chi = 2$~\cite{crosswhite2008fsa, motruk2016density, moudgalya2021review}, hence the half-subsystem entanglement entropies of $\ket{n-1,\pi}$ and $\ket{n,k}$ can differ at most of an additive term $\log 2$.
In other words, since the operator $J_k^+$ can be split in two terms, one acting on $j<L/2$ and one on $j \geq L/2$, it is possible to show~\cite{SM} that the total number of Schmidt states in $\ket{n, k}$ is at most twice than that in $\ket{n-1,\pi}$.
To further characterise the states in Eq.~\eqref{Eq:Scar:New}, we compute the variance of the energy $\Delta H^2$ under the Hamiltonian $H$ in PBC, and as we show in the SM~\cite{SM}, we obtain:
\begin{equation}
 \Delta H^2 =   4\left[J^2 \cos^2 \left(\frac{k}{2} \right) + J_3^2 \cos^2 \left( \frac{3k}{2} \right)\right].
\label{eq:variance}
\end{equation}
%
Among the states defined in Eq.~\eqref{Eq:Scar:New}, the $\ket{n, \pi}$ are the only eigenstates of the Hamiltonian, because $\Delta H^2=0$ only for $k=\pi$.
When $k \neq \pi$, $\ket{n,k}$ must be a linear superposition of the energy eigenstates of $H$, which are mostly in a window centered around the same energy of $\ket{n,\pi}$ and in a width of about $\Delta H$.
When $k \neq \pi$ is chosen to be an integer multiple of $\frac{2\pi}{L}$, $\ket{n,\pi}$ is not part of this set of states due to orthogonality.
Since $\ket{n, \pi}$ numerically appear to be the only exact QMBS of $H$~\cite{schecter2019weak}, we conclude that such states $\ket{n,k}$ must be a linear superposition of ``thermal" eigenstates, i.e., those that are typically said to satisfy ETH, having an entanglement entropy and expectation values of local observables that are smooth functions of energy.

\begin{figure}[t]
\includegraphics[width=\columnwidth]{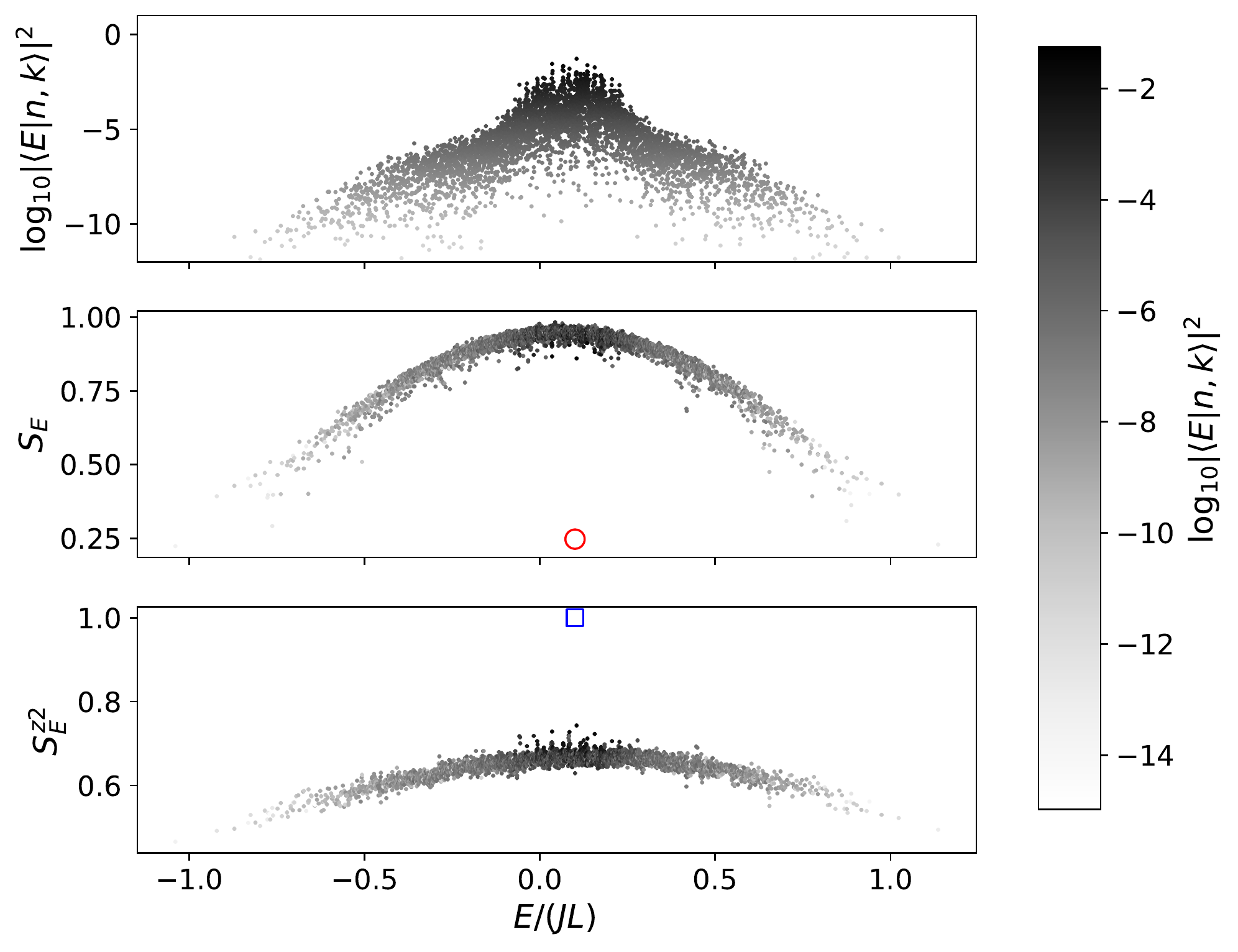}
 \caption{Top: Squared overlap of $\ket{n,k}$ for $n=L/2$ and $k = \pi - \frac{2\pi}{L}$ with the eigenstates $\ket{E_i}$ of Hamiltonian~\eqref{Eq:Ham:1D} with zero magnetisation, $S_z=0$; the parameters of the simulation are $\{J, h, D, J_3 \} = \{1, 0, 0.1, 0.1\}$ and $L=10$. 
 The information on $|\langle E_i\ket{n, k}|^2$ is also encoded in the color code of the marker of all panels using a logarithmic scale, see colorbar.
 Middle and bottom: We plot the data of the top panel in a diagram with the energy $E$ on the abscissa and the bipartition entanglement entropy $S_{E}$ or the average square magnetisation $S^{z2}(E)$ of the eigenstate on the ordinate, respectively.
 For the entanglement entropy, we use the natural logarithm and we divide the result by $L/2$ to obtain an intensive quantity.
 The state $\ket{n,k}$ has overlap only with states whose $S_{E_i}$ or $S^{z2}_{E_i}$ lies on the continuous ``thermal" curve.
 The red circle and the blue square highlight the regions of the plots where the QMBS $\ket{n=L/2,\pi}$ appear: the absence of any grey mark means that the scalar product is compatible with the numerical zero.
 }
 \label{Fig:Spectrum:Overlap}
\end{figure}

We have numerically verified this statement using the python-based package QuSpin~\cite{weinberg2017quspin}: we diagonalize the Hamiltonian~\eqref{Eq:Ham:1D} and compute the bipartition entanglement entropy $S_{E_i}$ and the average square magnetisation $S^{z2}_{E_i} = \frac{1}{L} \sum_j \langle (S^z_j)^2 \rangle$ of all eigenstates.
Subsequently, we compute the scalar product of the state $\ket{n, k}$ with all eigenstates for $n=L/2$ and $k = \pi - \frac{2 \pi}{L}$ and look at the properties of the eigenstates with whom the overlap is not zero.
The results are reported in Fig.~\ref{Fig:Spectrum:Overlap}, and support our thesis.
\paragraph{\textbf{Dynamics and asymptotic QMBS} ---}
The dynamical properties of the states $\ket{n,k}$ for large system sizes depend on how we approach $L \rightarrow \infty$.
If the limit is taken while the momentum $k$ is held fixed, then the variance is finite in the TL (see Ref.~\cite{lin2020slow} for 
examples in the PXP model).
Loosely speaking, we can invoke the well-known energy-time uncertainty relation, linking the typical timescale of the dynamics $\tau$ of a quantum state to the 
fluctuations
of the energy:
\begin{equation}
 \tau \geq \frac{\hbar}{2 \Delta H},
 \label{Eq:Time:Energy:Uncer}
\end{equation}
to claim that for these states the dynamics is frozen up to a given time-scale $\tau$ that is independent of $L$ and that afterwards an evolution towards thermal equilibration 
takes place~\cite{SM}.
%
%
To be more precise, the energy variance $\Delta H^2$ of the initial state determines the fidelity relaxation time $\tau \sim 1/\Delta H$~\cite{camposvenuti2010unitary}, since the fidelity $F(t) = |\bra{\Psi} e^{- i H t} \ket{\Psi}|^2$ of an initial state $\ket{\Psi}$ decays at short times as $ \sim \exp(- \Delta H^2 t^2)$; $\tau$ is a lower bound for the relaxation time of local observables~\cite{mori2017thermalization, mori2018thermalization}. 
%
%

%
Another class of states can be obtained by approaching the TL while letting $k$ flow to $\pi$.
This can be done by setting $k = \pi + \frac{2 \pi}{L} m$, with the coefficient $m\in \mathbb Z$ kept constant while $L \to \infty$.
In this case the energy variance scales as $\Delta H^2 \sim  (J^2 + 9 J_3^2) ( k- \pi)^2$ and tends to zero as $1/L^2$.
We refer to this second class of states as \textit{asymptotic QMBS} of the model, since according to \eqref{Eq:Time:Energy:Uncer}, the typical relaxation timescale of their dynamics scales as $\tau \sim L$, i.e., the system is frozen for timescales that \textit{increase} polynomially with the system size.
On the contrary, low entanglement states, by virtue of their diverging variance~\cite{banuls2020entanglement}, are typically expected to lose fidelity on timescales that decrease with system size, and the expectation values of typical observables relax in timescales that do not change drastically with system size~\cite{goldstein2013time, malabarba2014quantum, goldstein2015extremely, reimann2016typical, garciapintos2017equilibration, wilming2018equilibration, mori2018thermalization, riddell2021relaxation}.
Hence the dynamics of this class of states asymptotically approaches QMBS-like behavior even though they are \textit{not} exact QMBS of the system at finite size, and moreover they are \textit{orthogonal} to all the exact QMBS $\ket{n,  \pi}$.
To the best of our knowledge, this phenomenology has never been discussed before.

\begin{figure}[t]
 \includegraphics[width=\columnwidth]{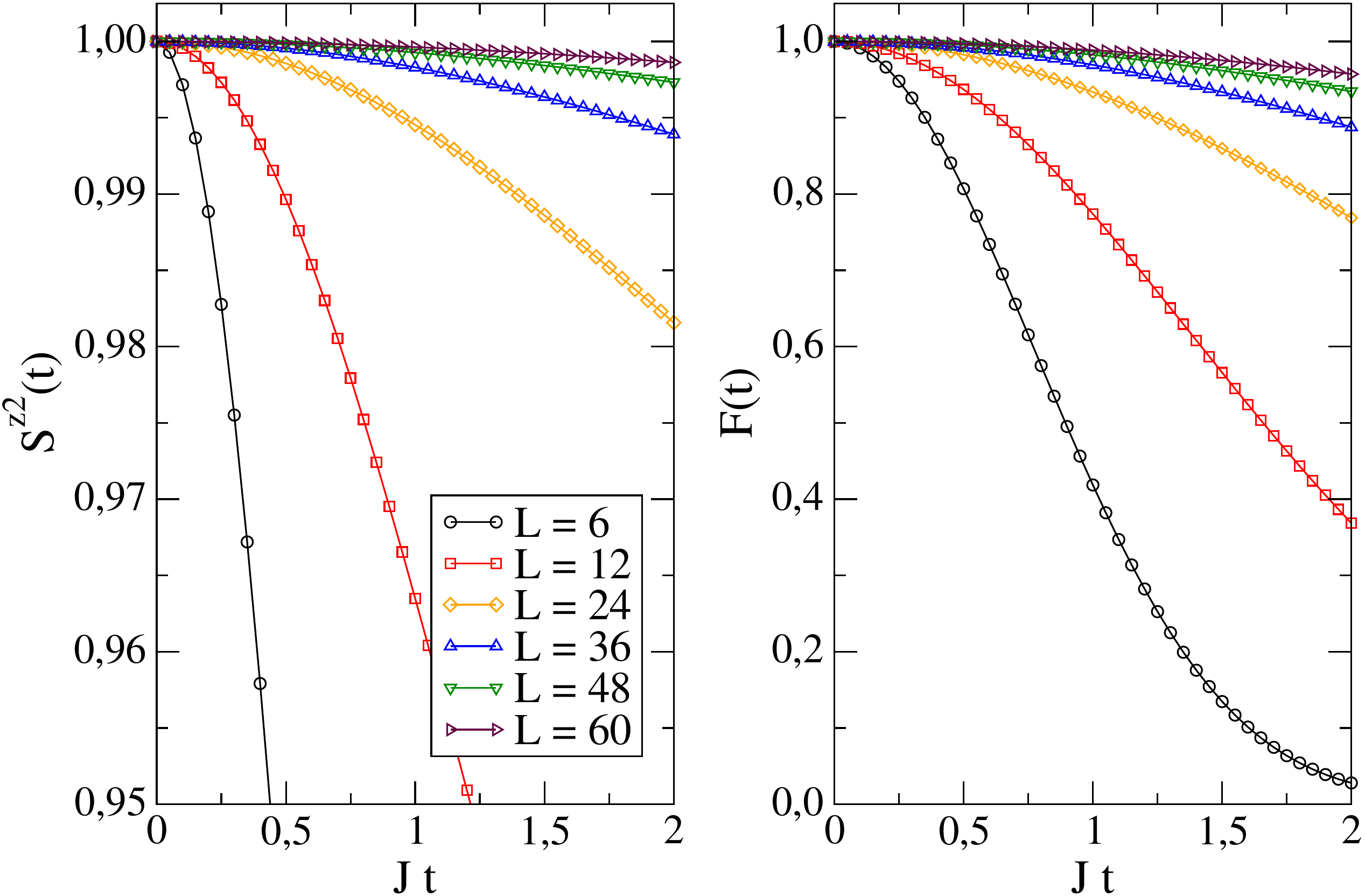}
 \caption{The properties of the state $e^{- i H t}\ket{n,k}$ for $n = L/2$ and $k=\pi - 2 \pi/L$ as a function of time for various system sizes $L$. Left: time evolution of the squared magnetisation $S^{z2}(t)$. Right: time evolution of the fidelity with the initial state $F(t)$. }
 \label{Fig:QuasiPI:TEV} 
\end{figure}

We support the previous statements with a numerical simulation of the dynamics of the states $\ket{n,k}$ under the action of $H$ using a time-evolving block decimation (TEBD) code based on a Matrix-Product-State (MPS) representation of the state obtained via the ITensor library~\cite{fishman2022itensor1, fishman2022itensor2}.
We consider in particular the state $\ket{n =L/2, k = \pi - 2\pi/L}$ for several system sizes up to $L=60$ and truncation error $10^{-12}$.
We then compute the observable $S^{z2}(t) = \frac{1}{L}\sum_j\langle \left(S^z_j \right)^2 \rangle_t$ and the fidelity of the time-evolved state with the initial state $F(t) = |\bra{n,k} e^{- i H t} \ket{n,k}|^2$. 
The results, reported in Fig.~\ref{Fig:QuasiPI:TEV}, show in both cases an important slow-down of the dynamics as the size increases.
In the SM we show that the data concerning the fidelity can be collapsed via a rescaling of time by a factor of $L$~\cite{SM}, which suggests the divergence of the relaxation time in the TL.
The result on the fidelity $F(t)$ 
shows undoubtedly that the time-evolved state maintains an overlap with the initial state that increases with $L$ and it implies the freezing of the state.
In the SM we complement this analysis by contrasting it with the typical dynamics of other states~\cite{SM};
%
we also analyze states obtained by acting on the exact QMBS with $(J^+_k)^m$, i.e., creating multiple quasiparticles of momenta close to $\pi$, and we argue that they should also be asymptotic QMBS as long as $m$ does not scale with $L$~\cite{SM}.
\paragraph{\textbf{Slow relaxation and non-thermalness in the middle of the energy spectrum} ---}
Two properties 
make the asymptotic QMBS particularly interesting:
(a) they have a limited amount of entanglement, i.e., a sub-volume law, but an extensive amount of energy; (b) they have an energy variance $\Delta H^2$ that drops fast enough to zero in the TL.
Any state that satisfies these conditions is guaranteed to have a long relaxation time, both in the fidelity and in the observables, while having an average energy that lies in the middle of the Hamiltonian spectrum.
Note that both (a) and (b) are necessary features that make the behavior of asymptotic QMBS atypical.
While any linear superposition of thermal eigenstates with small energy variance relaxes slowly, it typically has a large entanglement~\cite{banuls2020entanglement}.
%
On the other hand, a typical low-entanglement state has an energy variance that increases with system size~\cite{banuls2020entanglement}.
%

%
It is tempting to think that the existence of asymptotic QMBS should imply some kind of ``non-thermalness"~\cite{lin2020slow} or ETH-violation in the ``thermal" states orthogonal to the exact QMBS, even at finite system size.
Note that ETH consists of two parts~\cite{srednicki1994chaos, d2016quantum, shiraishimorireply2018}, pertaining to diagonal and off-diagonal matrix elements of a local operator in the energy eigenbasis. 
The diagonal matrix elements control the late-time expectation values of observables, and the existence of asymptotic QMBS does not imply any violation of diagonal ETH since we expect them to eventually thermalize for any finite system size.
On the other hand, the timescale of relaxation is controlled by both the energy variance of the initial state and the off-diagonal matrix elements~\cite{wilming2018equilibration}.
%
It is plausible that our result entails a violation of off-diagonal ETH at least in a part of the Hamiltonian spectrum.
\paragraph{\textbf{Asymptotic QMBS without exact QMBS} ---}
Our definition of asymptotic QMBS is based on a deformation of the tower of exact QMBS supported at finite size;
it is not clear whether asymptotic QMBS can exist in models without any exact QMBS or 
at energies distant from those of the exact QMBS. 
%

%
\begin{figure}[t]
 \includegraphics[width=\columnwidth]{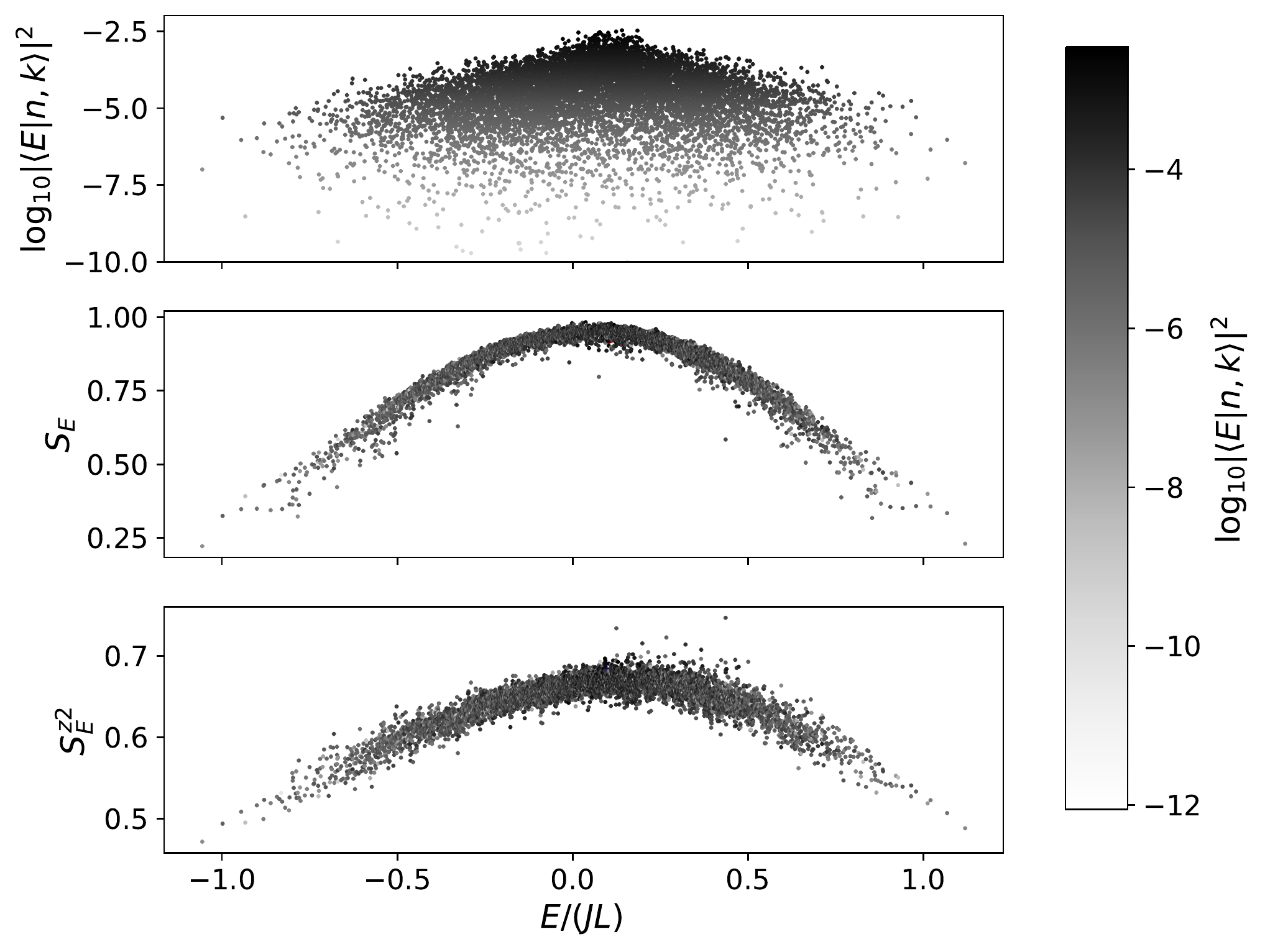}  
 \caption{Properties of the eigenstates of Hamiltonian $H'$ in the zero magnetization sector $S_z=0$; the parameters of the simulation are $\{J, h, D, J_3, J_z \} = \{1, 0, 0.1, 0.1, 1\}$ and $L=10$.
 Top: Squared overlap of $\ket{n,\pi}$ for $n=L/2$ with the eigenstates $\ket{E_i}$ of Hamiltonian $H'$ with zero magnetisation, $S_z=0$. 
 The information on $|\langle E_i\ket{n, \pi}|^2$ is also encoded in the color code of the marker of all panels using a logarithmic scale, see colorbar.
 Middle and bottom: We plot the data of the top panel in a diagram with the energy $E$ on the abscissa and the bipartition entanglement entropy $S_{E}$ or the average square magnetisation $S^{z2}(E)$ of the eigenstate on the ordinate, respectively. The state $\ket{n,\pi}$ has overlap only with states whose $S_{E_i}$ or $S^{z2}(E_i)$ lies on a continuous curve. In the SM~\cite{SM} we show the entire spectrum and show that the model does not have any QMBS (here the spectrum is incomplete because we plot only state that have a non-negligible overlap with $\ket{n,\pi}$).
 %
}\label{Fig:4}
\end{figure}

We now show that it is possible to weakly perturb the Hamiltonian $H$ in a way that destroys all exact QMBS, but such that the perturbed model maintains the asymptotic QMBS.
As an example, we consider  $H' = H + V$ with $V= (J_z/L) \sum_j S^z_j S^z_{j+1}$, which is still a non-trivial local perturbation since its spectral norm $\| V\|_\infty$ corresponding to its largest singular value is subextensive and scales as $O(1)$.
Using the python-based QuSpin package~\cite{weinberg2017quspin}, we numerically diagonalize $H'$ and compute the the entanglement entropy $S_{E_i}$ and the average square magnetisation $S^{z2}(E_i)$ for all eigenstates.
The plots, in Fig.~\ref{Fig:4},  do not indicate the presence of any exact QMBS.

\begin{figure}[t]
 \includegraphics[width=\columnwidth]{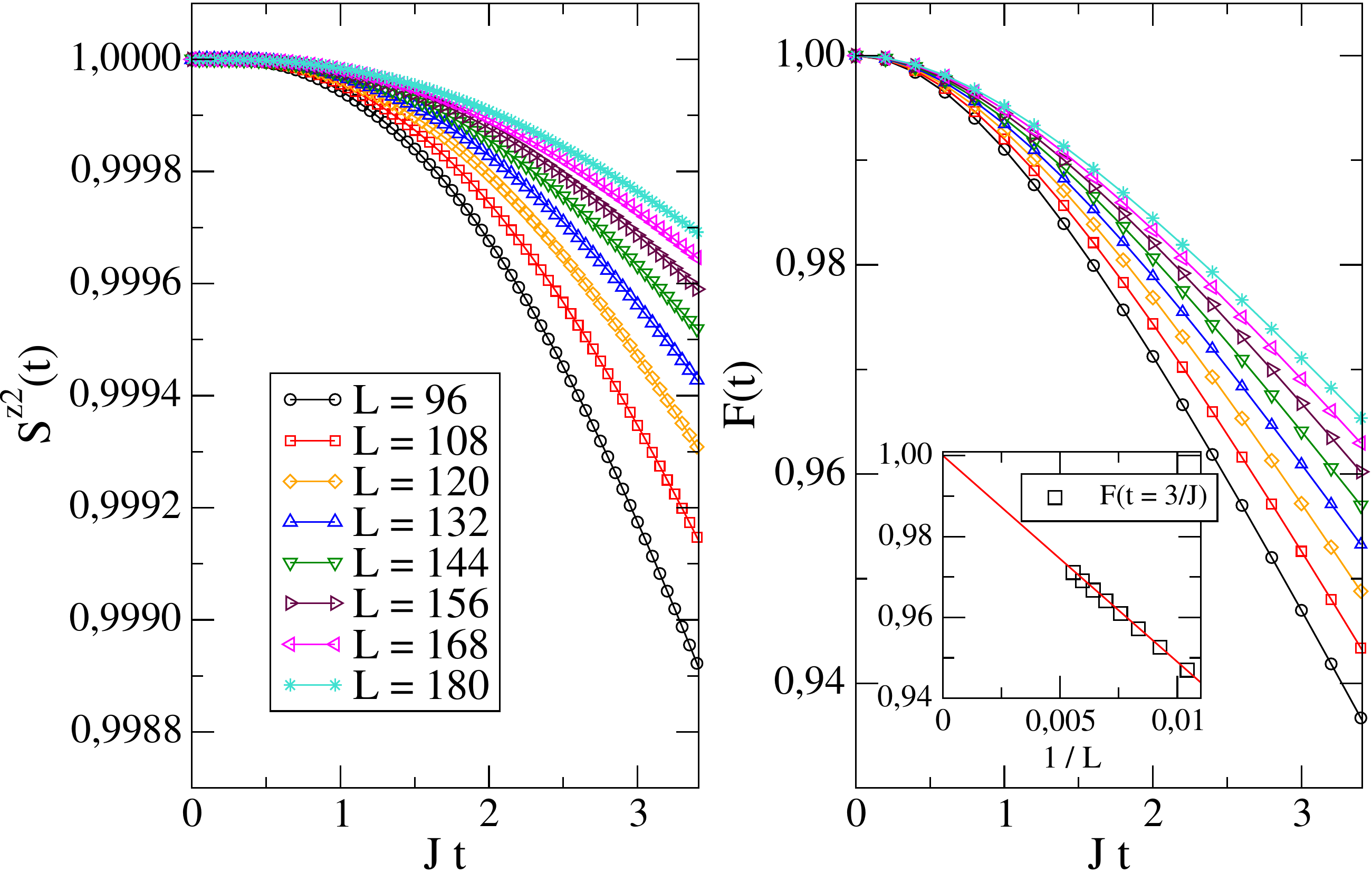}
 \caption{The properties of the state $e^{- i H' t}\ket{n,\pi}$ for $n = L/2$ as a function of time; the parameters of the Hamiltonian employed in the simulation are the same of Fig.~\ref{Fig:4}.
 Left: time evolution of the squared magnetisation $S^{z2}(t)$; right: time evolution of the fidelity with the initial state $F(t)$.
 The inset shows the scaling as a function of size of the values of $F(t = 3/J)$; we find a scaling to $1$ as $1/L \to 0$. 
 }
 \label{Fig:Perturbed:Hamiltonian}
\end{figure}

We now consider the state $\ket{n,\pi}$ of Eq.~(\ref{Eq:Scar:Old:1D}), which is an exact QMBS of $H$ but \textit{not} an eigenstate of $H'$.
Using the ITensor library~\cite{fishman2022itensor1, fishman2022itensor2},
we compute $S^{z2}(t)$ and the fidelity $F(t)$ for the time-evolved state $\ket{\Psi(t)} = e^{- i H' t} \ket{n, \pi}$; the results are in Fig.~\ref{Fig:Perturbed:Hamiltonian}.
The plots display the phenomenology of an asymptotic QMBS in a Hamiltonian that does not show any exact QMBS at finite size, and the $F(t)$ curves exhibit a collapse when time is rescaled by a factor $\sqrt{L}$~\cite{SM}, indicating a diverging relaxation time. 
This behavior can be directly attributed to the fact that the variance of the state $\ket{n, \pi}$ under the Hamiltonian $H'$ scales as $~\sim 1/L$ when $n$ is a finite fraction of $L$, as it is proven in the SM~\cite{SM}.  
\paragraph{\textbf{Conclusions} --- }
In this letter we 
revisited the paradigmatic one-dimensional spin-1 XY model that supports exact QMBS at finite size, and we explored the properties of the rest of the spectrum.
We showed that it is possible to construct other states, dubbed asymptotic QMBS, with little entanglement and whose relaxation time diverges polynomially in the thermodynamic limit. 
These asymptotic QMBS indicate the existence of slowly relaxing modes and novel long-lived quasiparticles in systems with exact QMBS; it would be interesting to understand their relations to analogous slowly relaxing modes of hydrodynamic origin.  

%
Remarkably, asymptotic QMBS are linear combinations of ``thermal" eigenstates whose entanglement entropy and average squared magnetization are ``smooth" functions of energy; 
we leave for future work the investigation of a possible violation of off-diagonal ETH~\cite{khatami2013fluctuation, steinigeweg2013eigenstate, beugeling2015offdiagonal, richter2020eigenstate, wurtz2020emergent, sugiura2021adiabatic, surace2021exact}.
%

%
%
%

%
Asymptotic QMBS with similar properties can also be constructed in higher dimensional spin-1 XY models~\cite{SM}, but other extensions would also be interesting, considering first the exhaustive algebra of local Hamiltonians that have the same exact QMBS $\ket{n, \pi}$~\cite{moudgalya2022exhaustive}.
%
%
Second, they likely can always be constructed in Hamiltonians with simple quasiparticle towers of exact QMBS~\cite{iadecola2020quantum, mark2020eta, moudgalya2020eta, pakrouski2020many, moudgalya2018exact, moudgalya2018entanglement, mark2020unified, moudgalya2020large}.
Third, there are many different types of exact QMBS~\cite{moudgalya2021review}, e.g., with non-local ``quasiparticles"~\cite{odea2020from, mark2020unified, langlett2022rainbow}, or with non-isolated states~\cite{shiraishi2017systematic, lin2019exact}; they could appear in gauge theories~\cite{biswas2022scars, banerjee2021quantum} or Floquet systems~\cite{mizuta2020exact, sugiura2021manybody, iadecola2020nonergodic, rozon2022constructing}. 
Are there asymptotic QMBS in these models?

Finally, one could also consider deformations of Hamiltonians with exact QMBS (a problem that we partially addressed in the final part of this letter), 
and ask what are the conditions for a Hamiltonian to display an asymptotic QMBS without any exact QMBS.

\begin{acknowledgments}

\paragraph{\textbf{Acknowledgements} ---}
We warmly acknowledge enlightening discussions with Saverio Bocini,  Xiangyu Cao, Maurizio Fagotti, David Huse, Michael Knap, Lesik Motrunich and Nicolas Regnault.
We also thank Lesik Motrunich for useful comments on a draft.
L.G. and L.M. also thank Guillaume Roux and Pascal Simon for discussions on previous shared projects.
This work is supported by the Walter Burke Institute for Theoretical Physics at Caltech and the Institute for Quantum Information and Matter, by LabEx PALM (ANR-10-LABX-0039-PALM) in Orsay, by Region Ile-de-France in the framework of the DIM Sirteq
and by the Swiss National Science Foundation under Division
II.
S.M. also acknowledges the hospitality of the Laboratoire de Physique Th\'{e}orique et Mod\`{e}les Statistiques (LPTMS) in Orsay, where this collaboration was initiated, and the Physik-Insitut of the University of Zurich, where some of this work was performed. 

\end{acknowledgments}
 

%


\newpage 


\renewcommand{\thefigure}{S\arabic{figure}}
\renewcommand{\thesection}{S\arabic{section}}
\renewcommand{\theequation}{S\arabic{equation}}
\renewcommand{\thepage}{S\arabic{page}}
\setcounter{figure}{0}
\setcounter{page}{0}
\setcounter{section}{0}
\setcounter{secnumdepth}{2}

\onecolumngrid
\begin{center}
{\large \textbf{Online supplementary material for: \\ \mytitle}}

\vspace{16pt}

Lorenzo Gotta$^{1,2}$, Sanjay Moudgalya$^{3,4}$, Leonardo Mazza$^2$

\begin{small}
$^1$\textit{ Department of Quantum Matter Physics, University of Geneva, 24 Quai Ernest-Ansermet, 1211 Geneva, Switzerland}

$^2$\textit{\it Universit\'e Paris-Saclay, CNRS, LPTMS, 91405, Orsay, France}\\

$^3$\textit{Department of Physics and Institute for Quantum Information and Matter,\\
California Institute of Technology, Pasadena, California 91125, USA}\\
 
$^4$\textit{Walter Burke Institute for Theoretical Physics, California Institute of Technology, Pasadena, California 91125, USA}\\
\end{small}

\vspace{10pt}

\today

\vspace{10pt}
\parbox{0.8\textwidth}{In this Supplementary Material we present the explicit calculations of the main relevant properties of the asymptotic QMBS presented in the main text:
\begin{itemize}
    \item[S1.] Orthogonality of the asymptotic QMBS with the exact QMBS
    \item[S2.] Average energy and energy variance for the asymptotic QMBS 
    \item[S3.] Entanglement entropy of the exact and asymptotic QMBS
    \item[S4.] Variance of the exact QMBS for the perturbed Hamiltonian
    \item[S5.] Dynamics of initial states that are not asymptotic QMBS
    \item[S6.] Spectral properties of the Hamiltonian $H'$
    \item[S7.] Universal rescaling of fidelities
    \item[S8.] Higher dimensional generalisations of asymptotic QMBS
\end{itemize}
}
\end{center}

\section{Orthogonality of the asymptotic QMBS with the exact QMBS}
In this section, we demonstrate the orthogonality of the states $\{\ket{n,k}\}$, defined in Eq.~(4) of the main text.
First, we note that $\ket{n,k}$ is orthogonal to $\ket{n',k'}$ when $n \neq n'$ because they have a different magnetisation $S^z =\sum_j S^z_j$, which is a simple function of $n$: $S^z = -L+2n$.
We now consider states with the same $n$ and take the system size $L$ to be even and $k$ to be an integer multiple of $\frac{2\pi}{L}$ for simplicity.
We then observe that $\langle n,k \ket{n,k'} \propto \bra{n-1, \pi} J^-_k J^+_{k'} \ket{n-1,\pi}$ for $n \geq 1$.
By definition of the operators $J_k^+$ in Eq.~(2) of the main text we have:
\begin{align}
    &\bra{n-1, \pi} J^-_k J^+_{k'} \ket{n-1,\pi} = \frac{1}{4}
    \sum_{j,j' = 1}^L{e^{-i (k j- k'j')}}\bra{n-1, \pi} (S^-_j)^2 (S^+_{j'})^2 \ket{n-1,\pi} \nonumber \\
    &= \frac{1}{4}
    \sum_{j = 1}^L{e^{-i (k - k')j}}\bra{n-1, \pi} (S^-_j)^2 (S^+_{j})^2 \ket{n-1,\pi} + \frac{1}{4}
    \sum_{j\neq j'}{e^{-i (k j- k'j')}}\bra{n-1, \pi} (S^-_j)^2 (S^+_{j'})^2 \ket{n-1,\pi} \nonumber \\
    &= \frac{1}{4} \alpha  \sum_{j = 1}^L e^{-i (k-k')j} 
    + \frac{1}{4} \beta \sum_{j=1}^L e^{i(\pi-k)j} \sum_{j'\neq j} e^{i  (k'-\pi)j'}
    = \frac{1}{4} L (\alpha - \beta) \delta_{k,k'}
    + \frac{1}{4} \beta L^2 \delta_{k,\pi} \delta_{k',\pi},
    \label{Eq:First:SupMat}
\end{align}
where $\alpha = 4\frac{\binom{L-1}{n-1}}{\binom{L}{n-1}} = 4\frac{L + 1 - n}{L}$ and $\beta = 4\frac{\binom{L-2}{n-2}}{\binom{L}{n-1}} = 4\frac{(n-1)(L-n+1)}{L(L-1)}$, and we have used the fact that $k$ and $k'$ are integer multiples of $\frac{2\pi}{L}$.
This calculation is done directly by using the expression of $\ket{n,\pi}$ as an equal amplitude superposition of ``fully-magnetised" product states
\begin{equation}
 \ket{n,\pi} = \sqrt{\frac{1}{2^{2n}\binom{L}{n}}} \sum_{1 \leq j_1 < j_2 < \ldots < j_n \leq L} e^{i \pi \sum_{i=1}^n j_i} 
 \left( S^+_{j_1} \right)^2
 \left( S^+_{j_2} \right)^2
 \ldots 
 \left( S^+_{j_n} \right)^2
 \ket{\Downarrow},
 \label{eq:qmbsproduct}
\end{equation}
and studying the action of the sandwiched operator on the basis states separately when $j = j'$ and when $j \neq j'$, and carefully accounting for the phase factors and normalization factors. 
It is important to visualise the combinatorial nature of this state, expanded on a basis of states where the bimagnons created by $\left( S^+_j\right)^2$ are equally distributed everywhere.
When $j = j'$, we obtain that $\alpha$ in Eq.~\eqref{Eq:First:SupMat} is simply related to the number of fully-magnetised product states that do not have a bimagnon at site $j$, or else the action of $(S^-_j)^2 (S^+_j)^2$ vanishes on such a basis state.
This number is $\binom{L-1}{n-1}$; if we consider the normalisation factor and the specific matrix elements of $\left(S^-_j\right)^2 (S^+_j)^2$, we obtain its expression, given after Eq.~\eqref{Eq:First:SupMat}.
Similarly, when $j \neq j'$ and $n > 1$, we obtain that $\beta$ in Eq.~\eqref{Eq:First:SupMat} is related to the number of fully-magnetised product states that have one bimagnon at site $j$, and no bimagnon at $j'$, which is $\binom{L-2}{n-2}$.
Its expression, given after Eq.~\eqref{Eq:First:SupMat}, then follows directly after taking into account the normalization factors and matrix elements.
Hence using Eq.~(\ref{Eq:First:SupMat}) for any $k \neq k'$ it is clear that we obtain $\braket{n, k'}{n, k} = 0$.
Given that we work with normalised states, we can combine the arguments above to conclude that $\langle n,k \ket{n', \pi} = \delta_{n,n'} \delta_{k,\pi}$ whenever $k$ is an integer multiple of $\frac{2\pi}{L}$ and $L$ is even. 
\section{Average energy and energy variance for the asymptotic QMBS}\label{sec:avgenergy}
In this section, we compute the average energy and variance of the asymptotic QMBS states $\{\ket{n, k}\}$ defined in Eq.~(4) of the main text.
\subsection{Rewriting the asymptotic QMBS}
For the convenience of explicit calculations, we propose the following rewriting of the asymptotic QMBS:
\begin{equation}
    \ket{n,k} = \frac{1}{\sqrt{\mathcal M_{n,k} }} J^+_k \ket{n-1,\pi}.
\label{eq:nkexpression}
\end{equation}
where the states $\ket{n-1,\pi}$ and $\ket{n,k}$ are normalised.
As a first step, we compute the normalization factor coefficient $\mathcal M_{n,k}$, which can be directly deduced from Eq.~(\ref{Eq:First:SupMat}).
That is, its expression reads
\begin{equation}
\mathcal M_{n,k} = \bra{n-1,\pi} J^-_{k} J^+_{k}\ket{n-1,\pi} = \frac{(L-n+1)(L-n)}{L-1} + \frac{L(L-n + 1)(n-1)}{L-1} \delta_{k,\pi}
\label{Eq:Normalization:SUPMAT}
\end{equation}

\subsection{Average energy}\label{SubSec:AverageEnergy}
To compute the average energy of the state $\ket{n,k}$, we first rewrite the OBC spin-1 XY Hamiltonian, along with the symmetry breaking perturbation [see discussion below Eq.~(1) in the main text], as:
\begin{equation}
 H = \frac{J}{2} \sum_{ j = 1}^{L-1}  \left( S^+_{j} S^-_{j+1} + S^-_{j} S^+_{j+1} \right) + h \sum_{j = 1}^L S_{j}^{z}+ D \sum_j \left( S_z^{z} \right)^2+
 \frac{J_3}{2} \sum_{j = 1}^{L-3} \left( S_{ j}^+ S^-_{j+3}+S_{ j}^+ S^-_{j+3}\right) . \label{Eq:Ham:H0:SMAT}
\end{equation}
In order to compute the average energy, we need to study the action of $[S^+_{j} S^-_{j+1}+ h.c.]$ onto the state $\ket{n,  k}$, and for this it is convenient to consider the decomposition of $\ket{n,k}$ over sites $j$ and $j+1$. 
For example, we can rewrite $\ket{n, \pi}$ as
\begin{equation}
     \ket{n, \pi} = \alpha_{n, \pi} \ket{+}_{j} \ket{+}_{j+1} 
 \ket{\psi_{n, \pi,1}} 
 + \beta_{n,  \pi} \ket{-}_{j} \ket{-}_{j+1} \ket{\psi_{n, \pi,2}}  +  \gamma_{n, \pi} \left( \frac{\ket{+}_{j} \ket{-}_{j+1}-\ket{-}_{j} \ket{+}_{j+1} }{\sqrt 2}\right) \ket{\psi_{n, \pi,3}};
\end{equation}
where $\alpha_{n, \pi}$, $\beta_{n,\pi}$, and $\gamma_{n,\pi}$ are numbers with $|\alpha_{n,\pi}|^2 + |\beta_{n,\pi}|^2 + |\gamma_{n,\pi}|^2 = 1$, and $\{\ket{\psi_{n, \pi, \ell}}\}$ for $1 \leq \ell \leq 3$ are some states with support on sites other than $j$ and $j+1$, and we have denoted the three spin-1 states on a site $j$ by $\ket{+}_j$, $\ket{-}_j$, and $\ket{0}_j$.
One can similarly rewrite the $\ket{n, k}$ as:
\begin{align}
     \ket{n, k} =& \alpha_{n, k} \ket{+}_{j} \ket{+}_{j+1} 
 \ket{\psi_{n, k,1}} 
 + \beta_{n,  k} \ket{-}_{j} \ket{-}_{j+1} \ket{\psi_{n, k,2}}  +  \nonumber \\ 
 &+\gamma_{n, k} \left( \frac{\ket{+}_{j} \ket{-}_{j+1}-\ket{-}_{j} \ket{+}_{j+1} }{\sqrt 2}\right) \ket{\psi_{n, k, 3}} + 
 \upsilon_{n, k} \left( \frac{\ket{+}_{j} \ket{-}_{j+1}+ e^{ik} \ket{-}_{j} \ket{+}_{j+1} }{\sqrt 2}\right) \ket{\psi_{n, k,4}},
 \label{eq:nkschmidt}
\end{align}
where $\alpha_{n,k}$, $\beta_{n,k}$, $\gamma_{n,k}$, and $v_{n,k}$ are numbers such that $\ket{n, k}$ is normalized and $\{\ket{\psi_{n, \pi, \ell}}\}$ for $1 \leq \ell \leq 4$ are some states without support on $j$ and $j+1$.
The action of the term $[S^+_{j} S^-_{j+1}+h.c.]$ can then be directly computed to be:
\begin{equation}
 \left( S^+_{j} S^-_{j+1} + h.c. \right) \ket{n,  k} = \sqrt{2}\upsilon_{n,  k} \left(1+ e^{i  k } \right) 
 \ket{0}_{j} \ket{0}_{j+1}
 \ket{\psi_{n, k, 4}}.
\end{equation}
Using Eq.~(\ref{eq:nkschmidt}), it then directly follows that $\bra{n,  k} \left( S^+_{j} S^-_{j+1} + h.c. \right) \ket{n,  k} = 0$.
A similar reasoning can be carried out for the interaction term proportional to $J_3$ to show that $\bra{n,  k} \left( S^+_{j} S^-_{j+3} + h.c. \right) \ket{n,  k} = 0$, hence in all we obtain $\bra{n,  k} H \ket{n,  k} =  h(-L+ 2n) + D L$ for all $k$. We conclude by noticing that the same result holds in PBC as well.
\subsection{Energy variance}
To compute the energy variance in any state, it is easy to see that the contribution of the terms in the Hamiltonian for which the state is an eigenstate simply vanishes.
Hence, in the computation of the variance of $\ket{n,k}$, we can simply ignore the magnetic field and anistropy terms in $H$ of Eq.~(\ref{Eq:Ham:H0:SMAT}), i.e., those that are proportional to $h$ and $D$, since $\ket{n,k}$ are their eigenstates.
For simplicity, we refer to the terms in $H$ proportional to $J$ and $J_3$ as $H_1$ and $H_3$, respectively, and work with PBC.
As we showed in the previous section, $\bra{n,k} H_1 \ket{n,k} =\bra{n,k} H_3 \ket{n,k} = 0$, and using similar ideas one can also show that $\bra{n,k} H_1\, H_3 \ket{n,k}=\bra{n,k} H_3\, H_1 \ket{n,k}=0$.
Hence the expression of the variance  of $\ket{n, k}$ in $H$ reduces to $\Delta H^2 = \bra{n,k} (H_1+H_3)^2 \ket{n,k} = \bra{n,k} (H_1^2+H_3^2) \ket{n,k}$.
We now propose a rewriting of each term:
\begin{align}
\bra{n,{k}} H_\ell^2 \ket{n,{k}}= &
\frac{1}{\mathcal M_{n,k}}\bra{n-1,\pi}  J^-_{{k}} H_\ell^2  J^+_{{k}}\ket{n-1,\pi} = \frac{1}{\mathcal M_{n,k}}
\bra{n-1,\pi} [ J^-_{{k}}, H_{\ell}]\, [ H_{\ell},  J^+_{{k}}]\ket{n-1,\pi} = \nonumber \\
= & \frac{1}{\mathcal M_{n,k}}\bra{{n-1},\pi} [ H_{\ell}, J^+_{{k}}]^{\dag} \, [ H_{\ell},  J^+_{{k}}]\ket{{n-1},\pi },
\label{Eq:S14}
\end{align}
where $\ell = 1, 3$, and we have exploited the fact that $ H_{\ell}\ket{n-1,\pi }=0$.
We a few straightforward algebraic passages, it is possible to show that:
\begin{align}
[ H_1, J^+_{{k}}] &= \frac{J}{2} \sum_{j=1}^{L}{e^{ikj} [S^+_j S^-_{j+1} + S^-_j S^+_{j+1}, \frac{1}{2}(S^+_j)^2 + \frac{e^{ik}}{2} (S^+_{j+1})^2]} \nonumber \\
&=-\frac{J}{2} \sum_{j =1}^{L}{e^{ikj}\left[ \{  S_{j}^z , S_{j}^+\} S_{j+1}^+ + e^{ik} S_{j}^+ \{  S_{j+1}^z , S_{j+1}^+\} \right]},
\label{Eq:S15}
\end{align}
where $\{\cdot, \cdot\}$ denotes the anti-commutator and we have used the identity $[ S_{m}^+ S_{n}^-,(S_{n}^+)^2]=-2  S_{m}^+ \{ S_{n}^z,  S_{n}^+ \}$.
The calculation proceeds by substituting Eq.~\eqref{Eq:S15} into Eq.~\eqref{Eq:S14} and it is greatly simplified by the fact that $S^z_j S^+_j\ket{n-1,\pi} = S_j^z S^{-}_j\ket{n-1,\pi} = 0$.
First, using this identity simplifies the action of $[H_1, J_k^+]$ on $\ket{n-1, \pi}$ to 
\begin{equation}
[ H_1, J^+_{{k}}]\ket{n-1, \pi} = -\frac{J}{2} \sum_{j =1}^{L}{e^{ikj}\left[S_{j}^+ S_{j}^z S_{j+1}^+ + e^{ik} S_{j}^+ S_{j+1}^+ S_{j+1}^z \right]}\ket{n-1, \pi},
\label{eq:actionsimp}
\end{equation}
and Eq.~(\ref{Eq:S14}) then reads
\begin{equation}
    \bra{n, k} H_1^2 \ket{n, k} = \frac{J^2}{4 \mathcal M_{n,k}} \sum_{j, j' =1}^{L}{e^{ik(j-j')}\bra{n-1,\pi}[S_{j'}^z S_{j'}^- S_{j'+1}^- + e^{-ik} S_{j'}^- S_{j'+1}^z S_{j'+1}^- ] [S_{j}^+ S_{j}^z S_{j+1}^+ + e^{ik} S_{j}^+ S_{j+1}^+ S_{j+1}^z]}\ket{n-1, \pi}.
\label{eq:actionsimp2}
\end{equation}
We then notice that in Eq.~\eqref{eq:actionsimp2}, all the terms with $j \neq j'$ in the sum vanish since the action of the sandwiched operator on $\ket{n-1,\pi}$ in such cases leads to inevitable appearance of spins with states $\ket{0}_m$ on certain sites $m$, which in turn have a vanishing overlap with $\bra{n-1,\pi}$.
Hence, we can simplify Eq.~(\ref{eq:actionsimp2}) to
\begin{align}
\bra{n, k} H_1^2 \ket{n, k} &= \frac{J^2}{4 \mathcal M_{n,k}} \sum_{j =1}^{L}{\bra{n-1, \pi} [S_{j}^z S_{j}^- S_{j+1}^- + e^{-ik} S_{j}^- S_{j+1}^z S_{j+1}^- ] [S_{j}^+ S_{j}^z S_{j+1}^+ + e^{ik} S_{j}^+ S_{j+1}^+ S_{j+1}^z]}\ket{n-1, \pi} \nonumber \\
&=\frac{J^2}{4 \mathcal M_{n,k}}\bra{n-1,\pi}\sum_{j=1}^L\Biggl[e^{-ik} S_{j}^-  S_{j}^+ S_{j}^z   S_{j+1}^z  S_{j+1}^- S_{j+1}^+ + S_{j}^-  S_{j}^+  S_{j+1}^z  S_{j+1}^-   S_{j+1}^+   S_{j+1}^z \nonumber \\
&+S_{j}^z  S_{j}^-   S_{j}^+  S_{j}^z   S_{j+1}^-  S_{j+1}^+ + e^{ik}  S_{j}^z  S^-_{j} S_{j}^+  S_{j+1}^-   S_{j+1}^+ S_{j+1}^z\Biggr]\ket{n-1,\pi},
\label{eq:varfinal}
\end{align}
Now we consider the expansion of $\ket{n-1, \pi}$ in the product state basis, as shown in Eq.~(\ref{eq:qmbsproduct}) and note that each of the terms in Eq.~(\ref{eq:varfinal}) vanish on the basis states unless there is no bimagnon on both sites $j$ and $j+1$.
Hence we can simply count the number of such states and incorporate the normalization factor to obtain:

\begin{equation} \label{Eq:SM:varfinal_2}
    \bra{n,{k}} H_1^2 \ket{n,{k}}=
    \frac{J^2}{\mathcal M_{n,k}} \sum_{j = 1}^L  \frac{\binom{L-2}{n-1}}{\binom{L}{n-1}}\left[e^{-i{k}} +1+1+e^{i{k}}\right]=\frac{4 J^2 \cos^2\left(\frac{k}{2} \right)}{1+\delta_{k,\pi}\frac{L(n-1)}{L-n}} = 4 J^2 \cos^2\left(\frac{k}{2} \right),
\end{equation}
where in the last step we have used the fact that the numerator anyway vanishes for $k = \pi$. The same calculation can be carried out in OBC and amounts to a multiplication of the result in Eq.~\eqref{Eq:SM:varfinal_2} by a factor $1-\frac{1}{L}$, which does not change the PBC result in the thermodynamic limit.
With similar arguments one can prove that:
\begin{equation}
    \bra{n,{k}} H_3^2 \ket{n,{k}}= 4 J_3^2 \cos^2 \left( \frac{3k}{2} \right),
\end{equation}
thus recovering the result in Eq.~(5) of the main text. Once again, the choice of OBC amounts to a correction factor $1-\frac{3}{L}$, which is irrelevant in the thermodynamic limit. 
\subsection{Considerations on multiparticle asymptotic QMBS}

\begin{figure}[t]
\centering 
\includegraphics[width=0.7\textwidth]{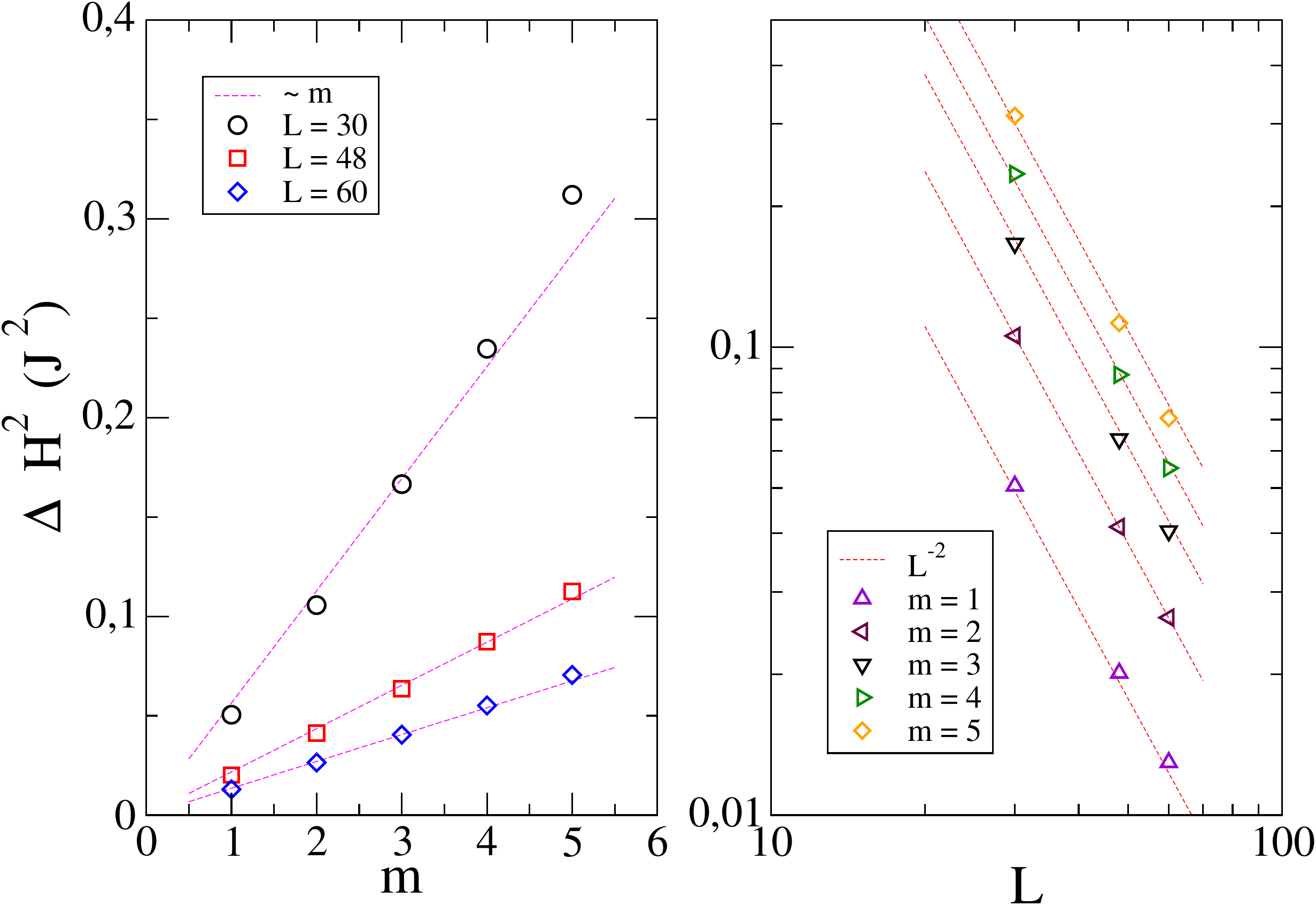}
\caption{Energy variance $\Delta H^2$ of the multiparticle QMBS obtained by acting $m$ times a bimagnon operator $J^+_k$ with $k = \pi - 2\pi/L$ on an exact scar $\ket{n, \pi}$. The state we are considering is thus proportional to $(J^+_{k})^m \ket{n, \pi}$. We study three different system sizes, $L=30$, $48$ and $60$, and five different values of $m$, from $1$ to $5$. The numerical data, obtained with a MPS representation of the states, yield a scaling of $\Delta H^2$ that is approximately linear in $m$ and proportional to $L^{-2}$. In the left panel the data are plotted versus $m$; in the right panel the same data are plotted versus $L$. The dashed lines are a guide to the eye to highlight the approximate behaviours as $\tilde m$ and as $\tilde L^{-2}$; note that the scalings are not precise at large $m$. }
\label{Fig:MultiParticle}
\end{figure}

The set of asymptotic QMBS is not limited to the single-particle asymptotic QMBS explicitly discussed above.
For instance, the action of an operator $(J_k^+)^m$ for $k = \pi - \epsilon$ (for $\epsilon \sim 1/L$ small) and $m \ll L$, on any exact QMBS eigenstate results in a state with variance scaling approximately as $\sim m \epsilon^2$;
a set of numerical results supporting this claim is given in Fig.~\ref{Fig:MultiParticle}. Based on these results, we can identify also the multiparticle QMBS as asymptotic QMBS.
In the rest of this section, we present an analytical calculation of the energy variance of the aforementioned state:
\begin{align}
\frac{\bra{n-1,\pi}(J^-_k)^m H^2 ( J^{+}_k)^m\ket{n-1,\pi}}{\bra{n-1,\pi}( J^-_k)^m(J^{+}_k)^m\ket{n-1,\pi}}=\frac{\norm{H (J^{+}_k)^m\ket{n-1,\pi}}^2}{\norm{(J^{+}_k)^m\ket{n-1,\pi}}^2},
\label{eq:varcalc}
\end{align}
where $\norm{\bullet}$ denotes the $L^2$ norm.
Although we are not able to compute the variance exactly, we will show that via some approximations we can reproduce the scalings obtained in Fig.~\ref{Fig:MultiParticle}.
Let us first remark that the formula in Eq.~\eqref{eq:varcalc} follows from the following facts:
(i) the state $(J_k^+)^m \ket{n-1,\pi}$ is an exact eigenstate of the Hamiltonian parts proportional to $h$ and $D$, with eigenvalue $-Lh + 2(n-1+m)h+LD$; 
(ii) it has zero expectation value of $H_1+H_3$. 
Both results follow from the fact that $(J_k^+)^m \ket{n-1,\pi}$ is only a linear superposition of $\ket{+}$ and $\ket{-}$ spin states, with $z$ the spin-quantisation axis: the action of $H_1$ and $H_3$ necessarily creates two $\ket{0}$ spin states, and thus make the state orthogonal to the initial one. A similar reasoning has been presented in Sec.~\ref{SubSec:AverageEnergy} for $m=1$.
As long as the energy variance is considered, we can thus simply focus on $H = H_1+ H_3$. Yet, for the sake of simplicity, in this Section we will only consider $H_1$. The results can be easily generalized to $H_3$.
We first focus on the denominator of the expression in Eq.~\eqref{eq:varcalc}: %
\begin{align}
    \norm{(J_k^+)^m &\ket{n-1,\pi}}^2  = \bra{n-1,\pi} (J_k^-)^m (J_k^+)^m \ket{n-1,\pi} = \nonumber \\
    =&\frac{(m!)^2}{2^{2m}} \sum_{j_1< \ldots < j_m} \sum_{l_1 < \ldots < l_m} e^{ik(l_1+ \ldots +l_m - j_1 - \ldots -j_m)} 
    \bra{n-1,\pi} (S_{j_1}^-)^2 \ldots (S_{j_m}^-)^2 (S_{l_1}^+)^2 \ldots (S_{l_m}^+)^2 \ket{n-1,\pi}.
\end{align}
The evaluation of this sum is a formidable task, and we approximate it by considering only the leading terms $j_i=l_i$, which are characterised by the fact that the phase is stationary. The factor $(m!)^2$ takes into account the possible orderings of the indexes. Other terms will be characterised by an oscillating phase and thus are expected to be inessential in the thermodynamic limit.
The denominator is then approximated by the following expression:
\begin{align}
    \norm{(J_k^+)^m \ket{n-1,\pi}}^2 ) \, \approx \, & \frac{(m!)^2}{2^{2m}} 
    \sum_{j_1< \ldots < j_m} \sum_{l_1 < \ldots < l_m} 
    \delta_{j_1, l_1} \ldots \delta_{j_m, l_m} 2^{2m} \frac{\binom{L-m}{n-1}}{\binom{L}{n-1}}= \nonumber \\
    \, = \, & (m!)^2 \binom{L}{m} \frac{\binom{L-m}{n-1}}{\binom{L}{n-1}} = m! \frac{(L-n+1)!}{(L-n-m+1)!}.
\end{align}
We now move to the numerator of Eq.~\eqref{eq:varcalc};
for its evaluation, the following relation is useful:
\begin{align}
[[H_1+H_3, J^+_k], J^+_k]=0.
\label{Eq:Double:Commutator:Zero}
\end{align}
Let us prove Eq.~\eqref{Eq:Double:Commutator:Zero} using the explicit expression of the commutator in Eq.~\eqref{Eq:S15}; we will only focus on the term $H_1$ of the Hamiltonian since the extension to $H_3$ is straightforward:
\begin{align}
    [[H_1, J^+_k], J^+_k]= - \frac{J}{2}
    \sum_j e^{i kj} \left[ 
    \{S^+_j, S_j^z \} S^+_{j+1} + e^{ik} S^+_j 
    \{S^+_{j+1}, S_{j+1}^z \}, \frac{1}{2} e^{ikj} (S^+_{j})^2+\frac{1}{2} e^{ik(j+1)} (S^+_{j+1})^2
    \right]
\end{align}
The commutator can be easily split into the sum of four commutators; let us begin by analysing the first:
\begin{align}
    \left[ 
    \{S^+_j, S_j^z \} S^+_{j+1} , (S^+_{j})^2
    \right] = 
    \left[ 
    (S^+_j S_j^z + S_j^z S^+_j  )  , (S^+_{j})^2
    \right] S^+_{j+1} =
    \left( S^+_j \left[ S_j^z , (S^+_{j})^2 \right] + \left[ S_j^z , (S^+_{j})^2 \right] S^+_j \right) S^+_{j+1}.
\end{align}
The commutator that appears in the last expression can be explicitly computed: $\left[ S_j^z , (S^+_{j})^2 \right] = 2 (S^+_{j})^2$. We thus obtain an expression proportional to $(S^+_{j})^3$, that for a spin-1 system is equal to zero. The thesis follows by applying similar calculations to the other three commutators.

With the help of Eq.~\eqref{Eq:Double:Commutator:Zero}, it is possible to show by induction that:
\begin{align}
H (J^+_k)^m \ket{n-1,\pi}=m (J^+_k)^{m-1}[H, J^+_k]\ket{n-1,\pi},\;\;\;1\leq m \leq L-n+1.
\end{align}
Hence we obtain that:
\begin{align}
\norm{H (J^{+}_k)^m\ket{n-1,\pi}}^2= \norm{m (J^+_k)^{m-1}[H, J^+_k]\ket{n-1,\pi}}^2
= m^2 \bra{n-1,\pi} [H,J_k^+]^\dagger (J^-_k)^{m-1}
(J^+_k)^{m-1} [H, J_k^+] \ket{n-1,\pi}
\end{align}
Using Eq.~\eqref{Eq:S15} we obtain:
\begin{align}
    \norm{H (J^{+}_k)^m\ket{n-1,\pi}}^2= 
    &  \frac{m^2 [(m-1)!]^2}{2^{2m-2}}  \left( \frac{J}{2} \right)^2 \sum_{j_1<\ldots <j_{m-1}}
    \sum_{l_1< \ldots < l_{m-1}}
    \sum_{r,s} e^{-ikr} (1+e^{-ik}) e^{iks} (1+e^{ik}) \times \nonumber \\ 
    & \times e^{ik(l_1+ \ldots +l_{m-1} - j_1 - \ldots -j_{m-1})}  \times \nonumber \\
    & \times \bra{n-1,\pi} S^-_r S^-_{r+1} (S^-_{j_1})^2 \ldots (S^-_{j_{m-1}})^2 
    (S^+_{l_1})^2 \ldots (S^+_{l_{m-1}})^2
    S^+_s S^+_{s+1} \ket{n-1,\pi}.
\end{align}
The evaluation of this expression can be performed using an approximation similar to that employed for the denominator: only the terms whose phase does not oscillate are retained, and namely those for which $j_i =l_i$ and $r=s$. 
The term inside the sum can then be evaluated analytically thanks to the special nature of exact quantum many-body scars: it reads
\begin{equation}
    2 (1+ \cos k) \delta_{r,s} \left[ \prod_{t=1}^{m-1} \delta_{j_t, l_t} (1- \delta_{j_t,r}) \right] 2^{2m} \frac{\binom{L-m-1}{n-1}}{\binom{L}{n-1}}.
\end{equation}
We can use the identities:
\begin{equation}
    \sum_r \sum_{j_1< \ldots <j_{m-1}} \prod_{t=1}^{m-1} (1- \delta_{j_t,r}) = L \binom{L-1}{m-1},
    \qquad \qquad 
    L (m-1)!\frac{\binom{L-m-1}{n-1} \binom{L-1}{m-1} }{ \binom{L}{n-1} } = \frac{(L-n+1)!}{(L-m-n)! (L-m)}
\end{equation}
and finally express:
\begin{equation}
   \norm{H (J^{+}_k)^m\ket{n-1,\pi}}^2=
   2 J^2 m^2 (m-1)! (1+ \cos k)\frac{(L-n+1)!}{(L-m-n)! (L-m)}.
\end{equation}
At this stage, we can compute the ratio of the numerator and of the denominator:
\begin{align}
    \Delta H_1^2 \approx& \frac{2 J^2 m^2 (1+ \cos k)}{L-m} \frac{(m-1)! (L-n+1)!}{(L-m-n)!} \frac{(L-m-n+1)!}{m! (L-n+1)!} = \nonumber \\
    =& m 4 J^2 \frac{L-m-n+1}{L-m} \cos^2 \left( \frac{k}{2} \right) \xrightarrow[m+n \ll L]{L \to + \infty} m \times 4 J^2 \cos^2 \left( \frac k2 \right)
\end{align}
We thus obtain that a state obtained by applying $m$ times the $J_k^+$ operator on an exact quantum many-body scars has an energy variance scaling linearly in $m$. Thus, as long as $m$ does not scale with the system size $L$, the state remains and asymptotic quantum many-body scar.

\subsection{Norm and variance of the localized bimagnon state}
In order to highlight the properties of the asymptotic QMBS states, we study here the properties of the localised bimagnon state:
\begin{equation}
\ket{\psi_j}= \frac{1}{\sqrt{4 \frac{L-n+1}{L}}} (S^+_j)^2 \ket{n-1,\pi} =\frac{1}{\sqrt{L (L-n+1)}} \sum_k e^{-ikj}J_k^+ \ket{n-1,\pi} = \sum_k e^{- i k j} \ket{\psi_k}  \qquad \text{for } 1\leq n \leq L.
\end{equation}
The localised bimagnon state is thus a linear superposition of the states $\ket{n,k}$, since expression in Eq.~\eqref{eq:nkexpression} allows us to write: 
\begin{equation}
    \ket{\psi_k} = \frac{1}{\sqrt{L(L-n+1)}} J^+_k \ket{n-1,\pi} = \sqrt{\frac{\mathcal M_{n,k}}{L(L-n+1)}} \ket{n,k}.
\end{equation}
Note that the scaling of the prefactor is $L^{-1/2}$.
This localised bimagnon state has average energy $0$ and thus its energy variance reads:
\begin{align}
    \bra{\psi_j} \left( H_1^2 + H_3^2 \right) \ket{\psi_j} &= \sum_k \bra{\psi_k} \left( H_1^2 + H_3^2 \right) \ket{\psi_k} = \sum_k \frac{\mathcal M_{n,k}}{L(L-n+1)} \bra{n,k} \left( H_1^2 + H_3^2 \right) \ket{n,k},\nonumber \\
    &= \frac{4 (J^2 + J_3^2)(L-n)}{L(L-1)}\sum_{k}{\cos^2(\frac{k}{2})} = \frac{2 (J^2 + J_3^2)(L-n)}{(L-1)}, 
\end{align}
where we have used PBC and hence $\bra{n, k'} H_1 \ket{n,k} = 0$ for $k \neq k'$; and also that $\bra{n,k} H_1 H_3 \ket{n,k} = 0$, and that $\ket{\psi_j}$ is an eigenstate of all the other terms of the Hamiltonian.
It is clear that this energy variance is finite in the thermodynamic limit for any $\frac{n}{L}
< 1$.
%
As a consequence, the fidelity relaxation time of this state is finite in the thermodynamic limit and the state cannot be considered as an asymptotic QMBS.

\section{Entanglement entropy of the exact and asymptotic QMBS}
In this section we review the calculation of the entanglement entropy for the states $\ket{n,k}$, which proceeds along the lines of calculations performed in \cite{vafek2017entanglement, schecter2019weak}.

We first divide the lattice into two parts, $A$ and $B$.
Typically, one considers $A$ as the set of lattice sites with $j\leq L/2$ and $B$ the rest, but this is not necessary.
The key observation is that it is always possible to split the $J^+_k$ operators as a sum of an operator acting on $A$ and of an operator acting on $B$:
\begin{equation}
    J^+_k = J_{k,A}^+ + J_{k,B}^+ = \frac 12 \sum_{j \in A}e^{ikj} \left( S^+_j\right)^2 + \frac 12 \sum_{j \in B}e^{ikj} \left( S^+_j\right)^2.
\end{equation}
The state $\ket{\Downarrow}$ is a product state: $\ket{\Downarrow}_A \otimes \ket{\Downarrow}_B$.
Hence, for $\ket{n, \pi}$, we obtain~\cite{vafek2017entanglement, schecter2019weak}
\begin{equation}
    \ket{n,\pi} = \frac{1}{\sqrt{N_{n,\pi}}}\left(J^+_{\pi,A} + J^+_{\pi,B} \right)^n 
    \ket{\Downarrow}_A \otimes \ket{\Downarrow}_B = 
    \frac{1}{\sqrt{N_{n,\pi}}} \sum_{m=0}^n \binom{n}{m}  \left( J_{\pi,A}^+ \right)^m \ket{\Downarrow}_A \otimes \left( J_{\pi,B}^+ \right)^{n-m} \ket{\Downarrow}_B, 
    \label{Eq:npiExpansion}
\end{equation}
where $N_{n,\pi}$ is the normalization factor for the state $\ket{n,\pi}$, given by $\binom{L}{n}$.
Additional care must be used in truncating the sum in the proper way: if $A$ is composed of $L_A$ lattice sites, it is not possible to apply the $J^+_{k,A}$ operator more than $L_A$ times; similarly for $L_B$.
Hence for simplicity, here we assume that $n < L_A, L_B$.
Therefore, the expansion in Eq.~\eqref{Eq:npiExpansion} gives the Schmidt decomposition of the state, which is composed of the $n+1$ orthogonal states $\{ J_{k,\ell}^m \ket{\Downarrow} \}_{m=0}^n$ for the $\ell \in \{A, B\}$ part.
In the presence of $n+1$ orthogonal states, the highest entropy state is the maximally mixed one, where they all have the same Schmidt coefficients; in that case $S_A =  \log(n+1)$.
If we consider a lattice of length $L$ and the bipartition with $L_A = L_B = L/2$, the states with an extensive number of bimagnons are those such that $n = \alpha L$, with $0<\alpha<1$, and thus these states satisfy the following $S_A \sim \log L + \log \alpha$.
As it is well-known, the quantum many-body scars have an entropy scaling with the logarithm of the volume.

Let us now consider the asymptotic QMBS states, $\ket{n,k}$.
In this case, we use Eq.~(\ref{eq:nkexpression}) to obtain
\begin{align}
    \ket{n,k} =  \frac{1}{\sqrt{\mathcal M_{n,k}}} J^+_{k} \ket{n-1,\pi} = &
    \frac{1}{\sqrt{\mathcal M_{n,k} N_{n-1,\pi}}} \sum_{m=0}^{n-1} 
    \binom{n-1}{m} J^+_{k,A} (J_{\pi,A}^+)^m \ket{\Downarrow}_A \otimes (J_{\pi,B}^+)^{n-1-m} \ket{\Downarrow}_B \nonumber \\ 
    &+ \frac{1}{\sqrt{\mathcal M_{n,k} N_{n-1,\pi}}} \sum_{m=0}^{n-1} 
    \binom{n-1}{m}  (J_{\pi,A}^+)^m \ket{\Downarrow}_A \otimes J^+_{k,B} (J_{\pi,B}^+)^{n-1-m} \ket{\Downarrow}_B.
\label{eq:nkalmostschmidt}
\end{align}
Note that unlike for the $\ket{n, \pi}$, Eq.~(\ref{eq:nkalmostschmidt}) is in general is not the Schmidt decomposition of the state.
Yet, if we consider one subsystem, say A,  the Schmidt states of a fixed magnetisation $-L_A + 2m$ are in the two-dimensional subspace spanned by the following linearly independent states:
\begin{equation}
 \left(J^+_{\pi,A} \right)^m\ket{\Downarrow}_A,
 \qquad 
 J_{k,A}^+\left(J^+_{\pi,A} \right)^{m-1} \ket{\Downarrow}_A.
\end{equation}
Hence we can conclude that the total number of Schmidt states is at most $2n$, and for an extensive number of bimagnons $n=\alpha L$, we obtain that in the highest entropy situation $S_{A} \sim \log 2 + \log \alpha + \log L$.
Thus, with respect to the exact QMBS $\ket{n-1, \pi}$, the asymptotic QMBS $\ket{n, k}$ has at most an additive correction of $\log 2$.
\section{Variance of the exact QMBS for the perturbed Hamiltonian}
We consider the perturbed Hamiltonian $H' = H + V$, where $H$ is the Hamiltonian~\eqref{Eq:Ham:H0:SMAT} with exact scars at finite size with PBC and $V = \frac{J_z}{L} \sum_j S_j^z S_{j+1}^z$.
Since $\ket{n,\pi}$ is an eigenstate of $H$, the variance can be computed focusing only on $V$:
\begin{align}
    \Delta H'^2 &= \Delta V^2 = \bra{n,\pi} V^2 \ket{n,\pi} - \bra{n,\pi} V \ket{n,\pi}^2 \nonumber \\
    &= \frac{J_z^2}{L^2} \sum_{j, j' = 1}^L\left(
    \bra{n,\pi}  S_j^z S_{j+1}^z S_{j'}^z S_{j'+1}^z \ket{n,\pi} - 
    \bra{n,\pi}  S_j^z S_{j+1}^z \ket{n,\pi} \bra{n,\pi} S_{j'}^z S_{j'+1}^z \ket{n,\pi} \right). 
\label{eq:fullvariance}
\end{align}
We can then use the structure of $\ket{n,\pi}$ to compute various correlation functions that appear in Eq.~(\ref{eq:fullvariance}).
We first compute the two point correlation function to be 
\begin{equation}
    \bra{n,\pi}  S_j^z S_{j+1}^z \ket{n,\pi} = \frac{\binom{L-2}{n-2}+\binom{L-2}{n}-2\binom{L-2}{n-1}}{\binom{L}{n}} \equiv F_2,
\label{eq:2pt}
\end{equation}
where we have used the action of $S^z_j S^z_{j+1}$ on the product basis states that compose $\ket{n, \pi}$, i.e., Eq.~(\ref{eq:qmbsproduct}), and noting that it takes the value of $+1$ if there are zero or two bimagnons on sites $j$ and $j+1$, and $-1$ if there is one bimagnon.
Using similar ideas, we obtain that when $j' \neq j -1, j, j+1$, the four point correlation function reads
\begin{equation}
  \bra{n, \pi} S_j^z S_{j+1}^z S_{j'}^z S_{j'+1}^z \ket{n, \pi} = \frac{\binom{L-4}{n}-4\binom{L-4}{n-1}+6\binom{L-4}{n-2}-4\binom{L-4}{n-3}+\binom{L-4}{n-4}}{\binom{L}{n}} \equiv F_4
\label{eq:4pt}
\end{equation}
Note that $F_2$ and $F_4$ in Eqs.~(\ref{eq:2pt}) and (\ref{eq:4pt}) are numbers that only depend on $L$ and $n$, and are independent of $j$; and we have assumed that $n \geq 4$ and PBC.   
When $j' = j-1, j,j+1$, we obtain the following expressions for the ``four point" correlation functions
\begin{equation}
    \bra{n,\pi} S_{j-1}^z \left(S_j^z\right)^2 S_{j+1}^z \ket{n,\pi} = \bra{n,\pi} S_{j}^z \left(S_{j+1}^z\right)^2 S_{j+2}^z \ket{n,\pi} = \bra{n,\pi}  S_j^z  S_{j+1}^z\ket{n,\pi} = F_2,\;\;\; \bra{n,\pi} \left(S_j^z\right)^2 \left(S_{j+1}^z\right)^2 \ket{n,\pi} = 1.
\label{eq:4ptoverlap}
\end{equation}
Combining Eqs.~(\ref{eq:fullvariance})-(\ref{eq:4ptoverlap}), and using translation invariance, we obtain that
\begin{equation}
\Delta H^{'2} = \frac{J_z^2}{L^2}\left[\sum_{j} \sum_{j'\neq j-1,j,j+1}(F_4-F_2^2) +\sum_j(1-F_2^2)+2\sum_j(F_2-F_2^2)\right]=J_z^2\left[F_4\left(1-\frac{3}{L}\right)-F_2^2+\frac{2}{L}F_2+\frac{1}{L}\right]
\label{eq:finalexp}
\end{equation}
Using Eq.~(\ref{eq:finalexp}), we find that when $n/L = \nu$, where $\nu$ is a constant, $\Delta H'^2$ asymptotically scales as $\sim \frac{16 \nu^2 (1 - \nu)^2}{L}$.
On the other hand, when $n$ is kept finite, $\Delta H'^2$ asymptotically scales as $\sim \frac{16 n (n-1)}{L^3}$.
\section{Dynamics of initial states that are not asymptotic QMBS}

\begin{figure}[t]
 \centering
 \includegraphics[width=0.47\columnwidth]{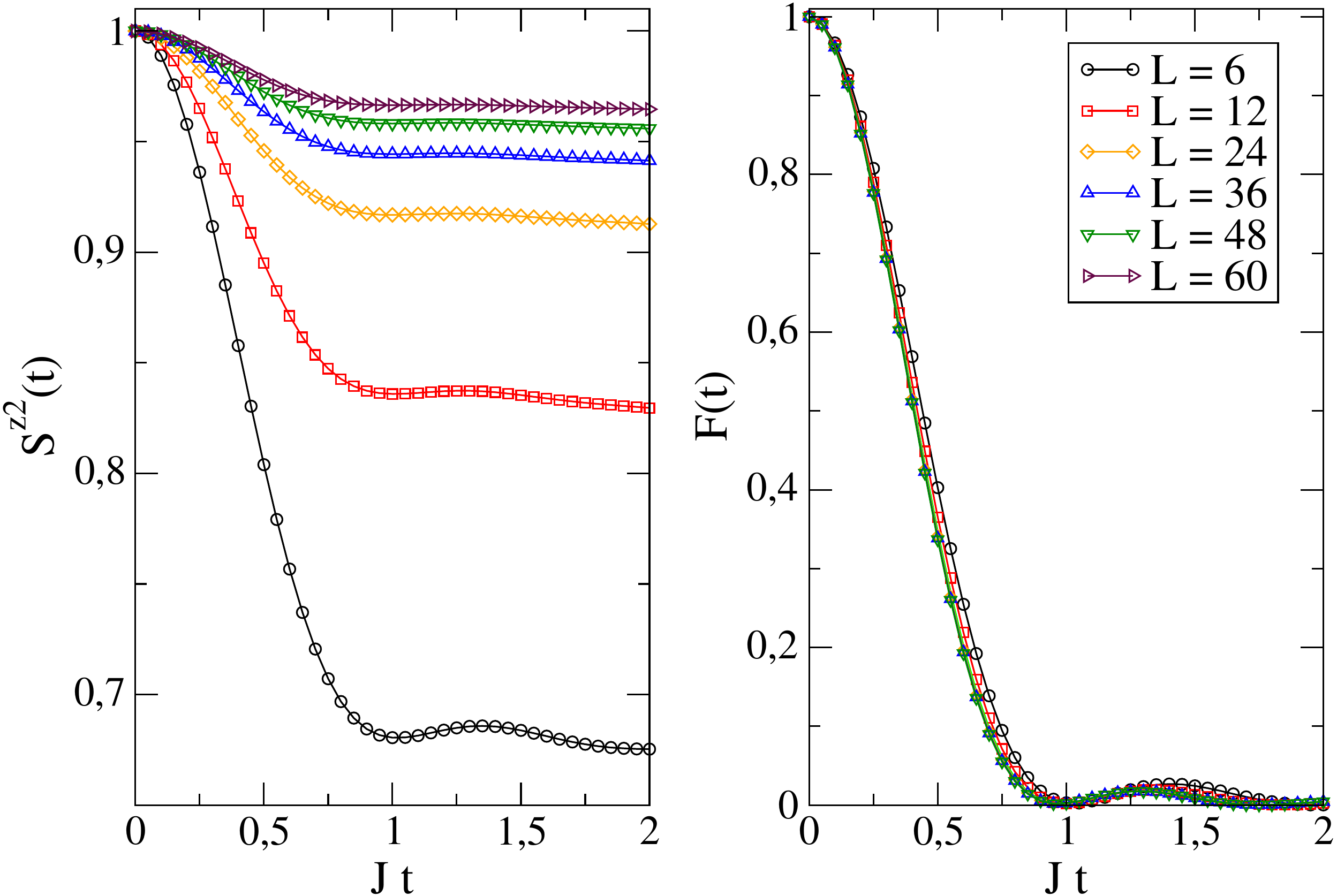}
 \hfill 
  \includegraphics[width=0.47\columnwidth]{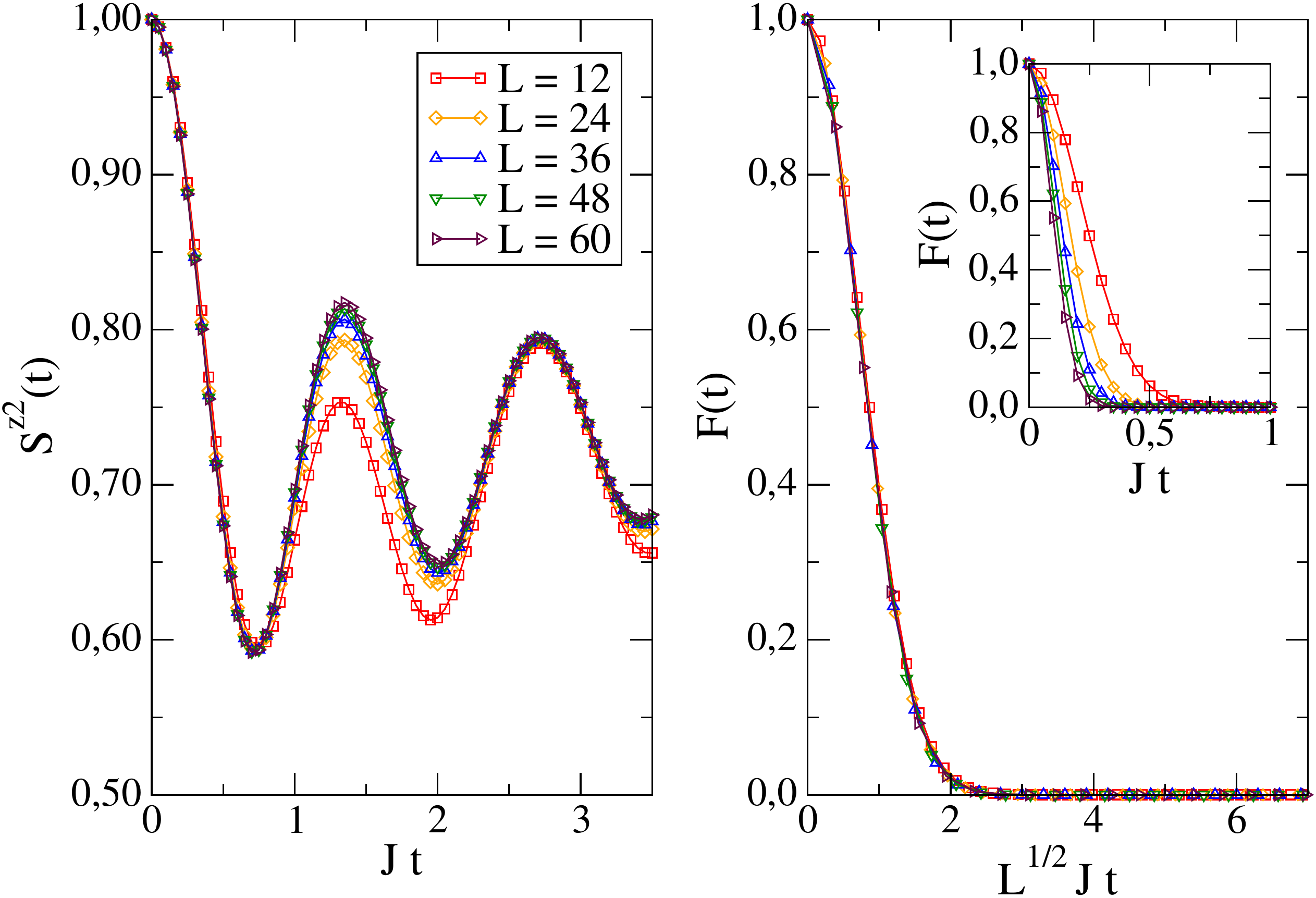}
 \caption{First and second panel: The properties of the state $e^{- i H t}\ket{n,k}$ for $n = L/2$ and $k=0$ as a function of time. First panel: time evolution of the squared magnetization $S^{z2}(t)$. Second panel: time evolution of the fidelity with the initial state $F(t)$.
 Third and fourth panel: The properties of the state $e^{- i H t}\ket{+-+-+-\ldots}$ as a function of time. Third panel: time evolution of the squared magnetization. Fourth panel: time evolution of the fidelity with the initial state $F(t)$; in the inset we show the bare data, whereas in the main plot we rescale time by a factor $\sqrt{L}$ to display a clear collapse.}
 \label{Fig:0:TEV}
\end{figure}
In this section, we study the dynamics of certain initial states, that are not asymptotic QMBS,  under the Hamiltonian $H$ in Eq.~(1) of the main text.
We present this study in order to further support our claim that the dynamics of $\ket{n, k = \pi - 2\pi/L}$ is special.
\subsection{Initial state with finite energy variance}
First, we consider the states $\ket{n, k = 0}$, which are in the family of states in Eq.~(4) of the main text, but are not asymptotic QMBS since they have a finite energy variance in the thermodynamic limit, as evident from Eq.~(5) of the main text.
Note that a state with finite energy variance was already discussed in Ref.~\cite{lin2020slow}, reaching similar conclusions.
In Fig.~\ref{Fig:0:TEV} we study the dynamics of $\ket{n ,k=0}$ by presenting similar numerical results for the time-evolution of the latter state.
The dynamics of the observable $S^{z2}(t)$ is ``activated" on a short time-scale of order $J^{-1}$ that does not depend on $L$ (see the first panel of Fig.~\ref{Fig:0:TEV}).
The dynamics reaches a ``pre-thermal" plateau~\cite{mori2017thermalization} that increases to the initial value for $L\to \infty$.
Note that this result does not contradict the fact that at finite size and in the long-time limit, observables should relax to their thermal value predicted by the diagonal ensemble.
However, the thermalization timescale is much longer than the typical times that we can probe numerically using MPS-based methods.
We have performed long-time simulations using exact diagonalization on small system sizes, and verified that this is indeed the case.
Although the apparently long thermalization time may lead one to consider these states as asymptotic QMBS,  the study of the fidelity with the initial state $F(t)$ is qualitatively very different.
This is shown in the second panel of Fig.~\ref{Fig:0:TEV}: on the same time-scale $J^{-1}$ the state becomes essentially orthogonal to the initial one, and the data for different sizes are basically indistinguishable.
The data on the fidelity relaxation time can be understood as a consequence of the finite energy-variance of the state $\ket{n = L/2, k = 0}$.
\subsection{Initial Product State}
It is also interesting to contrast the dynamics of the asymptotic QMBS with that of an uncorrelated product state; we consider here the staggered state $\ket{\ldots +-+-+- \ldots}$ which has the same zero magnetisation as the states considered in the main text and the same average squared magnetisation as the asymptotic QMBS, equal to one.
The data on the dynamics of $S^{z2}(t)$ collapse on the same curve for all $L$ considered (third panel of Fig.~\ref{Fig:0:TEV}); the fidelity relaxation time instead becomes shorter with increasing $L$ (fourth panel of Fig.~\ref{Fig:0:TEV}).
The behaviour is consistent with expectations for the time evolution of generic product states~\cite{wilming2018equilibration, camposvenuti2010unitary}, and is radically different from that of the asymptotic QMBS.
\section{Spectral properties of the Hamiltonian $H'$}
\begin{figure}[t]
 \includegraphics[width=\textwidth]{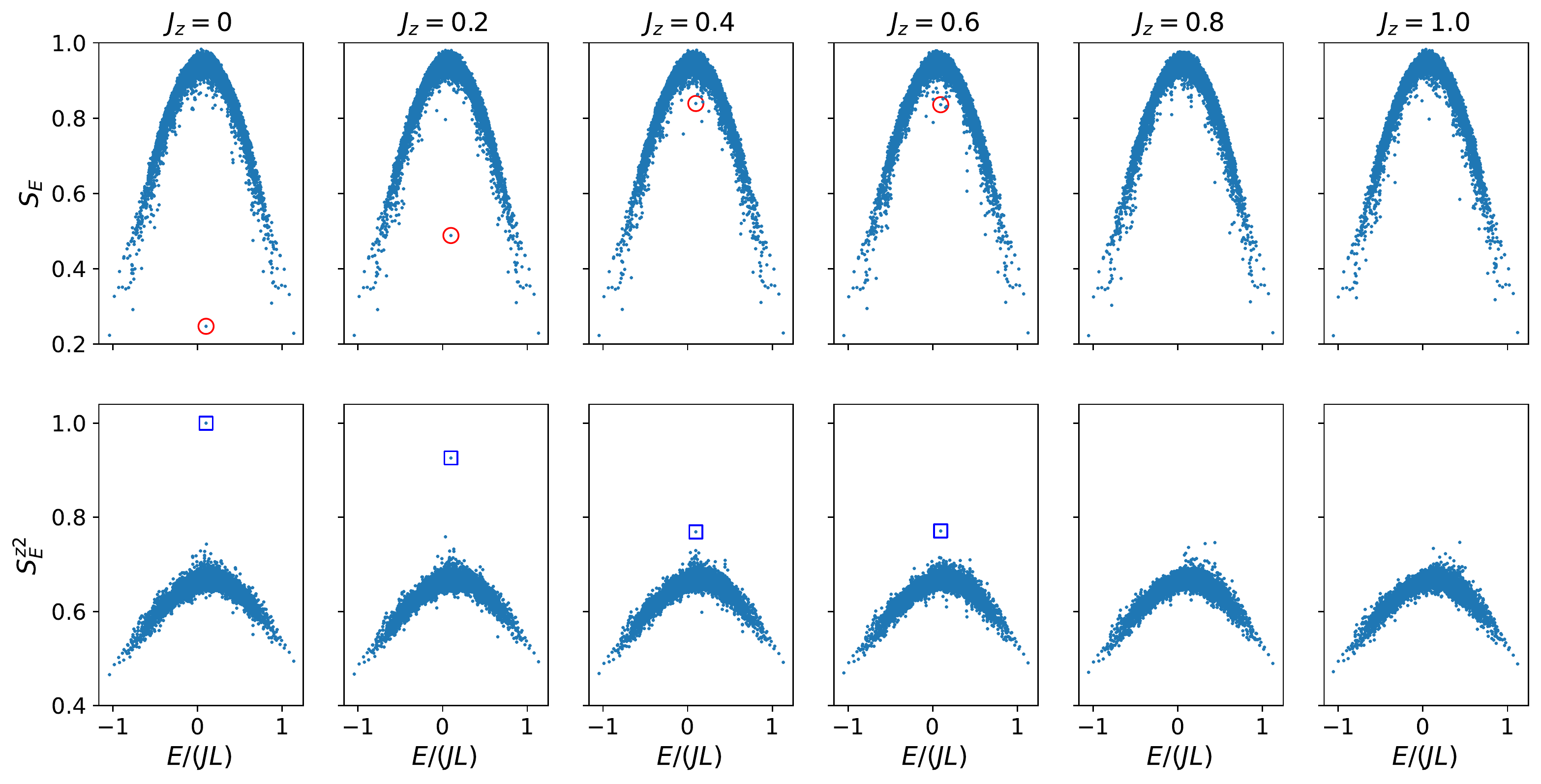}
 \caption{Spectral properties of the Hamiltonian $H' = H+V$ in the zero magnetization sector $S_z=0$ for several values of $J_z$, ranging for $J_z=0$ to $J_z=1$; results are obtained by performing exact diagonalization on a spin chain of length $L=10$. The parameters of the simulation are $\{J, h, D, J_3 \} = \{1, 0, 0.1, 0.1\}$ and $J_z$ is varied.
 In the first line we plot the bipartite entanglement entropy $S_E$ of the eigenstates as a function of their energy $E$; in the second line we focus instead on the expectation value $S^{z2}_E$ of the observable $S^{z2} = \sum_j (S^z_j)^2$ on the eigenstate with energy $E$. The plots highlight the behaviour of the exact scar of the model at $J_z=0$ and that disappears as $J_z$ increases.}
 \label{Fig:5:ScarFlow} 
 \label{Fig:Spectrum:HPrime}
\end{figure}

%
%
In this section, we analyze the spectrum of the Hamiltonian $H' = H + V$ discussed in the main text where $H$ is the spin-1 XY Hamiltonian exhibiting exact QMBS and the perturbation reads:
\begin{equation}V= \frac{J_z}{L} \sum_j S^z_j S^z_{j+1}.\end{equation}
Our goal is to better clarify the disappearance of the exact QMBS that is present for $J_z=0$ and that is absent for $J_z=1$. 
In Fig.~\ref{Fig:5:ScarFlow} we discuss the spectral properties of the model for several values of $J_z$ ranging from $0$ to $1$ for a spin chain of length $L=10$. The plots show the bipartite entanglement entropy of all eigenstates and the expectation value of $S^{z2} = \sum_j (S^z)^2$.
At these system sizes, we observe the presence of a clear outlying state for $J_z \lesssim 0.2$ in both the entanglement entropy and the observable.
For $J_z \lesssim 0.6$ we can observe a state that is an outlier in what concerns the expectation value of $S^{z2}$, but that has an elevated entanglement entropy, comparable to that of other eigenstates with the same energy.
For larger values of $J_z$ it is difficult to identify a unique outlier QMBS, although the spectrum maintains a few states that are not collapsed on the main curve. 
It is important to stress that these simulations have been performed at finite size and that a proper scaling towards the thermodynamic limit could make disappear the outliers that we have shown for $J_z \neq 0 $.
We also considered the case of negative values of $J_z$ and obtained results very similar to those in Fig.~\ref{Fig:5:ScarFlow}, which are not reported here for brevity.

\section{Universal rescaling of fidelities}
\begin{figure}[t]
\centering
\includegraphics[width=0.666666\textwidth]{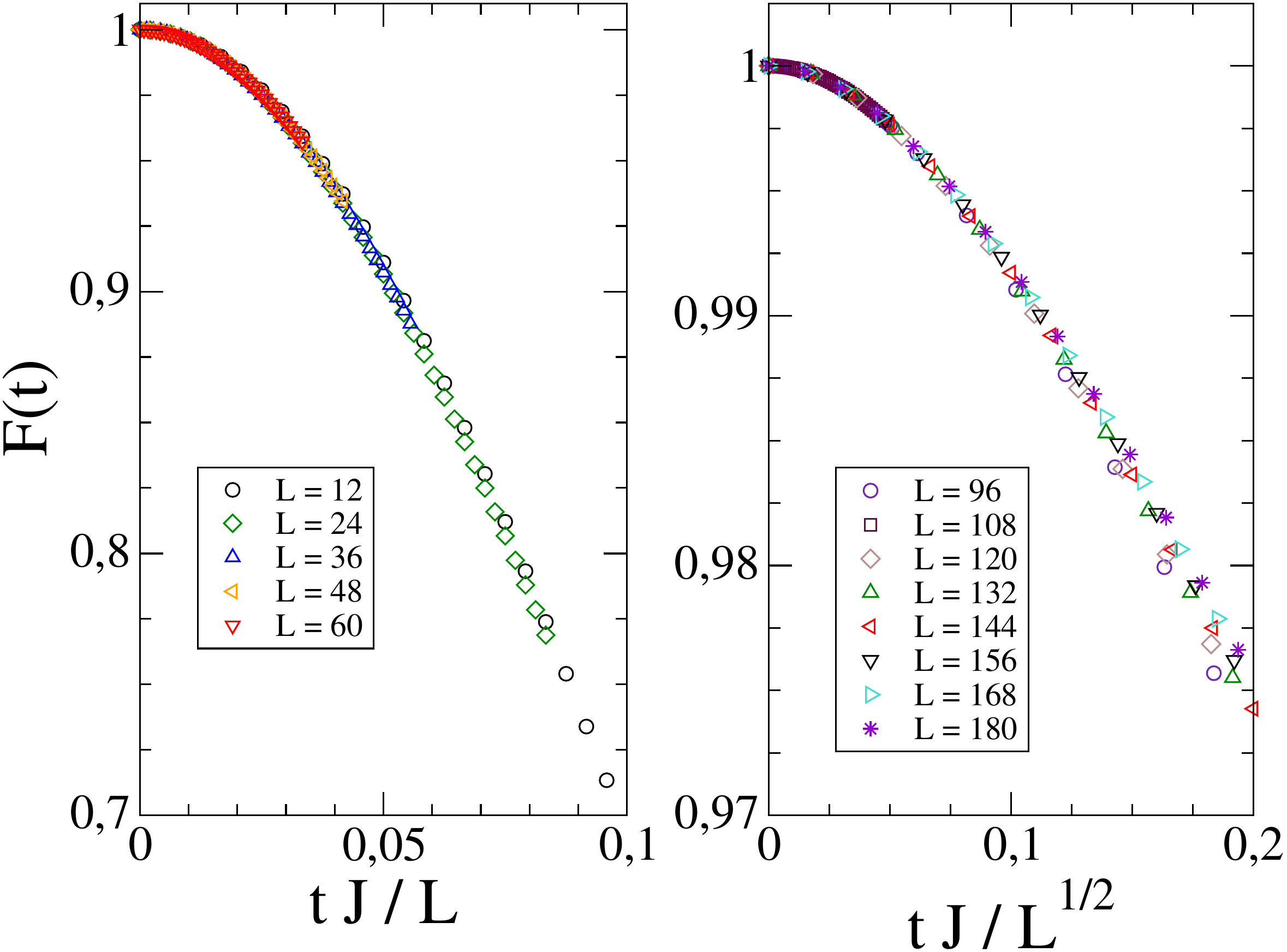}
\caption{Rescaling of the fidelities $F(t)$ plotted in Fig.~2 and Fig.~4 of the main text.}
\label{Fig:SM:Fid:Rescaling}
\end{figure}
In this section, we present the data collapse of the fidelities for the asymptotic QMBS for various system sizes presented in the main text.
Such a data collapse occurs at short times, once the time is rescaled by a factor that depends on the size of the system, as shown in Fig.~\ref{Fig:SM:Fid:Rescaling}.
In the left panel, we present data for the asymptotic QMBS $\ket{n, k = \pi - 2 \pi/L}$ time-evolved with the spin-1 XY Hamiltonian $H$ of Eq.~(1) of the main text, which includes the term proportional to $J_3$, and the collapse is obtained by rescaling the time as $\tau = t/L$.
In the right panel, we present data for the state $\ket{n, \pi}$ time-evolved with the Hamiltonian $H' = H + V$; the collapse is obtained by rescaling the time as $\tau = t / L^{1/2}$.
It is interesting to link these results to the energy-time uncertainty relation in Eq.~(6) of the main text, whose proof is presented in many quantum mechanics textbooks and will not be reviewed here.
The overlap of the time-evolved state with the initial one is related to the expectation value of the Hamiltonian and of its powers as~\cite{camposvenuti2010unitary}
\begin{equation}
    \bra \Psi e^{- i H t } \ket{\Psi} \approx 1 - i t \bra{\Psi} H \ket{\Psi} - \frac{1}{2}t^2 \bra{\Psi} H^2 \ket{\Psi} + \frac{i}{6}t^3 \bra{\Psi} H^3 \ket{\Psi} + \ldots
\end{equation}
and thus we can express the fidelity as
\begin{equation}
    F(t) = | \bra \Psi e^{- i H t } \ket{\Psi}  |^2 \approx 1 - t^2 \left( \bra{\Psi} H^2 \ket{\Psi} - \bra{\Psi} H \ket{\Psi}^2\right) + \ldots
\end{equation}
The short-time fidelity dynamics is thus completely dictated by the energy-variance of the initial state with respect to the Hamiltonian of the dynamics.

Note that the precise scaling of the relaxation time depends on the definition. The fidelity of an initial state at short times decays as $F(t) \sim \exp(- \Delta H^2 t^2)$~\cite{camposvenuti2010unitary}, where $\Delta H^2$ is the variance, this gives a timescale $\tau \sim 1/\Delta H$. On the other hand, one can define the fidelity relaxation time as the timescale at which fidelity decays to the the typical fidelity between two many body states, which scales as $\exp(-L)$~\cite{camposvenuti2010unitary}, this adds an extra factor of $\sqrt{L}$. In this work, we use the former definition, and are mostly interesting in the relative decay timescales between initial states of different variances.

It is interesting to study a state with a Gaussian energy spread, for which the calculation of the time-dynamics of the fidelity is exactly possible.
In fact here it is possible to show that it minimizes the inequality and has a fidelity $F(t)$ whose dynamics happens on the shortest possible timescale.
Consider indeed an initial state that is a Gaussian linear superposition of energy eigenstates with average energy $E_0$ and energy variance $\sigma^2$ (we introduce also a normalisation prefactor $\alpha \in \mathbb C$):
Assuming that the density of states in the energy window $[E_0 - \sigma, E_0+ \sigma]$ is approximately constant and takes the value $\rho(E_0)$,
the scalar product between the time-evolved state and the initial one is given by:
\begin{equation}
    \bra{\Psi_0} e^{- i H t} \ket{\Psi_0} \approx \int |\alpha|^2 e^{- \frac{(E-E_0)^2}{2 \sigma^2}} e^{- i Et} \rho(E) dE = |\alpha|^2 \sqrt{ 2 \pi \sigma^2 } e^{- i E_0 t} e^{- \frac{\sigma^2 t^2}{2}} \rho(E_0).
\end{equation}
The normalisation of the state, computed for $t=0$, requires that $|\alpha|^2 \sqrt{ 2 \pi \sigma^2 } \rho(E_0) = 1$.
The fidelity $F(t)$ is the squared modulus of this scalar product and hence $F(t) = \exp \left[-\sigma^2 t^2 \right]$;
we can define the typical time scale of the fidelity dynamics as $\tau = 1/(2 \sigma)$, and the energy-time inequality is satisfied and minimised.
In general terms, we thus expect that the dynamics of the fidelity at short times takes place on time-scales that are the shortest possible and minimize the energy-time inequality. This short-time behaviour is indeed verified by the numerics plotted in Fig.~\ref{Fig:SM:Fid:Rescaling}.
In the left panel we have $\tau \sim L$ and $\sigma \sim 1/L$; in the right panel we have $\tau \sim \sqrt{L}$ and $\sigma \sim 1/\sqrt{L}$. 
Note that this timescale also matches the rigorous lower bounds on relaxation times for weak perturbations of models with exact QMBS~\cite{lin2020slow} by setting the perturbation strength $\lambda = 1/L$, although we note the latter is the observable relaxation time, which we generically expect to be different from the fidelity relaxation time we discuss.
However, it is important to keep in mind that the numerics has been performed only at short times and that long-time behaviours would need further investigation.
\section{Higher dimensional generalisations of asymptotic QMBS}
Finally, we show that the existence of the asymptotic QMBS is not limited to one-dimensional systems, but can be easily generalised to higher-dimensional lattices.
As an example, we consider a simple cubic Bravais lattice in $d>1$ dimensions with primitive vectors $\mathbf t_i$ and $i=1, \ldots d$; the vectors are adimensional and orthonormal: $\mathbf t_i \cdot \mathbf t_j = \delta_{ij}$.
The lattice has linear dimension $L$ and is composed of $L^d$ sites; periodic boundary conditions (PBC) are applied.
On each site of the lattice there is a spin-1 degree of freedom and we define the spin-1 operators $S_{\mathbf r}^\alpha$, with $\alpha = x,y,z$.
We then consider a nearest-neighbor XY model with external magnetic field:
\begin{equation}
 H = J \sum_{ \bf r} \sum_{i=1}^d \left( S^x_{\bf r} S^x_{\mathbf r+ \mathbf t_i} + S^y_{\bf r} S^y_{\mathbf r+ \mathbf t_i} \right) +h \sum_{\bf r} S_{\bf r}^{z}. \label{Eq:Ham}
\end{equation}
As discussed in \cite{schecter2019weak, mark2020eta}, this model  in Eq.~\eqref{Eq:Ham} exhibits exact QMBS for any finite value of $L$ and for any dimension $d$.
Note that when $d > 1$, the model of Eq.~(\ref{Eq:Ham}) is non-integrable, and unlike in the one-dimensional case in Eq.~(1) of the main text, we need not add the anisotropy term proportional to $D$ or the longer range term proportional to $J_3$ to break integrability or unusual symmetries.
Starting from the fully-polarised state $\ket{\Downarrow}$, we define the quasiparticle creation operator $J^+_{\bk} = \frac{1}{2} \sum_{\br} e^{i \mathbf{k \cdot r}} \left( S^+_{\br}\right)^2$.
The exact QMBS states then read
$\ket{n, \boldsymbol \pi} = \frac{1}{\sqrt{N_{n, \boldsymbol \pi}}}\left( J^+_{\boldsymbol \pi} \right)^n \ket{\Downarrow}$
where $\boldsymbol \pi$ is the vector with all $d$ components equal to $\pi$.
It is easy to show that $H \ket{n, \boldsymbol \pi}=h (-L^d + 2n)\ket{n, \boldsymbol \pi}$, hence the state is an exact QMBS in the middle of the spectrum of the Hamiltonian~\cite{schecter2019weak, mark2020eta, wildeboer2022quantum}. 
The states that we are interested in are:
\begin{equation}
 \ket{n, \mathbf k} = \frac{1}{\sqrt{N_{n, \mathbf k}}}  J^+_{\mathbf k} (J^+_{\boldsymbol \pi})^{n-1} \ket{\Downarrow},
\end{equation}
where $\bk$ is any vector of the reciprocal space confined to the first Brillouin zone (1BZ).
Similar to the one-dimensional case, it is possible to show that as long as the momentum $\bk$ is chosen compatible with PBC in all directions, we can show that $\braket{n, \bk}{n', \bk'} = \delta_{n, n'} \delta_{\bk, \bk'}$.
With these states, we can directly repeat the proof in Sec.~\ref{sec:avgenergy} \textit{mutatis mutandis}.
We find that the average energy is given by $\bra{n, \mathbf k} H \ket{n,\mathbf k} = h (- L^d + 2n)$, and the energy variance is given by
\begin{equation}
 \Delta H^2 = \frac{4 J^2 \sum_{i = 1}^d \cos^2 \left( \frac{k_i}{2}\right)}{1+\frac{(n-1)L^d}{L^d-n}\delta_{\vec{k},\vec{\pi}}} = 4 J^2 \sum_{i = 1}^d \cos^2 \left( \frac{k_i}{2}\right).
\end{equation}
Thus, if we consider $\bk$ with components $k_i = \pi + \frac{2 \pi}{L}m_i$ and keep the $m_i\in \mathbb Z$ fixed while $L \to \infty$, the variance reduces to zero while being orthogonal to the exact QMBS. 
For such states, we expect the same phenomenology of asymptotic QMBS discussed for the one-dimensional case.


\begin{thebibliography}{78}%
\makeatletter
\providecommand \@ifxundefined [1]{%
 \@ifx{#1\undefined}
}%
\providecommand \@ifnum [1]{%
 \ifnum #1\expandafter \@firstoftwo
 \else \expandafter \@secondoftwo
 \fi
}%
\providecommand \@ifx [1]{%
 \ifx #1\expandafter \@firstoftwo
 \else \expandafter \@secondoftwo
 \fi
}%
\providecommand \natexlab [1]{#1}%
\providecommand \enquote  [1]{``#1''}%
\providecommand \bibnamefont  [1]{#1}%
\providecommand \bibfnamefont [1]{#1}%
\providecommand \citenamefont [1]{#1}%
\providecommand \href@noop [0]{\@secondoftwo}%
\providecommand \href [0]{\begingroup \@sanitize@url \@href}%
\providecommand \@href[1]{\@@startlink{#1}\@@href}%
\providecommand \@@href[1]{\endgroup#1\@@endlink}%
\providecommand \@sanitize@url [0]{\catcode `\\12\catcode `\$12\catcode
  `\&12\catcode `\#12\catcode `\^12\catcode `\_12\catcode `\%12\relax}%
\providecommand \@@startlink[1]{}%
\providecommand \@@endlink[0]{}%
\providecommand \url  [0]{\begingroup\@sanitize@url \@url }%
\providecommand \@url [1]{\endgroup\@href {#1}{\urlprefix }}%
\providecommand \urlprefix  [0]{URL }%
\providecommand \Eprint [0]{\href }%
\providecommand \doibase [0]{http://dx.doi.org/}%
\providecommand \selectlanguage [0]{\@gobble}%
\providecommand \bibinfo  [0]{\@secondoftwo}%
\providecommand \bibfield  [0]{\@secondoftwo}%
\providecommand \translation [1]{[#1]}%
\providecommand \BibitemOpen [0]{}%
\providecommand \bibitemStop [0]{}%
\providecommand \bibitemNoStop [0]{.\EOS\space}%
\providecommand \EOS [0]{\spacefactor3000\relax}%
\providecommand \BibitemShut  [1]{\csname bibitem#1\endcsname}%
\let\auto@bib@innerbib\@empty
\bibitem [{\citenamefont {Serbyn}\ \emph {et~al.}(2021)\citenamefont {Serbyn},
  \citenamefont {Abanin},\ and\ \citenamefont
  {Papi{\'{c}}}}]{serbyn2020review}%
  \BibitemOpen
  \bibfield  {author} {\bibinfo {author} {\bibfnamefont {Maksym}\ \bibnamefont
  {Serbyn}}, \bibinfo {author} {\bibfnamefont {Dmitry~A.}\ \bibnamefont
  {Abanin}}, \ and\ \bibinfo {author} {\bibfnamefont {Zlatko}\ \bibnamefont
  {Papi{\'{c}}}},\ }\bibfield  {title} {\enquote {\bibinfo {title} {Quantum
  many-body scars and weak breaking of ergodicity},}\ }\href {\doibase
  10.1038/s41567-021-01230-2} {\bibfield  {journal} {\bibinfo  {journal}
  {Nature Physics}\ }\textbf {\bibinfo {volume} {17}},\ \bibinfo {pages}
  {675--685} (\bibinfo {year} {2021})}\BibitemShut {NoStop}%
\bibitem [{\citenamefont {Papi{\'{c}}}(2022)}]{papic2021review}%
  \BibitemOpen
  \bibfield  {author} {\bibinfo {author} {\bibfnamefont {Zlatko}\ \bibnamefont
  {Papi{\'{c}}}},\ }\enquote {\bibinfo {title} {{Weak Ergodicity Breaking
  Through the Lens of Quantum Entanglement}},}\ in\ \href
  {https://doi.org/10.1007/978-3-031-03998-0_13} {\emph {\bibinfo {booktitle}
  {{Entanglement in Spin Chains: From Theory to Quantum Technology
  Applications}}}},\ \bibinfo {editor} {edited by\ \bibinfo {editor}
  {\bibfnamefont {Abolfazl}\ \bibnamefont {Bayat}}, \bibinfo {editor}
  {\bibfnamefont {Sougato}\ \bibnamefont {Bose}}, \ and\ \bibinfo {editor}
  {\bibfnamefont {Henrik}\ \bibnamefont {Johannesson}}}\ (\bibinfo  {publisher}
  {Springer International Publishing},\ \bibinfo {address} {Cham},\ \bibinfo
  {year} {2022})\ pp.\ \bibinfo {pages} {341--395}\BibitemShut {NoStop}%
\bibitem [{\citenamefont {Moudgalya}\ \emph {et~al.}(2022)\citenamefont
  {Moudgalya}, \citenamefont {Bernevig},\ and\ \citenamefont
  {Regnault}}]{moudgalya2021review}%
  \BibitemOpen
  \bibfield  {author} {\bibinfo {author} {\bibfnamefont {Sanjay}\ \bibnamefont
  {Moudgalya}}, \bibinfo {author} {\bibfnamefont {B~Andrei}\ \bibnamefont
  {Bernevig}}, \ and\ \bibinfo {author} {\bibfnamefont {Nicolas}\ \bibnamefont
  {Regnault}},\ }\bibfield  {title} {\enquote {\bibinfo {title} {Quantum
  many-body scars and hilbert space fragmentation: a review of exact
  results},}\ }\href {\doibase 10.1088/1361-6633/ac73a0} {\bibfield  {journal}
  {\bibinfo  {journal} {Reports on Progress in Physics}\ }\textbf {\bibinfo
  {volume} {85}},\ \bibinfo {pages} {086501} (\bibinfo {year}
  {2022})}\BibitemShut {NoStop}%
\bibitem [{\citenamefont {Chandran}\ \emph {et~al.}(2023)\citenamefont
  {Chandran}, \citenamefont {Iadecola}, \citenamefont {Khemani},\ and\
  \citenamefont {Moessner}}]{chandran2022review}%
  \BibitemOpen
  \bibfield  {author} {\bibinfo {author} {\bibfnamefont {Anushya}\ \bibnamefont
  {Chandran}}, \bibinfo {author} {\bibfnamefont {Thomas}\ \bibnamefont
  {Iadecola}}, \bibinfo {author} {\bibfnamefont {Vedika}\ \bibnamefont
  {Khemani}}, \ and\ \bibinfo {author} {\bibfnamefont {Roderich}\ \bibnamefont
  {Moessner}},\ }\bibfield  {title} {\enquote {\bibinfo {title} {{Quantum
  Many-Body Scars: A Quasiparticle Perspective}},}\ }\href {\doibase
  10.1146/annurev-conmatphys-031620-101617} {\bibfield  {journal} {\bibinfo
  {journal} {Annual Review of Condensed Matter Physics}\ }\textbf {\bibinfo
  {volume} {14}},\ \bibinfo {pages} {443--469} (\bibinfo {year}
  {2023})}\BibitemShut {NoStop}%
\bibitem [{\citenamefont {Deutsch}(1991)}]{deutsch1991quantum}%
  \BibitemOpen
  \bibfield  {author} {\bibinfo {author} {\bibfnamefont {J.~M.}\ \bibnamefont
  {Deutsch}},\ }\bibfield  {title} {\enquote {\bibinfo {title} {Quantum
  statistical mechanics in a closed system},}\ }\href {\doibase
  10.1103/PhysRevA.43.2046} {\bibfield  {journal} {\bibinfo  {journal}
  {Physical Review A}\ }\textbf {\bibinfo {volume} {43}},\ \bibinfo {pages}
  {2046--2049} (\bibinfo {year} {1991})}\BibitemShut {NoStop}%
\bibitem [{\citenamefont {Srednicki}(1994)}]{srednicki1994chaos}%
  \BibitemOpen
  \bibfield  {author} {\bibinfo {author} {\bibfnamefont {Mark}\ \bibnamefont
  {Srednicki}},\ }\bibfield  {title} {\enquote {\bibinfo {title} {Chaos and
  quantum thermalization},}\ }\href {\doibase 10.1103/PhysRevE.50.888}
  {\bibfield  {journal} {\bibinfo  {journal} {Physical Review E}\ }\textbf
  {\bibinfo {volume} {50}},\ \bibinfo {pages} {888--901} (\bibinfo {year}
  {1994})}\BibitemShut {NoStop}%
\bibitem [{\citenamefont {{Rigol}}\ \emph {et~al.}(2008)\citenamefont
  {{Rigol}}, \citenamefont {{Dunjko}},\ and\ \citenamefont
  {{Olshanii}}}]{rigol2008thermalization}%
  \BibitemOpen
  \bibfield  {author} {\bibinfo {author} {\bibfnamefont {Marcos}\ \bibnamefont
  {{Rigol}}}, \bibinfo {author} {\bibfnamefont {Vanja}\ \bibnamefont
  {{Dunjko}}}, \ and\ \bibinfo {author} {\bibfnamefont {Maxim}\ \bibnamefont
  {{Olshanii}}},\ }\bibfield  {title} {\enquote {\bibinfo {title}
  {{Thermalization and its mechanism for generic isolated quantum systems}},}\
  }\href {\doibase 10.1038/nature06838} {\bibfield  {journal} {\bibinfo
  {journal} {Nature}\ }\textbf {\bibinfo {volume} {452}},\ \bibinfo {pages}
  {854--858} (\bibinfo {year} {2008})}\BibitemShut {NoStop}%
\bibitem [{\citenamefont {Polkovnikov}\ \emph {et~al.}(2011)\citenamefont
  {Polkovnikov}, \citenamefont {Sengupta}, \citenamefont {Silva},\ and\
  \citenamefont {Vengalattore}}]{polkovnikov2011colloquium}%
  \BibitemOpen
  \bibfield  {author} {\bibinfo {author} {\bibfnamefont {Anatoli}\ \bibnamefont
  {Polkovnikov}}, \bibinfo {author} {\bibfnamefont {Krishnendu}\ \bibnamefont
  {Sengupta}}, \bibinfo {author} {\bibfnamefont {Alessandro}\ \bibnamefont
  {Silva}}, \ and\ \bibinfo {author} {\bibfnamefont {Mukund}\ \bibnamefont
  {Vengalattore}},\ }\bibfield  {title} {\enquote {\bibinfo {title}
  {Colloquium: Nonequilibrium dynamics of closed interacting quantum
  systems},}\ }\href {\doibase 10.1103/RevModPhys.83.863} {\bibfield  {journal}
  {\bibinfo  {journal} {Reviews of Modern Physics}\ }\textbf {\bibinfo {volume}
  {83}},\ \bibinfo {pages} {863--883} (\bibinfo {year} {2011})}\BibitemShut
  {NoStop}%
\bibitem [{\citenamefont {{D'Alessio}}\ \emph {et~al.}(2016)\citenamefont
  {{D'Alessio}}, \citenamefont {{Kafri}}, \citenamefont {{Polkovnikov}},\ and\
  \citenamefont {{Rigol}}}]{d2016quantum}%
  \BibitemOpen
  \bibfield  {author} {\bibinfo {author} {\bibfnamefont {Luca}\ \bibnamefont
  {{D'Alessio}}}, \bibinfo {author} {\bibfnamefont {Yariv}\ \bibnamefont
  {{Kafri}}}, \bibinfo {author} {\bibfnamefont {Anatoli}\ \bibnamefont
  {{Polkovnikov}}}, \ and\ \bibinfo {author} {\bibfnamefont {Marcos}\
  \bibnamefont {{Rigol}}},\ }\bibfield  {title} {\enquote {\bibinfo {title}
  {{From quantum chaos and eigenstate thermalization to statistical mechanics
  and thermodynamics}},}\ }\href {\doibase 10.1080/00018732.2016.1198134}
  {\bibfield  {journal} {\bibinfo  {journal} {Advances in Physics}\ }\textbf
  {\bibinfo {volume} {65}},\ \bibinfo {pages} {239--362} (\bibinfo {year}
  {2016})}\BibitemShut {NoStop}%
\bibitem [{\citenamefont {Mori}\ \emph {et~al.}(2018)\citenamefont {Mori},
  \citenamefont {Ikeda}, \citenamefont {Kaminishi},\ and\ \citenamefont
  {Ueda}}]{mori2018thermalization}%
  \BibitemOpen
  \bibfield  {author} {\bibinfo {author} {\bibfnamefont {Takashi}\ \bibnamefont
  {Mori}}, \bibinfo {author} {\bibfnamefont {Tatsuhiko~N}\ \bibnamefont
  {Ikeda}}, \bibinfo {author} {\bibfnamefont {Eriko}\ \bibnamefont
  {Kaminishi}}, \ and\ \bibinfo {author} {\bibfnamefont {Masahito}\
  \bibnamefont {Ueda}},\ }\bibfield  {title} {\enquote {\bibinfo {title}
  {Thermalization and prethermalization in isolated quantum systems: a
  theoretical overview},}\ }\href {\doibase 10.1088/1361-6455/aabcdf}
  {\bibfield  {journal} {\bibinfo  {journal} {Journal of Physics B: Atomic,
  Molecular and Optical Physics}\ }\textbf {\bibinfo {volume} {51}},\ \bibinfo
  {pages} {112001} (\bibinfo {year} {2018})}\BibitemShut {NoStop}%
\bibitem [{\citenamefont {Moudgalya}\ \emph
  {et~al.}(2018{\natexlab{a}})\citenamefont {Moudgalya}, \citenamefont
  {Rachel}, \citenamefont {Bernevig},\ and\ \citenamefont
  {Regnault}}]{moudgalya2018exact}%
  \BibitemOpen
  \bibfield  {author} {\bibinfo {author} {\bibfnamefont {Sanjay}\ \bibnamefont
  {Moudgalya}}, \bibinfo {author} {\bibfnamefont {Stephan}\ \bibnamefont
  {Rachel}}, \bibinfo {author} {\bibfnamefont {B.~Andrei}\ \bibnamefont
  {Bernevig}}, \ and\ \bibinfo {author} {\bibfnamefont {Nicolas}\ \bibnamefont
  {Regnault}},\ }\bibfield  {title} {\enquote {\bibinfo {title} {Exact excited
  states of nonintegrable models},}\ }\href {\doibase
  10.1103/PhysRevB.98.235155} {\bibfield  {journal} {\bibinfo  {journal}
  {Physical Review B}\ }\textbf {\bibinfo {volume} {98}},\ \bibinfo {pages}
  {235155} (\bibinfo {year} {2018}{\natexlab{a}})}\BibitemShut {NoStop}%
\bibitem [{\citenamefont {Moudgalya}\ \emph
  {et~al.}(2018{\natexlab{b}})\citenamefont {Moudgalya}, \citenamefont
  {Regnault},\ and\ \citenamefont {Bernevig}}]{moudgalya2018entanglement}%
  \BibitemOpen
  \bibfield  {author} {\bibinfo {author} {\bibfnamefont {Sanjay}\ \bibnamefont
  {Moudgalya}}, \bibinfo {author} {\bibfnamefont {Nicolas}\ \bibnamefont
  {Regnault}}, \ and\ \bibinfo {author} {\bibfnamefont {B.~Andrei}\
  \bibnamefont {Bernevig}},\ }\bibfield  {title} {\enquote {\bibinfo {title}
  {Entanglement of exact excited states of affleck-kennedy-lieb-tasaki models:
  Exact results, many-body scars, and violation of the strong eigenstate
  thermalization hypothesis},}\ }\href {\doibase 10.1103/PhysRevB.98.235156}
  {\bibfield  {journal} {\bibinfo  {journal} {Phys. Rev. B}\ }\textbf {\bibinfo
  {volume} {98}},\ \bibinfo {pages} {235156} (\bibinfo {year}
  {2018}{\natexlab{b}})}\BibitemShut {NoStop}%
\bibitem [{\citenamefont {Mark}\ \emph {et~al.}(2020)\citenamefont {Mark},
  \citenamefont {Lin},\ and\ \citenamefont {Motrunich}}]{mark2020unified}%
  \BibitemOpen
  \bibfield  {author} {\bibinfo {author} {\bibfnamefont {Daniel~K.}\
  \bibnamefont {Mark}}, \bibinfo {author} {\bibfnamefont {Cheng-Ju}\
  \bibnamefont {Lin}}, \ and\ \bibinfo {author} {\bibfnamefont {Olexei~I.}\
  \bibnamefont {Motrunich}},\ }\bibfield  {title} {\enquote {\bibinfo {title}
  {Unified structure for exact towers of scar states in the
  affleck-kennedy-lieb-tasaki and other models},}\ }\href {\doibase
  10.1103/PhysRevB.101.195131} {\bibfield  {journal} {\bibinfo  {journal}
  {Phys. Rev. B}\ }\textbf {\bibinfo {volume} {101}},\ \bibinfo {pages}
  {195131} (\bibinfo {year} {2020})}\BibitemShut {NoStop}%
\bibitem [{\citenamefont {Schecter}\ and\ \citenamefont
  {Iadecola}(2019)}]{schecter2019weak}%
  \BibitemOpen
  \bibfield  {author} {\bibinfo {author} {\bibfnamefont {Michael}\ \bibnamefont
  {Schecter}}\ and\ \bibinfo {author} {\bibfnamefont {Thomas}\ \bibnamefont
  {Iadecola}},\ }\bibfield  {title} {\enquote {\bibinfo {title} {Weak
  ergodicity breaking and quantum many-body scars in spin-1 $xy$ magnets},}\
  }\href {\doibase 10.1103/PhysRevLett.123.147201} {\bibfield  {journal}
  {\bibinfo  {journal} {Phys. Rev. Lett.}\ }\textbf {\bibinfo {volume} {123}},\
  \bibinfo {pages} {147201} (\bibinfo {year} {2019})}\BibitemShut {NoStop}%
\bibitem [{\citenamefont {Wildeboer}\ \emph {et~al.}(2022)\citenamefont
  {Wildeboer}, \citenamefont {Langlett}, \citenamefont {Yang}, \citenamefont
  {Gorshkov}, \citenamefont {Iadecola},\ and\ \citenamefont
  {Xu}}]{wildeboer2022quantum}%
  \BibitemOpen
  \bibfield  {author} {\bibinfo {author} {\bibfnamefont {Julia}\ \bibnamefont
  {Wildeboer}}, \bibinfo {author} {\bibfnamefont {Christopher~M.}\ \bibnamefont
  {Langlett}}, \bibinfo {author} {\bibfnamefont {Zhi-Cheng}\ \bibnamefont
  {Yang}}, \bibinfo {author} {\bibfnamefont {Alexey~V.}\ \bibnamefont
  {Gorshkov}}, \bibinfo {author} {\bibfnamefont {Thomas}\ \bibnamefont
  {Iadecola}}, \ and\ \bibinfo {author} {\bibfnamefont {Shenglong}\
  \bibnamefont {Xu}},\ }\bibfield  {title} {\enquote {\bibinfo {title} {Quantum
  many-body scars from einstein-podolsky-rosen states in bilayer systems},}\
  }\href {\doibase 10.1103/PhysRevB.106.205142} {\bibfield  {journal} {\bibinfo
   {journal} {Phys. Rev. B}\ }\textbf {\bibinfo {volume} {106}},\ \bibinfo
  {pages} {205142} (\bibinfo {year} {2022})}\BibitemShut {NoStop}%
\bibitem [{\citenamefont {Yang}(1989)}]{yang1989eta}%
  \BibitemOpen
  \bibfield  {author} {\bibinfo {author} {\bibfnamefont {Chen~Ning}\
  \bibnamefont {Yang}},\ }\bibfield  {title} {\enquote {\bibinfo {title}
  {$\eta$ pairing and off-diagonal long-range order in a hubbard model},}\
  }\href@noop {} {\bibfield  {journal} {\bibinfo  {journal} {Physical Review
  letters}\ }\textbf {\bibinfo {volume} {63}},\ \bibinfo {pages} {2144}
  (\bibinfo {year} {1989})}\BibitemShut {NoStop}%
\bibitem [{\citenamefont {Moudgalya}\ \emph
  {et~al.}(2020{\natexlab{a}})\citenamefont {Moudgalya}, \citenamefont
  {Regnault},\ and\ \citenamefont {Bernevig}}]{moudgalya2020eta}%
  \BibitemOpen
  \bibfield  {author} {\bibinfo {author} {\bibfnamefont {Sanjay}\ \bibnamefont
  {Moudgalya}}, \bibinfo {author} {\bibfnamefont {Nicolas}\ \bibnamefont
  {Regnault}}, \ and\ \bibinfo {author} {\bibfnamefont {B.~Andrei}\
  \bibnamefont {Bernevig}},\ }\bibfield  {title} {\enquote {\bibinfo {title}
  {$\ensuremath{\eta}$-pairing in hubbard models: From spectrum generating
  algebras to quantum many-body scars},}\ }\href {\doibase
  10.1103/PhysRevB.102.085140} {\bibfield  {journal} {\bibinfo  {journal}
  {Phys. Rev. B}\ }\textbf {\bibinfo {volume} {102}},\ \bibinfo {pages}
  {085140} (\bibinfo {year} {2020}{\natexlab{a}})}\BibitemShut {NoStop}%
\bibitem [{\citenamefont {Mark}\ and\ \citenamefont
  {Motrunich}(2020)}]{mark2020eta}%
  \BibitemOpen
  \bibfield  {author} {\bibinfo {author} {\bibfnamefont {Daniel~K.}\
  \bibnamefont {Mark}}\ and\ \bibinfo {author} {\bibfnamefont {Olexei~I.}\
  \bibnamefont {Motrunich}},\ }\bibfield  {title} {\enquote {\bibinfo {title}
  {$\ensuremath{\eta}$-pairing states as true scars in an extended hubbard
  model},}\ }\href {\doibase 10.1103/PhysRevB.102.075132} {\bibfield  {journal}
  {\bibinfo  {journal} {Phys. Rev. B}\ }\textbf {\bibinfo {volume} {102}},\
  \bibinfo {pages} {075132} (\bibinfo {year} {2020})}\BibitemShut {NoStop}%
\bibitem [{\citenamefont {Pakrouski}\ \emph {et~al.}(2020)\citenamefont
  {Pakrouski}, \citenamefont {Pallegar}, \citenamefont {Popov},\ and\
  \citenamefont {Klebanov}}]{pakrouski2020many}%
  \BibitemOpen
  \bibfield  {author} {\bibinfo {author} {\bibfnamefont {K.}~\bibnamefont
  {Pakrouski}}, \bibinfo {author} {\bibfnamefont {P.~N.}\ \bibnamefont
  {Pallegar}}, \bibinfo {author} {\bibfnamefont {F.~K.}\ \bibnamefont {Popov}},
  \ and\ \bibinfo {author} {\bibfnamefont {I.~R.}\ \bibnamefont {Klebanov}},\
  }\bibfield  {title} {\enquote {\bibinfo {title} {Many-body scars as a group
  invariant sector of hilbert space},}\ }\href {\doibase
  10.1103/PhysRevLett.125.230602} {\bibfield  {journal} {\bibinfo  {journal}
  {Phys. Rev. Lett.}\ }\textbf {\bibinfo {volume} {125}},\ \bibinfo {pages}
  {230602} (\bibinfo {year} {2020})}\BibitemShut {NoStop}%
\bibitem [{\citenamefont {Pakrouski}\ \emph {et~al.}(2021)\citenamefont
  {Pakrouski}, \citenamefont {Pallegar}, \citenamefont {Popov},\ and\
  \citenamefont {Klebanov}}]{pakrouski2021group}%
  \BibitemOpen
  \bibfield  {author} {\bibinfo {author} {\bibfnamefont {K.}~\bibnamefont
  {Pakrouski}}, \bibinfo {author} {\bibfnamefont {P.~N.}\ \bibnamefont
  {Pallegar}}, \bibinfo {author} {\bibfnamefont {F.~K.}\ \bibnamefont {Popov}},
  \ and\ \bibinfo {author} {\bibfnamefont {I.~R.}\ \bibnamefont {Klebanov}},\
  }\bibfield  {title} {\enquote {\bibinfo {title} {Group theoretic approach to
  many-body scar states in fermionic lattice models},}\ }\href {\doibase
  10.1103/PhysRevResearch.3.043156} {\bibfield  {journal} {\bibinfo  {journal}
  {Phys. Rev. Research}\ }\textbf {\bibinfo {volume} {3}},\ \bibinfo {pages}
  {043156} (\bibinfo {year} {2021})}\BibitemShut {NoStop}%
\bibitem [{\citenamefont {Yoshida}\ and\ \citenamefont
  {Katsura}(2022)}]{yoshida2022exact}%
  \BibitemOpen
  \bibfield  {author} {\bibinfo {author} {\bibfnamefont {Hironobu}\
  \bibnamefont {Yoshida}}\ and\ \bibinfo {author} {\bibfnamefont {Hosho}\
  \bibnamefont {Katsura}},\ }\bibfield  {title} {\enquote {\bibinfo {title}
  {Exact eigenstates of extended $\mathrm{SU}(n)$ hubbard models:
  Generalization of $\ensuremath{\eta}$-pairing states with $n$-particle
  off-diagonal long-range order},}\ }\href {\doibase
  10.1103/PhysRevB.105.024520} {\bibfield  {journal} {\bibinfo  {journal}
  {Phys. Rev. B}\ }\textbf {\bibinfo {volume} {105}},\ \bibinfo {pages}
  {024520} (\bibinfo {year} {2022})}\BibitemShut {NoStop}%
\bibitem [{\citenamefont {Gotta}\ \emph {et~al.}(2022)\citenamefont {Gotta},
  \citenamefont {Mazza}, \citenamefont {Simon},\ and\ \citenamefont
  {Roux}}]{Gotta_2022}%
  \BibitemOpen
  \bibfield  {author} {\bibinfo {author} {\bibfnamefont {Lorenzo}\ \bibnamefont
  {Gotta}}, \bibinfo {author} {\bibfnamefont {Leonardo}\ \bibnamefont {Mazza}},
  \bibinfo {author} {\bibfnamefont {Pascal}\ \bibnamefont {Simon}}, \ and\
  \bibinfo {author} {\bibfnamefont {Guillaume}\ \bibnamefont {Roux}},\
  }\bibfield  {title} {\enquote {\bibinfo {title} {Exact many-body scars based
  on pairs or multimers in a chain of spinless fermions},}\ }\href {\doibase
  10.1103/PhysRevB.106.235147} {\bibfield  {journal} {\bibinfo  {journal}
  {Phys. Rev. B}\ }\textbf {\bibinfo {volume} {106}},\ \bibinfo {pages}
  {235147} (\bibinfo {year} {2022})}\BibitemShut {NoStop}%
\bibitem [{\citenamefont {{Nakagawa}}\ \emph {et~al.}(2022)\citenamefont
  {{Nakagawa}}, \citenamefont {{Katsura}},\ and\ \citenamefont
  {{Ueda}}}]{nakagawa2022exact}%
  \BibitemOpen
  \bibfield  {author} {\bibinfo {author} {\bibfnamefont {Masaya}\ \bibnamefont
  {{Nakagawa}}}, \bibinfo {author} {\bibfnamefont {Hosho}\ \bibnamefont
  {{Katsura}}}, \ and\ \bibinfo {author} {\bibfnamefont {Masahito}\
  \bibnamefont {{Ueda}}},\ }\bibfield  {title} {\enquote {\bibinfo {title}
  {{Exact eigenstates of multicomponent Hubbard models: SU($N$) magnetic $\eta$
  pairing, weak ergodicity breaking, and partial integrability}},}\ }\href@noop
  {} {\bibfield  {journal} {\bibinfo  {journal} {arXiv e-prints}\ } (\bibinfo
  {year} {2022})},\ \Eprint {http://arxiv.org/abs/2205.07235} {arXiv:2205.07235
  [cond-mat.str-el]} \BibitemShut {NoStop}%
\bibitem [{\citenamefont {Shiraishi}\ and\ \citenamefont
  {Mori}(2017)}]{shiraishi2017systematic}%
  \BibitemOpen
  \bibfield  {author} {\bibinfo {author} {\bibfnamefont {Naoto}\ \bibnamefont
  {Shiraishi}}\ and\ \bibinfo {author} {\bibfnamefont {Takashi}\ \bibnamefont
  {Mori}},\ }\bibfield  {title} {\enquote {\bibinfo {title} {Systematic
  construction of counterexamples to the eigenstate thermalization
  hypothesis},}\ }\href {\doibase 10.1103/PhysRevLett.119.030601} {\bibfield
  {journal} {\bibinfo  {journal} {Phys. Rev. Lett.}\ }\textbf {\bibinfo
  {volume} {119}},\ \bibinfo {pages} {030601} (\bibinfo {year}
  {2017})}\BibitemShut {NoStop}%
\bibitem [{\citenamefont {Moudgalya}\ \emph
  {et~al.}(2020{\natexlab{b}})\citenamefont {Moudgalya}, \citenamefont
  {O'Brien}, \citenamefont {Bernevig}, \citenamefont {Fendley},\ and\
  \citenamefont {Regnault}}]{moudgalya2020large}%
  \BibitemOpen
  \bibfield  {author} {\bibinfo {author} {\bibfnamefont {Sanjay}\ \bibnamefont
  {Moudgalya}}, \bibinfo {author} {\bibfnamefont {Edward}\ \bibnamefont
  {O'Brien}}, \bibinfo {author} {\bibfnamefont {B.~Andrei}\ \bibnamefont
  {Bernevig}}, \bibinfo {author} {\bibfnamefont {Paul}\ \bibnamefont
  {Fendley}}, \ and\ \bibinfo {author} {\bibfnamefont {Nicolas}\ \bibnamefont
  {Regnault}},\ }\bibfield  {title} {\enquote {\bibinfo {title} {Large classes
  of quantum scarred hamiltonians from matrix product states},}\ }\href
  {\doibase 10.1103/PhysRevB.102.085120} {\bibfield  {journal} {\bibinfo
  {journal} {Phys. Rev. B}\ }\textbf {\bibinfo {volume} {102}},\ \bibinfo
  {pages} {085120} (\bibinfo {year} {2020}{\natexlab{b}})}\BibitemShut
  {NoStop}%
\bibitem [{\citenamefont {Ren}\ \emph {et~al.}(2021)\citenamefont {Ren},
  \citenamefont {Liang},\ and\ \citenamefont {Fang}}]{ren2020quasisymmetry}%
  \BibitemOpen
  \bibfield  {author} {\bibinfo {author} {\bibfnamefont {Jie}\ \bibnamefont
  {Ren}}, \bibinfo {author} {\bibfnamefont {Chenguang}\ \bibnamefont {Liang}},
  \ and\ \bibinfo {author} {\bibfnamefont {Chen}\ \bibnamefont {Fang}},\
  }\bibfield  {title} {\enquote {\bibinfo {title} {Quasisymmetry groups and
  many-body scar dynamics},}\ }\href {\doibase 10.1103/PhysRevLett.126.120604}
  {\bibfield  {journal} {\bibinfo  {journal} {Phys. Rev. Lett.}\ }\textbf
  {\bibinfo {volume} {126}},\ \bibinfo {pages} {120604} (\bibinfo {year}
  {2021})}\BibitemShut {NoStop}%
\bibitem [{\citenamefont {O'Dea}\ \emph {et~al.}(2020)\citenamefont {O'Dea},
  \citenamefont {Burnell}, \citenamefont {Chandran},\ and\ \citenamefont
  {Khemani}}]{odea2020from}%
  \BibitemOpen
  \bibfield  {author} {\bibinfo {author} {\bibfnamefont {Nicholas}\
  \bibnamefont {O'Dea}}, \bibinfo {author} {\bibfnamefont {Fiona}\ \bibnamefont
  {Burnell}}, \bibinfo {author} {\bibfnamefont {Anushya}\ \bibnamefont
  {Chandran}}, \ and\ \bibinfo {author} {\bibfnamefont {Vedika}\ \bibnamefont
  {Khemani}},\ }\bibfield  {title} {\enquote {\bibinfo {title} {From tunnels to
  towers: Quantum scars from lie algebras and $q$-deformed lie algebras},}\
  }\href {\doibase 10.1103/PhysRevResearch.2.043305} {\bibfield  {journal}
  {\bibinfo  {journal} {Phys. Rev. Research}\ }\textbf {\bibinfo {volume}
  {2}},\ \bibinfo {pages} {043305} (\bibinfo {year} {2020})}\BibitemShut
  {NoStop}%
\bibitem [{\citenamefont {Rozon}\ \emph {et~al.}(2022)\citenamefont {Rozon},
  \citenamefont {Gullans},\ and\ \citenamefont
  {Agarwal}}]{rozon2022constructing}%
  \BibitemOpen
  \bibfield  {author} {\bibinfo {author} {\bibfnamefont {Pierre-Gabriel}\
  \bibnamefont {Rozon}}, \bibinfo {author} {\bibfnamefont {Michael~J.}\
  \bibnamefont {Gullans}}, \ and\ \bibinfo {author} {\bibfnamefont {Kartiek}\
  \bibnamefont {Agarwal}},\ }\bibfield  {title} {\enquote {\bibinfo {title}
  {Constructing quantum many-body scar hamiltonians from floquet automata},}\
  }\href {\doibase 10.1103/PhysRevB.106.184304} {\bibfield  {journal} {\bibinfo
   {journal} {Phys. Rev. B}\ }\textbf {\bibinfo {volume} {106}},\ \bibinfo
  {pages} {184304} (\bibinfo {year} {2022})}\BibitemShut {NoStop}%
\bibitem [{\citenamefont {{Moudgalya}}\ and\ \citenamefont
  {{Motrunich}}(2022)}]{moudgalya2022exhaustive}%
  \BibitemOpen
  \bibfield  {author} {\bibinfo {author} {\bibfnamefont {Sanjay}\ \bibnamefont
  {{Moudgalya}}}\ and\ \bibinfo {author} {\bibfnamefont {Olexei~I.}\
  \bibnamefont {{Motrunich}}},\ }\bibfield  {title} {\enquote {\bibinfo {title}
  {{Exhaustive Characterization of Quantum Many-Body Scars using Commutant
  Algebras}},}\ }\href@noop {} {\bibfield  {journal} {\bibinfo  {journal}
  {arXiv e-prints}\ } (\bibinfo {year} {2022})},\ \Eprint
  {http://arxiv.org/abs/2209.03377} {arXiv:2209.03377 [cond-mat.str-el]}
  \BibitemShut {NoStop}%
\bibitem [{\citenamefont {{Rozon}}\ and\ \citenamefont
  {{Agarwal}}(2023)}]{rozon2023broken}%
  \BibitemOpen
  \bibfield  {author} {\bibinfo {author} {\bibfnamefont {Pierre-Gabriel}\
  \bibnamefont {{Rozon}}}\ and\ \bibinfo {author} {\bibfnamefont {Kartiek}\
  \bibnamefont {{Agarwal}}},\ }\bibfield  {title} {\enquote {\bibinfo {title}
  {{Broken unitary picture of dynamics in quantum many-body scars}},}\ }\href
  {\doibase 10.48550/arXiv.2302.04885} {\bibfield  {journal} {\bibinfo
  {journal} {arXiv e-prints}\ ,\ \bibinfo {pages} {arXiv:2302.04885}} (\bibinfo
  {year} {2023})}\BibitemShut {NoStop}%
\bibitem [{\citenamefont {Lin}\ \emph {et~al.}(2020)\citenamefont {Lin},
  \citenamefont {Chandran},\ and\ \citenamefont {Motrunich}}]{lin2020slow}%
  \BibitemOpen
  \bibfield  {author} {\bibinfo {author} {\bibfnamefont {Cheng-Ju}\
  \bibnamefont {Lin}}, \bibinfo {author} {\bibfnamefont {Anushya}\ \bibnamefont
  {Chandran}}, \ and\ \bibinfo {author} {\bibfnamefont {Olexei~I.}\
  \bibnamefont {Motrunich}},\ }\bibfield  {title} {\enquote {\bibinfo {title}
  {Slow thermalization of exact quantum many-body scar states under
  perturbations},}\ }\href {\doibase 10.1103/PhysRevResearch.2.033044}
  {\bibfield  {journal} {\bibinfo  {journal} {Phys. Rev. Res.}\ }\textbf
  {\bibinfo {volume} {2}},\ \bibinfo {pages} {033044} (\bibinfo {year}
  {2020})}\BibitemShut {NoStop}%
\bibitem [{\citenamefont {Lin}\ and\ \citenamefont
  {Motrunich}(2019)}]{lin2019exact}%
  \BibitemOpen
  \bibfield  {author} {\bibinfo {author} {\bibfnamefont {Cheng-Ju}\
  \bibnamefont {Lin}}\ and\ \bibinfo {author} {\bibfnamefont {Olexei~I.}\
  \bibnamefont {Motrunich}},\ }\bibfield  {title} {\enquote {\bibinfo {title}
  {Exact quantum many-body scar states in the {Rydberg}-blockaded atom
  chain},}\ }\href {\doibase 10.1103/PhysRevLett.122.173401} {\bibfield
  {journal} {\bibinfo  {journal} {Phys. Rev. Lett.}\ }\textbf {\bibinfo
  {volume} {122}},\ \bibinfo {pages} {173401} (\bibinfo {year}
  {2019})}\BibitemShut {NoStop}%
\bibitem [{\citenamefont {Ba\~nuls}\ \emph {et~al.}(2020)\citenamefont
  {Ba\~nuls}, \citenamefont {Huse},\ and\ \citenamefont
  {Cirac}}]{banuls2020entanglement}%
  \BibitemOpen
  \bibfield  {author} {\bibinfo {author} {\bibfnamefont {Mari~Carmen}\
  \bibnamefont {Ba\~nuls}}, \bibinfo {author} {\bibfnamefont {David~A.}\
  \bibnamefont {Huse}}, \ and\ \bibinfo {author} {\bibfnamefont {J.~Ignacio}\
  \bibnamefont {Cirac}},\ }\bibfield  {title} {\enquote {\bibinfo {title}
  {Entanglement and its relation to energy variance for local one-dimensional
  hamiltonians},}\ }\href {\doibase 10.1103/PhysRevB.101.144305} {\bibfield
  {journal} {\bibinfo  {journal} {Phys. Rev. B}\ }\textbf {\bibinfo {volume}
  {101}},\ \bibinfo {pages} {144305} (\bibinfo {year} {2020})}\BibitemShut
  {NoStop}%
\bibitem [{\citenamefont {Thouless}(1977)}]{thouless1977maximum}%
  \BibitemOpen
  \bibfield  {author} {\bibinfo {author} {\bibfnamefont {D.~J.}\ \bibnamefont
  {Thouless}},\ }\bibfield  {title} {\enquote {\bibinfo {title} {Maximum
  metallic resistance in thin wires},}\ }\href {\doibase
  10.1103/PhysRevLett.39.1167} {\bibfield  {journal} {\bibinfo  {journal}
  {Phys. Rev. Lett.}\ }\textbf {\bibinfo {volume} {39}},\ \bibinfo {pages}
  {1167--1169} (\bibinfo {year} {1977})}\BibitemShut {NoStop}%
\bibitem [{\citenamefont {Chan}\ \emph {et~al.}(2018)\citenamefont {Chan},
  \citenamefont {De~Luca},\ and\ \citenamefont {Chalker}}]{chan2018spectral}%
  \BibitemOpen
  \bibfield  {author} {\bibinfo {author} {\bibfnamefont {Amos}\ \bibnamefont
  {Chan}}, \bibinfo {author} {\bibfnamefont {Andrea}\ \bibnamefont {De~Luca}},
  \ and\ \bibinfo {author} {\bibfnamefont {J.~T.}\ \bibnamefont {Chalker}},\
  }\bibfield  {title} {\enquote {\bibinfo {title} {Spectral statistics in
  spatially extended chaotic quantum many-body systems},}\ }\href {\doibase
  10.1103/PhysRevLett.121.060601} {\bibfield  {journal} {\bibinfo  {journal}
  {Phys. Rev. Lett.}\ }\textbf {\bibinfo {volume} {121}},\ \bibinfo {pages}
  {060601} (\bibinfo {year} {2018})}\BibitemShut {NoStop}%
\bibitem [{\citenamefont {Schiulaz}\ \emph {et~al.}(2019)\citenamefont
  {Schiulaz}, \citenamefont {Torres-Herrera},\ and\ \citenamefont
  {Santos}}]{schiulaz2019thouless}%
  \BibitemOpen
  \bibfield  {author} {\bibinfo {author} {\bibfnamefont {Mauro}\ \bibnamefont
  {Schiulaz}}, \bibinfo {author} {\bibfnamefont {E.~Jonathan}\ \bibnamefont
  {Torres-Herrera}}, \ and\ \bibinfo {author} {\bibfnamefont {Lea~F.}\
  \bibnamefont {Santos}},\ }\bibfield  {title} {\enquote {\bibinfo {title}
  {Thouless and relaxation time scales in many-body quantum systems},}\ }\href
  {\doibase 10.1103/PhysRevB.99.174313} {\bibfield  {journal} {\bibinfo
  {journal} {Phys. Rev. B}\ }\textbf {\bibinfo {volume} {99}},\ \bibinfo
  {pages} {174313} (\bibinfo {year} {2019})}\BibitemShut {NoStop}%
\bibitem [{\citenamefont {Dymarsky}(2022)}]{dymarsky2022bound}%
  \BibitemOpen
  \bibfield  {author} {\bibinfo {author} {\bibfnamefont {Anatoly}\ \bibnamefont
  {Dymarsky}},\ }\bibfield  {title} {\enquote {\bibinfo {title} {Bound on
  eigenstate thermalization from transport},}\ }\href {\doibase
  10.1103/PhysRevLett.128.190601} {\bibfield  {journal} {\bibinfo  {journal}
  {Phys. Rev. Lett.}\ }\textbf {\bibinfo {volume} {128}},\ \bibinfo {pages}
  {190601} (\bibinfo {year} {2022})}\BibitemShut {NoStop}%
\bibitem [{\citenamefont {Chaikin}\ \emph {et~al.}(1995)\citenamefont
  {Chaikin}, \citenamefont {Lubensky},\ and\ \citenamefont
  {Witten}}]{chaikin1995principles}%
  \BibitemOpen
  \bibfield  {author} {\bibinfo {author} {\bibfnamefont {Paul~M}\ \bibnamefont
  {Chaikin}}, \bibinfo {author} {\bibfnamefont {Tom~C}\ \bibnamefont
  {Lubensky}}, \ and\ \bibinfo {author} {\bibfnamefont {Thomas~A}\ \bibnamefont
  {Witten}},\ }\href@noop {} {\emph {\bibinfo {title} {Principles of condensed
  matter physics}}},\ Vol.~\bibinfo {volume} {10}\ (\bibinfo  {publisher}
  {Cambridge university press Cambridge},\ \bibinfo {year} {1995})\BibitemShut
  {NoStop}%
\bibitem [{\citenamefont {Mukerjee}\ \emph {et~al.}(2006)\citenamefont
  {Mukerjee}, \citenamefont {Oganesyan},\ and\ \citenamefont
  {Huse}}]{mukerjee2006statistical}%
  \BibitemOpen
  \bibfield  {author} {\bibinfo {author} {\bibfnamefont {Subroto}\ \bibnamefont
  {Mukerjee}}, \bibinfo {author} {\bibfnamefont {Vadim}\ \bibnamefont
  {Oganesyan}}, \ and\ \bibinfo {author} {\bibfnamefont {David}\ \bibnamefont
  {Huse}},\ }\bibfield  {title} {\enquote {\bibinfo {title} {Statistical theory
  of transport by strongly interacting lattice fermions},}\ }\href {\doibase
  10.1103/PhysRevB.73.035113} {\bibfield  {journal} {\bibinfo  {journal} {Phys.
  Rev. B}\ }\textbf {\bibinfo {volume} {73}},\ \bibinfo {pages} {035113}
  (\bibinfo {year} {2006})}\BibitemShut {NoStop}%
\bibitem [{\citenamefont {Lux}\ \emph {et~al.}(2014)\citenamefont {Lux},
  \citenamefont {M\"uller}, \citenamefont {Mitra},\ and\ \citenamefont
  {Rosch}}]{lux2014hydrodynamic}%
  \BibitemOpen
  \bibfield  {author} {\bibinfo {author} {\bibfnamefont {Jonathan}\
  \bibnamefont {Lux}}, \bibinfo {author} {\bibfnamefont {Jan}\ \bibnamefont
  {M\"uller}}, \bibinfo {author} {\bibfnamefont {Aditi}\ \bibnamefont {Mitra}},
  \ and\ \bibinfo {author} {\bibfnamefont {Achim}\ \bibnamefont {Rosch}},\
  }\bibfield  {title} {\enquote {\bibinfo {title} {Hydrodynamic long-time tails
  after a quantum quench},}\ }\href {\doibase 10.1103/PhysRevA.89.053608}
  {\bibfield  {journal} {\bibinfo  {journal} {Phys. Rev. A}\ }\textbf {\bibinfo
  {volume} {89}},\ \bibinfo {pages} {053608} (\bibinfo {year}
  {2014})}\BibitemShut {NoStop}%
\bibitem [{\citenamefont {Gromov}\ \emph {et~al.}(2020)\citenamefont {Gromov},
  \citenamefont {Lucas},\ and\ \citenamefont
  {Nandkishore}}]{gromov2020fracton}%
  \BibitemOpen
  \bibfield  {author} {\bibinfo {author} {\bibfnamefont {Andrey}\ \bibnamefont
  {Gromov}}, \bibinfo {author} {\bibfnamefont {Andrew}\ \bibnamefont {Lucas}},
  \ and\ \bibinfo {author} {\bibfnamefont {Rahul~M.}\ \bibnamefont
  {Nandkishore}},\ }\bibfield  {title} {\enquote {\bibinfo {title} {Fracton
  hydrodynamics},}\ }\href {\doibase 10.1103/PhysRevResearch.2.033124}
  {\bibfield  {journal} {\bibinfo  {journal} {Phys. Rev. Res.}\ }\textbf
  {\bibinfo {volume} {2}},\ \bibinfo {pages} {033124} (\bibinfo {year}
  {2020})}\BibitemShut {NoStop}%
\bibitem [{\citenamefont {Feldmeier}\ \emph {et~al.}(2020)\citenamefont
  {Feldmeier}, \citenamefont {Sala}, \citenamefont {De~Tomasi}, \citenamefont
  {Pollmann},\ and\ \citenamefont {Knap}}]{feldmeier2020anomalous}%
  \BibitemOpen
  \bibfield  {author} {\bibinfo {author} {\bibfnamefont {Johannes}\
  \bibnamefont {Feldmeier}}, \bibinfo {author} {\bibfnamefont {Pablo}\
  \bibnamefont {Sala}}, \bibinfo {author} {\bibfnamefont {Giuseppe}\
  \bibnamefont {De~Tomasi}}, \bibinfo {author} {\bibfnamefont {Frank}\
  \bibnamefont {Pollmann}}, \ and\ \bibinfo {author} {\bibfnamefont {Michael}\
  \bibnamefont {Knap}},\ }\bibfield  {title} {\enquote {\bibinfo {title}
  {Anomalous diffusion in dipole- and higher-moment-conserving systems},}\
  }\href {\doibase 10.1103/PhysRevLett.125.245303} {\bibfield  {journal}
  {\bibinfo  {journal} {Phys. Rev. Lett.}\ }\textbf {\bibinfo {volume} {125}},\
  \bibinfo {pages} {245303} (\bibinfo {year} {2020})}\BibitemShut {NoStop}%
\bibitem [{\citenamefont {Moudgalya}\ \emph {et~al.}(2021)\citenamefont
  {Moudgalya}, \citenamefont {Prem}, \citenamefont {Huse},\ and\ \citenamefont
  {Chan}}]{moudgalya2020spectral}%
  \BibitemOpen
  \bibfield  {author} {\bibinfo {author} {\bibfnamefont {Sanjay}\ \bibnamefont
  {Moudgalya}}, \bibinfo {author} {\bibfnamefont {Abhinav}\ \bibnamefont
  {Prem}}, \bibinfo {author} {\bibfnamefont {David~A.}\ \bibnamefont {Huse}}, \
  and\ \bibinfo {author} {\bibfnamefont {Amos}\ \bibnamefont {Chan}},\
  }\bibfield  {title} {\enquote {\bibinfo {title} {Spectral statistics in
  constrained many-body quantum chaotic systems},}\ }\href {\doibase
  10.1103/PhysRevResearch.3.023176} {\bibfield  {journal} {\bibinfo  {journal}
  {Phys. Rev. Res.}\ }\textbf {\bibinfo {volume} {3}},\ \bibinfo {pages}
  {023176} (\bibinfo {year} {2021})}\BibitemShut {NoStop}%
\bibitem [{\citenamefont {{Bu{\v{c}}a}}(2023)}]{buca2023unified}%
  \BibitemOpen
  \bibfield  {author} {\bibinfo {author} {\bibfnamefont {Berislav}\
  \bibnamefont {{Bu{\v{c}}a}}},\ }\bibfield  {title} {\enquote {\bibinfo
  {title} {{Unified theory of local quantum many-body dynamics: Eigenoperator
  thermalization theorems}},}\ }\href@noop {} {\bibfield  {journal} {\bibinfo
  {journal} {arXiv e-prints}\ } (\bibinfo {year} {2023})},\ \Eprint
  {http://arxiv.org/abs/2301.07091} {arXiv:2301.07091 [cond-mat.stat-mech]}
  \BibitemShut {NoStop}%
\bibitem [{\citenamefont {Pitaevskii}\ and\ \citenamefont
  {Stringari}(2016)}]{stringari}%
  \BibitemOpen
  \bibfield  {author} {\bibinfo {author} {\bibfnamefont {L.}~\bibnamefont
  {Pitaevskii}}\ and\ \bibinfo {author} {\bibfnamefont {S.}~\bibnamefont
  {Stringari}},\ }\href@noop {} {\emph {\bibinfo {title} {Bose-Einstein
  Condensation}}}\ (\bibinfo  {publisher} {Oxford University Press},\ \bibinfo
  {year} {2016})\BibitemShut {NoStop}%
\bibitem [{\citenamefont {Kitazawa}\ \emph {et~al.}(2003)\citenamefont
  {Kitazawa}, \citenamefont {Hijii},\ and\ \citenamefont
  {Nomura}}]{kitazawa2003su2}%
  \BibitemOpen
  \bibfield  {author} {\bibinfo {author} {\bibfnamefont {Atsuhiro}\
  \bibnamefont {Kitazawa}}, \bibinfo {author} {\bibfnamefont {Keigo}\
  \bibnamefont {Hijii}}, \ and\ \bibinfo {author} {\bibfnamefont {Kiyohide}\
  \bibnamefont {Nomura}},\ }\bibfield  {title} {\enquote {\bibinfo {title} {An
  su(2) symmetry of the one-dimensional spin-1 xy model},}\ }\href {\doibase
  10.1088/0305-4470/36/23/104} {\bibfield  {journal} {\bibinfo  {journal}
  {Journal of Physics A: Mathematical and General}\ }\textbf {\bibinfo {volume}
  {36}},\ \bibinfo {pages} {L351} (\bibinfo {year} {2003})}\BibitemShut
  {NoStop}%
\bibitem [{SM()}]{SM}%
  \BibitemOpen
  \href@noop {} {\enquote {\bibinfo {title} {See the supplemental material},}\
  }\BibitemShut {NoStop}%
\bibitem [{\citenamefont {Vafek}\ \emph {et~al.}(2017)\citenamefont {Vafek},
  \citenamefont {Regnault},\ and\ \citenamefont
  {Bernevig}}]{vafek2017entanglement}%
  \BibitemOpen
  \bibfield  {author} {\bibinfo {author} {\bibfnamefont {Oskar}\ \bibnamefont
  {Vafek}}, \bibinfo {author} {\bibfnamefont {Nicolas}\ \bibnamefont
  {Regnault}}, \ and\ \bibinfo {author} {\bibfnamefont {B.~Andrei}\
  \bibnamefont {Bernevig}},\ }\bibfield  {title} {\enquote {\bibinfo {title}
  {Entanglement of exact excited eigenstates of the {Hubbard} model in
  arbitrary dimension},}\ }\href {\doibase 10.21468/SciPostPhys.3.6.043}
  {\bibfield  {journal} {\bibinfo  {journal} {SciPost Phys.}\ }\textbf
  {\bibinfo {volume} {3}},\ \bibinfo {pages} {043} (\bibinfo {year}
  {2017})}\BibitemShut {NoStop}%
\bibitem [{\citenamefont {Tang}\ \emph {et~al.}(2022)\citenamefont {Tang},
  \citenamefont {O'Dea},\ and\ \citenamefont {Chandran}}]{tang2021multimagnon}%
  \BibitemOpen
  \bibfield  {author} {\bibinfo {author} {\bibfnamefont {Long-Hin}\
  \bibnamefont {Tang}}, \bibinfo {author} {\bibfnamefont {Nicholas}\
  \bibnamefont {O'Dea}}, \ and\ \bibinfo {author} {\bibfnamefont {Anushya}\
  \bibnamefont {Chandran}},\ }\bibfield  {title} {\enquote {\bibinfo {title}
  {Multimagnon quantum many-body scars from tensor operators},}\ }\href
  {\doibase 10.1103/PhysRevResearch.4.043006} {\bibfield  {journal} {\bibinfo
  {journal} {Phys. Rev. Res.}\ }\textbf {\bibinfo {volume} {4}},\ \bibinfo
  {pages} {043006} (\bibinfo {year} {2022})}\BibitemShut {NoStop}%
\bibitem [{\citenamefont {Crosswhite}\ and\ \citenamefont
  {Bacon}(2008)}]{crosswhite2008fsa}%
  \BibitemOpen
  \bibfield  {author} {\bibinfo {author} {\bibfnamefont {Gregory~M.}\
  \bibnamefont {Crosswhite}}\ and\ \bibinfo {author} {\bibfnamefont {Dave}\
  \bibnamefont {Bacon}},\ }\bibfield  {title} {\enquote {\bibinfo {title}
  {Finite automata for caching in matrix product algorithms},}\ }\href
  {\doibase 10.1103/PhysRevA.78.012356} {\bibfield  {journal} {\bibinfo
  {journal} {Phys. Rev. A}\ }\textbf {\bibinfo {volume} {78}},\ \bibinfo
  {pages} {012356} (\bibinfo {year} {2008})}\BibitemShut {NoStop}%
\bibitem [{\citenamefont {Motruk}\ \emph {et~al.}(2016)\citenamefont {Motruk},
  \citenamefont {Zaletel}, \citenamefont {Mong},\ and\ \citenamefont
  {Pollmann}}]{motruk2016density}%
  \BibitemOpen
  \bibfield  {author} {\bibinfo {author} {\bibfnamefont {Johannes}\
  \bibnamefont {Motruk}}, \bibinfo {author} {\bibfnamefont {Michael~P.}\
  \bibnamefont {Zaletel}}, \bibinfo {author} {\bibfnamefont {Roger S.~K.}\
  \bibnamefont {Mong}}, \ and\ \bibinfo {author} {\bibfnamefont {Frank}\
  \bibnamefont {Pollmann}},\ }\bibfield  {title} {\enquote {\bibinfo {title}
  {Density matrix renormalization group on a cylinder in mixed real and
  momentum space},}\ }\href {\doibase 10.1103/PhysRevB.93.155139} {\bibfield
  {journal} {\bibinfo  {journal} {Phys. Rev. B}\ }\textbf {\bibinfo {volume}
  {93}},\ \bibinfo {pages} {155139} (\bibinfo {year} {2016})}\BibitemShut
  {NoStop}%
\bibitem [{\citenamefont {Weinberg}\ and\ \citenamefont
  {Bukov}(2017)}]{weinberg2017quspin}%
  \BibitemOpen
  \bibfield  {author} {\bibinfo {author} {\bibfnamefont {Phillip}\ \bibnamefont
  {Weinberg}}\ and\ \bibinfo {author} {\bibfnamefont {Marin}\ \bibnamefont
  {Bukov}},\ }\bibfield  {title} {\enquote {\bibinfo {title} {{QuSpin: a Python
  package for dynamics and exact diagonalisation of quantum many body systems
  part I: spin chains}},}\ }\href {\doibase 10.21468/SciPostPhys.2.1.003}
  {\bibfield  {journal} {\bibinfo  {journal} {SciPost Phys.}\ }\textbf
  {\bibinfo {volume} {2}},\ \bibinfo {pages} {003} (\bibinfo {year}
  {2017})}\BibitemShut {NoStop}%
\bibitem [{\citenamefont {Campos~Venuti}\ and\ \citenamefont
  {Zanardi}(2010)}]{camposvenuti2010unitary}%
  \BibitemOpen
  \bibfield  {author} {\bibinfo {author} {\bibfnamefont {Lorenzo}\ \bibnamefont
  {Campos~Venuti}}\ and\ \bibinfo {author} {\bibfnamefont {Paolo}\ \bibnamefont
  {Zanardi}},\ }\bibfield  {title} {\enquote {\bibinfo {title} {Unitary
  equilibrations: Probability distribution of the loschmidt echo},}\ }\href
  {\doibase 10.1103/PhysRevA.81.022113} {\bibfield  {journal} {\bibinfo
  {journal} {Phys. Rev. A}\ }\textbf {\bibinfo {volume} {81}},\ \bibinfo
  {pages} {022113} (\bibinfo {year} {2010})}\BibitemShut {NoStop}%
\bibitem [{\citenamefont {Mori}\ and\ \citenamefont
  {Shiraishi}(2017)}]{mori2017thermalization}%
  \BibitemOpen
  \bibfield  {author} {\bibinfo {author} {\bibfnamefont {Takashi}\ \bibnamefont
  {Mori}}\ and\ \bibinfo {author} {\bibfnamefont {Naoto}\ \bibnamefont
  {Shiraishi}},\ }\bibfield  {title} {\enquote {\bibinfo {title}
  {Thermalization without eigenstate thermalization hypothesis after a quantum
  quench},}\ }\href {\doibase 10.1103/PhysRevE.96.022153} {\bibfield  {journal}
  {\bibinfo  {journal} {Phys. Rev. E}\ }\textbf {\bibinfo {volume} {96}},\
  \bibinfo {pages} {022153} (\bibinfo {year} {2017})}\BibitemShut {NoStop}%
\bibitem [{\citenamefont {Goldstein}\ \emph {et~al.}(2013)\citenamefont
  {Goldstein}, \citenamefont {Hara},\ and\ \citenamefont
  {Tasaki}}]{goldstein2013time}%
  \BibitemOpen
  \bibfield  {author} {\bibinfo {author} {\bibfnamefont {Sheldon}\ \bibnamefont
  {Goldstein}}, \bibinfo {author} {\bibfnamefont {Takashi}\ \bibnamefont
  {Hara}}, \ and\ \bibinfo {author} {\bibfnamefont {Hal}\ \bibnamefont
  {Tasaki}},\ }\bibfield  {title} {\enquote {\bibinfo {title} {Time scales in
  the approach to equilibrium of macroscopic quantum systems},}\ }\href
  {\doibase 10.1103/PhysRevLett.111.140401} {\bibfield  {journal} {\bibinfo
  {journal} {Phys. Rev. Lett.}\ }\textbf {\bibinfo {volume} {111}},\ \bibinfo
  {pages} {140401} (\bibinfo {year} {2013})}\BibitemShut {NoStop}%
\bibitem [{\citenamefont {Malabarba}\ \emph {et~al.}(2014)\citenamefont
  {Malabarba}, \citenamefont {Garc\'{\i}a-Pintos}, \citenamefont {Linden},
  \citenamefont {Farrelly},\ and\ \citenamefont
  {Short}}]{malabarba2014quantum}%
  \BibitemOpen
  \bibfield  {author} {\bibinfo {author} {\bibfnamefont {Artur S.~L.}\
  \bibnamefont {Malabarba}}, \bibinfo {author} {\bibfnamefont {Luis~Pedro}\
  \bibnamefont {Garc\'{\i}a-Pintos}}, \bibinfo {author} {\bibfnamefont {Noah}\
  \bibnamefont {Linden}}, \bibinfo {author} {\bibfnamefont {Terence~C.}\
  \bibnamefont {Farrelly}}, \ and\ \bibinfo {author} {\bibfnamefont
  {Anthony~J.}\ \bibnamefont {Short}},\ }\bibfield  {title} {\enquote {\bibinfo
  {title} {Quantum systems equilibrate rapidly for most observables},}\ }\href
  {\doibase 10.1103/PhysRevE.90.012121} {\bibfield  {journal} {\bibinfo
  {journal} {Phys. Rev. E}\ }\textbf {\bibinfo {volume} {90}},\ \bibinfo
  {pages} {012121} (\bibinfo {year} {2014})}\BibitemShut {NoStop}%
\bibitem [{\citenamefont {Goldstein}\ \emph {et~al.}(2015)\citenamefont
  {Goldstein}, \citenamefont {Hara},\ and\ \citenamefont
  {Tasaki}}]{goldstein2015extremely}%
  \BibitemOpen
  \bibfield  {author} {\bibinfo {author} {\bibfnamefont {Sheldon}\ \bibnamefont
  {Goldstein}}, \bibinfo {author} {\bibfnamefont {Takashi}\ \bibnamefont
  {Hara}}, \ and\ \bibinfo {author} {\bibfnamefont {Hal}\ \bibnamefont
  {Tasaki}},\ }\bibfield  {title} {\enquote {\bibinfo {title} {Extremely quick
  thermalization in a macroscopic quantum system for a typical nonequilibrium
  subspace},}\ }\href {\doibase 10.1088/1367-2630/17/4/045002} {\bibfield
  {journal} {\bibinfo  {journal} {New Journal of Physics}\ }\textbf {\bibinfo
  {volume} {17}},\ \bibinfo {pages} {045002} (\bibinfo {year}
  {2015})}\BibitemShut {NoStop}%
\bibitem [{\citenamefont {Reimann}(2016)}]{reimann2016typical}%
  \BibitemOpen
  \bibfield  {author} {\bibinfo {author} {\bibfnamefont {Peter}\ \bibnamefont
  {Reimann}},\ }\bibfield  {title} {\enquote {\bibinfo {title} {Typical fast
  thermalization processes in closed many-body systems},}\ }\href {\doibase
  10.1038/ncomms10821} {\bibfield  {journal} {\bibinfo  {journal} {Nature
  Communications}\ }\textbf {\bibinfo {volume} {7}},\ \bibinfo {pages} {10821}
  (\bibinfo {year} {2016})}\BibitemShut {NoStop}%
\bibitem [{\citenamefont {Garc\'{\i}a-Pintos}\ \emph
  {et~al.}(2017)\citenamefont {Garc\'{\i}a-Pintos}, \citenamefont {Linden},
  \citenamefont {Malabarba}, \citenamefont {Short},\ and\ \citenamefont
  {Winter}}]{garciapintos2017equilibration}%
  \BibitemOpen
  \bibfield  {author} {\bibinfo {author} {\bibfnamefont {Luis~Pedro}\
  \bibnamefont {Garc\'{\i}a-Pintos}}, \bibinfo {author} {\bibfnamefont {Noah}\
  \bibnamefont {Linden}}, \bibinfo {author} {\bibfnamefont {Artur S.~L.}\
  \bibnamefont {Malabarba}}, \bibinfo {author} {\bibfnamefont {Anthony~J.}\
  \bibnamefont {Short}}, \ and\ \bibinfo {author} {\bibfnamefont {Andreas}\
  \bibnamefont {Winter}},\ }\bibfield  {title} {\enquote {\bibinfo {title}
  {Equilibration time scales of physically relevant observables},}\ }\href
  {\doibase 10.1103/PhysRevX.7.031027} {\bibfield  {journal} {\bibinfo
  {journal} {Phys. Rev. X}\ }\textbf {\bibinfo {volume} {7}},\ \bibinfo {pages}
  {031027} (\bibinfo {year} {2017})}\BibitemShut {NoStop}%
\bibitem [{\citenamefont {Wilming}\ \emph {et~al.}(2018)\citenamefont
  {Wilming}, \citenamefont {de~Oliveira}, \citenamefont {Short},\ and\
  \citenamefont {Eisert}}]{wilming2018equilibration}%
  \BibitemOpen
  \bibfield  {author} {\bibinfo {author} {\bibfnamefont {Henrik}\ \bibnamefont
  {Wilming}}, \bibinfo {author} {\bibfnamefont {Thiago~R.}\ \bibnamefont
  {de~Oliveira}}, \bibinfo {author} {\bibfnamefont {Anthony~J.}\ \bibnamefont
  {Short}}, \ and\ \bibinfo {author} {\bibfnamefont {Jens}\ \bibnamefont
  {Eisert}},\ }\enquote {\bibinfo {title} {Equilibration times in closed
  quantum many-body systems},}\ in\ \href {\doibase
  10.1007/978-3-319-99046-0_18} {\emph {\bibinfo {booktitle} {Thermodynamics in
  the Quantum Regime: Fundamental Aspects and New Directions}}},\ \bibinfo
  {editor} {edited by\ \bibinfo {editor} {\bibfnamefont {Felix}\ \bibnamefont
  {Binder}}, \bibinfo {editor} {\bibfnamefont {Luis~A.}\ \bibnamefont
  {Correa}}, \bibinfo {editor} {\bibfnamefont {Christian}\ \bibnamefont
  {Gogolin}}, \bibinfo {editor} {\bibfnamefont {Janet}\ \bibnamefont {Anders}},
  \ and\ \bibinfo {editor} {\bibfnamefont {Gerardo}\ \bibnamefont {Adesso}}}\
  (\bibinfo  {publisher} {Springer International Publishing},\ \bibinfo
  {address} {Cham},\ \bibinfo {year} {2018})\ pp.\ \bibinfo {pages}
  {435--455}\BibitemShut {NoStop}%
\bibitem [{\citenamefont {{Riddell}}\ \emph {et~al.}(2021)\citenamefont
  {{Riddell}}, \citenamefont {{Garc{\'\i}a-Pintos}},\ and\ \citenamefont
  {{Alhambra}}}]{riddell2021relaxation}%
  \BibitemOpen
  \bibfield  {author} {\bibinfo {author} {\bibfnamefont {Jonathon}\
  \bibnamefont {{Riddell}}}, \bibinfo {author} {\bibfnamefont {Luis~Pedro}\
  \bibnamefont {{Garc{\'\i}a-Pintos}}}, \ and\ \bibinfo {author} {\bibfnamefont
  {{\'A}lvaro~M.}\ \bibnamefont {{Alhambra}}},\ }\bibfield  {title} {\enquote
  {\bibinfo {title} {{Relaxation of non-integrable systems and correlation
  functions}},}\ }\href {\doibase 10.48550/arXiv.2112.09475} {\bibfield
  {journal} {\bibinfo  {journal} {arXiv e-prints}\ ,\ \bibinfo {pages}
  {arXiv:2112.09475}} (\bibinfo {year} {2021})}\BibitemShut {NoStop}%
\bibitem [{\citenamefont {Fishman}\ \emph
  {et~al.}(2022{\natexlab{a}})\citenamefont {Fishman}, \citenamefont {White},\
  and\ \citenamefont {Stoudenmire}}]{fishman2022itensor1}%
  \BibitemOpen
  \bibfield  {author} {\bibinfo {author} {\bibfnamefont {Matthew}\ \bibnamefont
  {Fishman}}, \bibinfo {author} {\bibfnamefont {Steven~R.}\ \bibnamefont
  {White}}, \ and\ \bibinfo {author} {\bibfnamefont {E.~Miles}\ \bibnamefont
  {Stoudenmire}},\ }\bibfield  {title} {\enquote {\bibinfo {title} {{The
  ITensor Software Library for Tensor Network Calculations}},}\ }\href
  {\doibase 10.21468/SciPostPhysCodeb.4} {\bibfield  {journal} {\bibinfo
  {journal} {SciPost Phys. Codebases}\ ,\ \bibinfo {pages} {4}} (\bibinfo
  {year} {2022}{\natexlab{a}})}\BibitemShut {NoStop}%
\bibitem [{\citenamefont {Fishman}\ \emph
  {et~al.}(2022{\natexlab{b}})\citenamefont {Fishman}, \citenamefont {White},\
  and\ \citenamefont {Stoudenmire}}]{fishman2022itensor2}%
  \BibitemOpen
  \bibfield  {author} {\bibinfo {author} {\bibfnamefont {Matthew}\ \bibnamefont
  {Fishman}}, \bibinfo {author} {\bibfnamefont {Steven~R.}\ \bibnamefont
  {White}}, \ and\ \bibinfo {author} {\bibfnamefont {E.~Miles}\ \bibnamefont
  {Stoudenmire}},\ }\bibfield  {title} {\enquote {\bibinfo {title} {{Codebase
  release 0.3 for ITensor}},}\ }\href {\doibase
  10.21468/SciPostPhysCodeb.4-r0.3} {\bibfield  {journal} {\bibinfo  {journal}
  {SciPost Phys. Codebases}\ ,\ \bibinfo {pages} {4--r0.3}} (\bibinfo {year}
  {2022}{\natexlab{b}})}\BibitemShut {NoStop}%
\bibitem [{\citenamefont {Shiraishi}\ and\ \citenamefont
  {Mori}(2018)}]{shiraishimorireply2018}%
  \BibitemOpen
  \bibfield  {author} {\bibinfo {author} {\bibfnamefont {Naoto}\ \bibnamefont
  {Shiraishi}}\ and\ \bibinfo {author} {\bibfnamefont {Takashi}\ \bibnamefont
  {Mori}},\ }\bibfield  {title} {\enquote {\bibinfo {title} {Shiraishi and mori
  reply},}\ }\href {\doibase 10.1103/PhysRevLett.121.038902} {\bibfield
  {journal} {\bibinfo  {journal} {Phys. Rev. Lett.}\ }\textbf {\bibinfo
  {volume} {121}},\ \bibinfo {pages} {038902} (\bibinfo {year}
  {2018})}\BibitemShut {NoStop}%
\bibitem [{\citenamefont {Khatami}\ \emph {et~al.}(2013)\citenamefont
  {Khatami}, \citenamefont {Pupillo}, \citenamefont {Srednicki},\ and\
  \citenamefont {Rigol}}]{khatami2013fluctuation}%
  \BibitemOpen
  \bibfield  {author} {\bibinfo {author} {\bibfnamefont {Ehsan}\ \bibnamefont
  {Khatami}}, \bibinfo {author} {\bibfnamefont {Guido}\ \bibnamefont
  {Pupillo}}, \bibinfo {author} {\bibfnamefont {Mark}\ \bibnamefont
  {Srednicki}}, \ and\ \bibinfo {author} {\bibfnamefont {Marcos}\ \bibnamefont
  {Rigol}},\ }\bibfield  {title} {\enquote {\bibinfo {title}
  {Fluctuation-dissipation theorem in an isolated system of quantum dipolar
  bosons after a quench},}\ }\href {\doibase 10.1103/PhysRevLett.111.050403}
  {\bibfield  {journal} {\bibinfo  {journal} {Phys. Rev. Lett.}\ }\textbf
  {\bibinfo {volume} {111}},\ \bibinfo {pages} {050403} (\bibinfo {year}
  {2013})}\BibitemShut {NoStop}%
\bibitem [{\citenamefont {Steinigeweg}\ \emph {et~al.}(2013)\citenamefont
  {Steinigeweg}, \citenamefont {Herbrych},\ and\ \citenamefont
  {Prelov\ifmmode~\check{s}\else \v{s}\fi{}ek}}]{steinigeweg2013eigenstate}%
  \BibitemOpen
  \bibfield  {author} {\bibinfo {author} {\bibfnamefont {R.}~\bibnamefont
  {Steinigeweg}}, \bibinfo {author} {\bibfnamefont {J.}~\bibnamefont
  {Herbrych}}, \ and\ \bibinfo {author} {\bibfnamefont {P.}~\bibnamefont
  {Prelov\ifmmode~\check{s}\else \v{s}\fi{}ek}},\ }\bibfield  {title} {\enquote
  {\bibinfo {title} {Eigenstate thermalization within isolated spin-chain
  systems},}\ }\href {\doibase 10.1103/PhysRevE.87.012118} {\bibfield
  {journal} {\bibinfo  {journal} {Phys. Rev. E}\ }\textbf {\bibinfo {volume}
  {87}},\ \bibinfo {pages} {012118} (\bibinfo {year} {2013})}\BibitemShut
  {NoStop}%
\bibitem [{\citenamefont {Beugeling}\ \emph {et~al.}(2015)\citenamefont
  {Beugeling}, \citenamefont {Moessner},\ and\ \citenamefont
  {Haque}}]{beugeling2015offdiagonal}%
  \BibitemOpen
  \bibfield  {author} {\bibinfo {author} {\bibfnamefont {Wouter}\ \bibnamefont
  {Beugeling}}, \bibinfo {author} {\bibfnamefont {Roderich}\ \bibnamefont
  {Moessner}}, \ and\ \bibinfo {author} {\bibfnamefont {Masudul}\ \bibnamefont
  {Haque}},\ }\bibfield  {title} {\enquote {\bibinfo {title} {Off-diagonal
  matrix elements of local operators in many-body quantum systems},}\ }\href
  {\doibase 10.1103/PhysRevE.91.012144} {\bibfield  {journal} {\bibinfo
  {journal} {Phys. Rev. E}\ }\textbf {\bibinfo {volume} {91}},\ \bibinfo
  {pages} {012144} (\bibinfo {year} {2015})}\BibitemShut {NoStop}%
\bibitem [{\citenamefont {Richter}\ \emph {et~al.}(2020)\citenamefont
  {Richter}, \citenamefont {Dymarsky}, \citenamefont {Steinigeweg},\ and\
  \citenamefont {Gemmer}}]{richter2020eigenstate}%
  \BibitemOpen
  \bibfield  {author} {\bibinfo {author} {\bibfnamefont {Jonas}\ \bibnamefont
  {Richter}}, \bibinfo {author} {\bibfnamefont {Anatoly}\ \bibnamefont
  {Dymarsky}}, \bibinfo {author} {\bibfnamefont {Robin}\ \bibnamefont
  {Steinigeweg}}, \ and\ \bibinfo {author} {\bibfnamefont {Jochen}\
  \bibnamefont {Gemmer}},\ }\bibfield  {title} {\enquote {\bibinfo {title}
  {Eigenstate thermalization hypothesis beyond standard indicators: Emergence
  of random-matrix behavior at small frequencies},}\ }\href {\doibase
  10.1103/PhysRevE.102.042127} {\bibfield  {journal} {\bibinfo  {journal}
  {Phys. Rev. E}\ }\textbf {\bibinfo {volume} {102}},\ \bibinfo {pages}
  {042127} (\bibinfo {year} {2020})}\BibitemShut {NoStop}%
\bibitem [{\citenamefont {Wurtz}\ and\ \citenamefont
  {Polkovnikov}(2020)}]{wurtz2020emergent}%
  \BibitemOpen
  \bibfield  {author} {\bibinfo {author} {\bibfnamefont {Jonathan}\
  \bibnamefont {Wurtz}}\ and\ \bibinfo {author} {\bibfnamefont {Anatoli}\
  \bibnamefont {Polkovnikov}},\ }\bibfield  {title} {\enquote {\bibinfo {title}
  {Emergent conservation laws and nonthermal states in the mixed-field ising
  model},}\ }\href {\doibase 10.1103/PhysRevB.101.195138} {\bibfield  {journal}
  {\bibinfo  {journal} {Phys. Rev. B}\ }\textbf {\bibinfo {volume} {101}},\
  \bibinfo {pages} {195138} (\bibinfo {year} {2020})}\BibitemShut {NoStop}%
\bibitem [{\citenamefont {Sugiura}\ \emph
  {et~al.}(2021{\natexlab{a}})\citenamefont {Sugiura}, \citenamefont {Claeys},
  \citenamefont {Dymarsky},\ and\ \citenamefont
  {Polkovnikov}}]{sugiura2021adiabatic}%
  \BibitemOpen
  \bibfield  {author} {\bibinfo {author} {\bibfnamefont {Sho}\ \bibnamefont
  {Sugiura}}, \bibinfo {author} {\bibfnamefont {Pieter~W.}\ \bibnamefont
  {Claeys}}, \bibinfo {author} {\bibfnamefont {Anatoly}\ \bibnamefont
  {Dymarsky}}, \ and\ \bibinfo {author} {\bibfnamefont {Anatoli}\ \bibnamefont
  {Polkovnikov}},\ }\bibfield  {title} {\enquote {\bibinfo {title} {Adiabatic
  landscape and optimal paths in ergodic systems},}\ }\href {\doibase
  10.1103/PhysRevResearch.3.013102} {\bibfield  {journal} {\bibinfo  {journal}
  {Phys. Rev. Res.}\ }\textbf {\bibinfo {volume} {3}},\ \bibinfo {pages}
  {013102} (\bibinfo {year} {2021}{\natexlab{a}})}\BibitemShut {NoStop}%
\bibitem [{\citenamefont {Surace}\ \emph {et~al.}(2021)\citenamefont {Surace},
  \citenamefont {Votto}, \citenamefont {Lazo}, \citenamefont {Silva},
  \citenamefont {Dalmonte},\ and\ \citenamefont {Giudici}}]{surace2021exact}%
  \BibitemOpen
  \bibfield  {author} {\bibinfo {author} {\bibfnamefont {Federica~Maria}\
  \bibnamefont {Surace}}, \bibinfo {author} {\bibfnamefont {Matteo}\
  \bibnamefont {Votto}}, \bibinfo {author} {\bibfnamefont {Eduardo~Gonzalez}\
  \bibnamefont {Lazo}}, \bibinfo {author} {\bibfnamefont {Alessandro}\
  \bibnamefont {Silva}}, \bibinfo {author} {\bibfnamefont {Marcello}\
  \bibnamefont {Dalmonte}}, \ and\ \bibinfo {author} {\bibfnamefont {Giuliano}\
  \bibnamefont {Giudici}},\ }\bibfield  {title} {\enquote {\bibinfo {title}
  {Exact many-body scars and their stability in constrained quantum chains},}\
  }\href {\doibase 10.1103/PhysRevB.103.104302} {\bibfield  {journal} {\bibinfo
   {journal} {Phys. Rev. B}\ }\textbf {\bibinfo {volume} {103}},\ \bibinfo
  {pages} {104302} (\bibinfo {year} {2021})}\BibitemShut {NoStop}%
\bibitem [{\citenamefont {Iadecola}\ and\ \citenamefont
  {Schecter}(2020)}]{iadecola2020quantum}%
  \BibitemOpen
  \bibfield  {author} {\bibinfo {author} {\bibfnamefont {Thomas}\ \bibnamefont
  {Iadecola}}\ and\ \bibinfo {author} {\bibfnamefont {Michael}\ \bibnamefont
  {Schecter}},\ }\bibfield  {title} {\enquote {\bibinfo {title} {Quantum
  many-body scar states with emergent kinetic constraints and
  finite-entanglement revivals},}\ }\href {\doibase
  10.1103/PhysRevB.101.024306} {\bibfield  {journal} {\bibinfo  {journal}
  {Phys. Rev. B}\ }\textbf {\bibinfo {volume} {101}},\ \bibinfo {pages}
  {024306} (\bibinfo {year} {2020})}\BibitemShut {NoStop}%
\bibitem [{\citenamefont {Langlett}\ \emph {et~al.}(2022)\citenamefont
  {Langlett}, \citenamefont {Yang}, \citenamefont {Wildeboer}, \citenamefont
  {Gorshkov}, \citenamefont {Iadecola},\ and\ \citenamefont
  {Xu}}]{langlett2022rainbow}%
  \BibitemOpen
  \bibfield  {author} {\bibinfo {author} {\bibfnamefont {Christopher~M.}\
  \bibnamefont {Langlett}}, \bibinfo {author} {\bibfnamefont {Zhi-Cheng}\
  \bibnamefont {Yang}}, \bibinfo {author} {\bibfnamefont {Julia}\ \bibnamefont
  {Wildeboer}}, \bibinfo {author} {\bibfnamefont {Alexey~V.}\ \bibnamefont
  {Gorshkov}}, \bibinfo {author} {\bibfnamefont {Thomas}\ \bibnamefont
  {Iadecola}}, \ and\ \bibinfo {author} {\bibfnamefont {Shenglong}\
  \bibnamefont {Xu}},\ }\bibfield  {title} {\enquote {\bibinfo {title} {Rainbow
  scars: From area to volume law},}\ }\href {\doibase
  10.1103/PhysRevB.105.L060301} {\bibfield  {journal} {\bibinfo  {journal}
  {Phys. Rev. B}\ }\textbf {\bibinfo {volume} {105}},\ \bibinfo {pages}
  {L060301} (\bibinfo {year} {2022})}\BibitemShut {NoStop}%
\bibitem [{\citenamefont {Biswas}\ \emph {et~al.}(2022)\citenamefont {Biswas},
  \citenamefont {Banerjee},\ and\ \citenamefont {Sen}}]{biswas2022scars}%
  \BibitemOpen
  \bibfield  {author} {\bibinfo {author} {\bibfnamefont {Saptarshi}\
  \bibnamefont {Biswas}}, \bibinfo {author} {\bibfnamefont {Debasish}\
  \bibnamefont {Banerjee}}, \ and\ \bibinfo {author} {\bibfnamefont {Arnab}\
  \bibnamefont {Sen}},\ }\bibfield  {title} {\enquote {\bibinfo {title} {{Scars
  from protected zero modes and beyond in $U(1)$ quantum link and quantum dimer
  models}},}\ }\href {\doibase 10.21468/SciPostPhys.12.5.148} {\bibfield
  {journal} {\bibinfo  {journal} {SciPost Phys.}\ }\textbf {\bibinfo {volume}
  {12}},\ \bibinfo {pages} {148} (\bibinfo {year} {2022})}\BibitemShut
  {NoStop}%
\bibitem [{\citenamefont {Banerjee}\ and\ \citenamefont
  {Sen}(2021)}]{banerjee2021quantum}%
  \BibitemOpen
  \bibfield  {author} {\bibinfo {author} {\bibfnamefont {Debasish}\
  \bibnamefont {Banerjee}}\ and\ \bibinfo {author} {\bibfnamefont {Arnab}\
  \bibnamefont {Sen}},\ }\bibfield  {title} {\enquote {\bibinfo {title}
  {Quantum scars from zero modes in an abelian lattice gauge theory on
  ladders},}\ }\href {\doibase 10.1103/PhysRevLett.126.220601} {\bibfield
  {journal} {\bibinfo  {journal} {Phys. Rev. Lett.}\ }\textbf {\bibinfo
  {volume} {126}},\ \bibinfo {pages} {220601} (\bibinfo {year}
  {2021})}\BibitemShut {NoStop}%
\bibitem [{\citenamefont {Mizuta}\ \emph {et~al.}(2020)\citenamefont {Mizuta},
  \citenamefont {Takasan},\ and\ \citenamefont {Kawakami}}]{mizuta2020exact}%
  \BibitemOpen
  \bibfield  {author} {\bibinfo {author} {\bibfnamefont {Kaoru}\ \bibnamefont
  {Mizuta}}, \bibinfo {author} {\bibfnamefont {Kazuaki}\ \bibnamefont
  {Takasan}}, \ and\ \bibinfo {author} {\bibfnamefont {Norio}\ \bibnamefont
  {Kawakami}},\ }\bibfield  {title} {\enquote {\bibinfo {title} {Exact floquet
  quantum many-body scars under rydberg blockade},}\ }\href {\doibase
  10.1103/PhysRevResearch.2.033284} {\bibfield  {journal} {\bibinfo  {journal}
  {Phys. Rev. Research}\ }\textbf {\bibinfo {volume} {2}},\ \bibinfo {pages}
  {033284} (\bibinfo {year} {2020})}\BibitemShut {NoStop}%
\bibitem [{\citenamefont {Sugiura}\ \emph
  {et~al.}(2021{\natexlab{b}})\citenamefont {Sugiura}, \citenamefont
  {Kuwahara},\ and\ \citenamefont {Saito}}]{sugiura2021manybody}%
  \BibitemOpen
  \bibfield  {author} {\bibinfo {author} {\bibfnamefont {Sho}\ \bibnamefont
  {Sugiura}}, \bibinfo {author} {\bibfnamefont {Tomotaka}\ \bibnamefont
  {Kuwahara}}, \ and\ \bibinfo {author} {\bibfnamefont {Keiji}\ \bibnamefont
  {Saito}},\ }\bibfield  {title} {\enquote {\bibinfo {title} {Many-body scar
  state intrinsic to periodically driven system},}\ }\href {\doibase
  10.1103/PhysRevResearch.3.L012010} {\bibfield  {journal} {\bibinfo  {journal}
  {Phys. Rev. Research}\ }\textbf {\bibinfo {volume} {3}},\ \bibinfo {pages}
  {L012010} (\bibinfo {year} {2021}{\natexlab{b}})}\BibitemShut {NoStop}%
\bibitem [{\citenamefont {Iadecola}\ and\ \citenamefont
  {Vijay}(2020)}]{iadecola2020nonergodic}%
  \BibitemOpen
  \bibfield  {author} {\bibinfo {author} {\bibfnamefont {Thomas}\ \bibnamefont
  {Iadecola}}\ and\ \bibinfo {author} {\bibfnamefont {Sagar}\ \bibnamefont
  {Vijay}},\ }\bibfield  {title} {\enquote {\bibinfo {title} {Nonergodic
  quantum dynamics from deformations of classical cellular automata},}\ }\href
  {\doibase 10.1103/PhysRevB.102.180302} {\bibfield  {journal} {\bibinfo
  {journal} {Phys. Rev. B}\ }\textbf {\bibinfo {volume} {102}},\ \bibinfo
  {pages} {180302} (\bibinfo {year} {2020})}\BibitemShut {NoStop}%
\end{thebibliography}

\begin{thebibliography}{8}%
\makeatletter
\providecommand \@ifxundefined [1]{%
 \@ifx{#1\undefined}
}%
\providecommand \@ifnum [1]{%
 \ifnum #1\expandafter \@firstoftwo
 \else \expandafter \@secondoftwo
 \fi
}%
\providecommand \@ifx [1]{%
 \ifx #1\expandafter \@firstoftwo
 \else \expandafter \@secondoftwo
 \fi
}%
\providecommand \natexlab [1]{#1}%
\providecommand \enquote  [1]{``#1''}%
\providecommand \bibnamefont  [1]{#1}%
\providecommand \bibfnamefont [1]{#1}%
\providecommand \citenamefont [1]{#1}%
\providecommand \href@noop [0]{\@secondoftwo}%
\providecommand \href [0]{\begingroup \@sanitize@url \@href}%
\providecommand \@href[1]{\@@startlink{#1}\@@href}%
\providecommand \@@href[1]{\endgroup#1\@@endlink}%
\providecommand \@sanitize@url [0]{\catcode `\\12\catcode `\$12\catcode
  `\&12\catcode `\#12\catcode `\^12\catcode `\_12\catcode `\%12\relax}%
\providecommand \@@startlink[1]{}%
\providecommand \@@endlink[0]{}%
\providecommand \url  [0]{\begingroup\@sanitize@url \@url }%
\providecommand \@url [1]{\endgroup\@href {#1}{\urlprefix }}%
\providecommand \urlprefix  [0]{URL }%
\providecommand \Eprint [0]{\href }%
\providecommand \doibase [0]{http://dx.doi.org/}%
\providecommand \selectlanguage [0]{\@gobble}%
\providecommand \bibinfo  [0]{\@secondoftwo}%
\providecommand \bibfield  [0]{\@secondoftwo}%
\providecommand \translation [1]{[#1]}%
\providecommand \BibitemOpen [0]{}%
\providecommand \bibitemStop [0]{}%
\providecommand \bibitemNoStop [0]{.\EOS\space}%
\providecommand \EOS [0]{\spacefactor3000\relax}%
\providecommand \BibitemShut  [1]{\csname bibitem#1\endcsname}%
\let\auto@bib@innerbib\@empty
\bibitem [{\citenamefont {Vafek}\ \emph {et~al.}(2017)\citenamefont {Vafek},
  \citenamefont {Regnault},\ and\ \citenamefont
  {Bernevig}}]{vafek2017entanglement}%
  \BibitemOpen
  \bibfield  {author} {\bibinfo {author} {\bibfnamefont {O.}~\bibnamefont
  {Vafek}}, \bibinfo {author} {\bibfnamefont {N.}~\bibnamefont {Regnault}}, \
  and\ \bibinfo {author} {\bibfnamefont {B.~A.}\ \bibnamefont {Bernevig}},\
  }\href {\doibase 10.21468/SciPostPhys.3.6.043} {\bibfield  {journal}
  {\bibinfo  {journal} {SciPost Phys.}\ }\textbf {\bibinfo {volume} {3}},\
  \bibinfo {pages} {043} (\bibinfo {year} {2017})}\BibitemShut {NoStop}%
\bibitem [{\citenamefont {Schecter}\ and\ \citenamefont
  {Iadecola}(2019)}]{schecter2019weak}%
  \BibitemOpen
  \bibfield  {author} {\bibinfo {author} {\bibfnamefont {M.}~\bibnamefont
  {Schecter}}\ and\ \bibinfo {author} {\bibfnamefont {T.}~\bibnamefont
  {Iadecola}},\ }\href {\doibase 10.1103/PhysRevLett.123.147201} {\bibfield
  {journal} {\bibinfo  {journal} {Phys. Rev. Lett.}\ }\textbf {\bibinfo
  {volume} {123}},\ \bibinfo {pages} {147201} (\bibinfo {year}
  {2019})}\BibitemShut {NoStop}%
\bibitem [{\citenamefont {Lin}\ \emph {et~al.}(2020)\citenamefont {Lin},
  \citenamefont {Chandran},\ and\ \citenamefont {Motrunich}}]{lin2020slow}%
  \BibitemOpen
  \bibfield  {author} {\bibinfo {author} {\bibfnamefont {C.-J.}\ \bibnamefont
  {Lin}}, \bibinfo {author} {\bibfnamefont {A.}~\bibnamefont {Chandran}}, \
  and\ \bibinfo {author} {\bibfnamefont {O.~I.}\ \bibnamefont {Motrunich}},\
  }\href {\doibase 10.1103/PhysRevResearch.2.033044} {\bibfield  {journal}
  {\bibinfo  {journal} {Phys. Rev. Res.}\ }\textbf {\bibinfo {volume} {2}},\
  \bibinfo {pages} {033044} (\bibinfo {year} {2020})}\BibitemShut {NoStop}%
\bibitem [{\citenamefont {Mori}\ and\ \citenamefont
  {Shiraishi}(2017)}]{mori2017thermalization}%
  \BibitemOpen
  \bibfield  {author} {\bibinfo {author} {\bibfnamefont {T.}~\bibnamefont
  {Mori}}\ and\ \bibinfo {author} {\bibfnamefont {N.}~\bibnamefont
  {Shiraishi}},\ }\href {\doibase 10.1103/PhysRevE.96.022153} {\bibfield
  {journal} {\bibinfo  {journal} {Phys. Rev. E}\ }\textbf {\bibinfo {volume}
  {96}},\ \bibinfo {pages} {022153} (\bibinfo {year} {2017})}\BibitemShut
  {NoStop}%
\bibitem [{\citenamefont {Wilming}\ \emph {et~al.}(2018)\citenamefont
  {Wilming}, \citenamefont {de~Oliveira}, \citenamefont {Short},\ and\
  \citenamefont {Eisert}}]{wilming2018equilibration}%
  \BibitemOpen
  \bibfield  {author} {\bibinfo {author} {\bibfnamefont {H.}~\bibnamefont
  {Wilming}}, \bibinfo {author} {\bibfnamefont {T.~R.}\ \bibnamefont
  {de~Oliveira}}, \bibinfo {author} {\bibfnamefont {A.~J.}\ \bibnamefont
  {Short}}, \ and\ \bibinfo {author} {\bibfnamefont {J.}~\bibnamefont
  {Eisert}},\ }\enquote {\bibinfo {title} {Equilibration times in closed
  quantum many-body systems},}\ in\ \href {\doibase
  10.1007/978-3-319-99046-0_18} {\emph {\bibinfo {booktitle} {Thermodynamics in
  the Quantum Regime: Fundamental Aspects and New Directions}}},\ \bibinfo
  {editor} {edited by\ \bibinfo {editor} {\bibfnamefont {F.}~\bibnamefont
  {Binder}}, \bibinfo {editor} {\bibfnamefont {L.~A.}\ \bibnamefont {Correa}},
  \bibinfo {editor} {\bibfnamefont {C.}~\bibnamefont {Gogolin}}, \bibinfo
  {editor} {\bibfnamefont {J.}~\bibnamefont {Anders}}, \ and\ \bibinfo {editor}
  {\bibfnamefont {G.}~\bibnamefont {Adesso}}}\ (\bibinfo  {publisher} {Springer
  International Publishing},\ \bibinfo {address} {Cham},\ \bibinfo {year}
  {2018})\ pp.\ \bibinfo {pages} {435--455}\BibitemShut {NoStop}%
\bibitem [{\citenamefont {Campos~Venuti}\ and\ \citenamefont
  {Zanardi}(2010)}]{camposvenuti2010unitary}%
  \BibitemOpen
  \bibfield  {author} {\bibinfo {author} {\bibfnamefont {L.}~\bibnamefont
  {Campos~Venuti}}\ and\ \bibinfo {author} {\bibfnamefont {P.}~\bibnamefont
  {Zanardi}},\ }\href {\doibase 10.1103/PhysRevA.81.022113} {\bibfield
  {journal} {\bibinfo  {journal} {Phys. Rev. A}\ }\textbf {\bibinfo {volume}
  {81}},\ \bibinfo {pages} {022113} (\bibinfo {year} {2010})}\BibitemShut
  {NoStop}%
\bibitem [{\citenamefont {Mark}\ and\ \citenamefont
  {Motrunich}(2020)}]{mark2020eta}%
  \BibitemOpen
  \bibfield  {author} {\bibinfo {author} {\bibfnamefont {D.~K.}\ \bibnamefont
  {Mark}}\ and\ \bibinfo {author} {\bibfnamefont {O.~I.}\ \bibnamefont
  {Motrunich}},\ }\href {\doibase 10.1103/PhysRevB.102.075132} {\bibfield
  {journal} {\bibinfo  {journal} {Phys. Rev. B}\ }\textbf {\bibinfo {volume}
  {102}},\ \bibinfo {pages} {075132} (\bibinfo {year} {2020})}\BibitemShut
  {NoStop}%
\bibitem [{\citenamefont {Wildeboer}\ \emph {et~al.}(2022)\citenamefont
  {Wildeboer}, \citenamefont {Langlett}, \citenamefont {Yang}, \citenamefont
  {Gorshkov}, \citenamefont {Iadecola},\ and\ \citenamefont
  {Xu}}]{wildeboer2022quantum}%
  \BibitemOpen
  \bibfield  {author} {\bibinfo {author} {\bibfnamefont {J.}~\bibnamefont
  {Wildeboer}}, \bibinfo {author} {\bibfnamefont {C.~M.}\ \bibnamefont
  {Langlett}}, \bibinfo {author} {\bibfnamefont {Z.-C.}\ \bibnamefont {Yang}},
  \bibinfo {author} {\bibfnamefont {A.~V.}\ \bibnamefont {Gorshkov}}, \bibinfo
  {author} {\bibfnamefont {T.}~\bibnamefont {Iadecola}}, \ and\ \bibinfo
  {author} {\bibfnamefont {S.}~\bibnamefont {Xu}},\ }\href {\doibase
  10.1103/PhysRevB.106.205142} {\bibfield  {journal} {\bibinfo  {journal}
  {Phys. Rev. B}\ }\textbf {\bibinfo {volume} {106}},\ \bibinfo {pages}
  {205142} (\bibinfo {year} {2022})}\BibitemShut {NoStop}%
\end{thebibliography}
\end{document}